\documentclass{article}

\usepackage[utf8]{inputenc}
\usepackage[sort&compress,numbers]{natbib}
\usepackage{amsmath}
\usepackage{mathtools}
\usepackage{amssymb}
\usepackage{hyperref}
\usepackage{cleveref}
\usepackage{caption}
\usepackage{subcaption}
\usepackage{soul}
\usepackage{fullpage}
\usepackage{afterpage}
\usepackage{graphicx}
\usepackage{fancyvrb}
\usepackage{booktabs}
\usepackage{placeins}
\usepackage{soul}
\usepackage{xcolor}
\usepackage[a4paper, total={6.5in, 9.5in}]{geometry}
\usepackage{setspace}
\usepackage[version=4]{mhchem}
\usepackage[titletoc,title]{appendix}
\usepackage{makecell}
\usepackage{authblk}
 
\newcommand{\de}{\mbox{d}}
\newcommand{\area}{\mathcal{A}}
\newcommand{\rad}{\mathcal{R}}
\newcommand{\vol}{\mathcal{V}}
\newcommand{\helm}{\mbox{A}}
\newcommand{\energy}{\mbox{E}}
\newcommand{\strain}{u}
\newcommand{\press}{P}

\newcommand{\helmA}{\widetilde{\helm}}
\newcommand{\etal}{\textit{ et al. }}

\title{Solid-liquid interfacial free energy from computer simulations: Challenges and recent advances}
\author[1]{Nicodemo Di Pasquale \thanks{corresponding author: nicodemo.dipasquale@unibo.it}}
\author[2]{Jesús Algaba}
\author[3]{Pablo Montero de Hijes}
\author[10]{Ignacio Sanchez-Burgos}
\author[4,5]{Andres R. Tejedor}
\author[6]{Stephen R. Yeandel}
\author[2]{Felipe J. Blas}
\author[7]{Ruslan L. Davidchack}
\author[4,5]{Jorge R. Espinosa}
\author[6]{Colin L. Freeman}
\author[6]{John H. Harding}
\author[8]{Brian B. Laird}
\author[5]{Eduardo Sanz}
\author[5]{Carlos Vega}
\author[9]{Lorenzo Rovigatti}

\affil[1]{{\small Department of Industrial Chemistry ``T. Montanari'', Università di Bologna, via Gobetti 85, 40129 Bologna}}

\affil[2]{{\small Laboratorio de Simulación Molecular y Química Computacional, CIQSO-Centro de Investigación en Química Sostenible and Departamento de Ciencias Integradas, Universidad de Huelva, 21006 Huelva, Spain}}

\affil[3]{{\small Faculty of Physics, University of Vienna, A-1090 Vienna, Austria}}

\affil[4]{{\small Yusuf Hamied Department of Chemistry, University of Cambridge, Lensfield Road, Cambridge, CB2 1EW, UK}}

\affil[5]{{\small Department of Physical Chemistry, Complutense University of Madrid, Avenida Complutense, Madrid, 28040, Spain }}

\affil[6]{{\small Department of Materials Science and Engineering, University of Sheffield, Sheffield, S1 3JD, UK}}

\affil[7]{{\small School of Computing and Mathematical Sciences, University of Leicester, Leicester, LE1 7RH, UK}}

\affil[8]{{\small Department of Chemistry, University of Kansas, Lawrence, KS 66045, USA}}

\affil[9]{{\small Physics Department, Sapienza University of Rome, P.le A. Moro 5, 00185, Rome, Italy}}

\affil[10]{{\small Maxwell Centre, Cavendish Laboratory, Department of Physics, University of Cambridge, J J Thomson Avenue, Cambridge CB3 0HE, United Kingdom}}

\date{}

\begin{document}

\maketitle

\begin{abstract}
\noindent
    The theory of interfacial properties in liquid-liquid or liquid-vapour systems is nearly 200 years old, and includes the contributions of great personalities like Young and Gibbs. The advent of computational tools has greatly advanced the field, mainly through the use of Molecular Dynamics simulations, which allows the precise study of interfaces between different systems at the atomistic level. Despite the successes and advances in the theory of interfacial phenomena for liquid-liquid systems, the study of solid-liquid interfaces still remains a challenge both theoretically and experimentally. The main reason why the treatment of solid-liquid systems has fallen behind that of liquid-liquid systems is that there are complications that arise whenever an interface involving solid systems is considered. These complications involve both the theory of the solid-liquid interface and the calculations of related properties using Molecular Dynamics simulations. An example of the former is that, contrary to the liquid-liquid case, the interfacial properties of solids depend on the lattice orientation. The main complications in the calculation of interfacial properties arise from the fact that for solids the ``mechanical route'', which requires a single calculation to determine the distribution of stresses in the sample considered, cannot be used. To overcome this problem, several numerical approaches were proposed to tackle the calculation of interfacial properties in solid-liquid systems. The main purpose of this review is to provide an overview of these different methodologies and to discuss their strengths and weaknesses. We classify these methodologies into two main groups: direct and indirect methods. Direct methods are those that are able to calculate directly (although in a more convoluted way) the properties of interfaces, while in indirect approaches the properties of the interface are not the primary result of the simulations. In order to provide a comprehensive understanding of solid-liquid interfaces, we also included a discussion on the origin of the difficulties in considering solid interfaces from a thermodynamic point of view. In the second part of the review, we discuss two key related topics: nucleation theory and curved interfaces. They both represent an important problem in the study of interfaces and in particular in the context of solid-liquid ones for which the research is still extremely active.
\end{abstract}

\newpage

\tableofcontents

\section{Introduction}

One of the famous quotes attributed to Pauli describes a surface in a really unflattering manner, portraying it as aligned with chaos and darkness, unlike the order and rationality of the bulk \footnote{As reported in \citep{Brantley1999}, the quote reads ``\textit{God made the bulk, surfaces were invented by the devil}''}. Although it is true that surfaces can be difficult to deal with (and we will describe why in this review), it is also true that they are often more interesting to describe and study than the bulk because several interesting phenomena occur only at the interface between two (or more) phases. This is especially true for solid-liquid interfaces. The detailed knowledge of the structure and thermodynamic properties of interfaces that separate a solid and other coexisting phases is the basis for many physical phenomena and technological processes, making it a matter of primary interest in several different fields. We dedicate the first part of this review to describing these problems and applications and, as will be made clear in the rest of this work, to showing why so much effort has been dedicated to the creation of reliable methods for the determination of interfacial properties. A central challenge (and the theme of this review) is the accurate calculation of the interfacial free energy (IFE).

Among the physical phenomena dependent on the interfacial properties, the formation of a new (solid) phase from a liquid one has several different and important applications. The formation of a new solid phase from a liquid describes two different but related phenomena: in \textit{freezing} (or \textit{solidification}) we have the formation of a new solid phase from its melt, whereas in \textit{crystallization} the new solid phase is created from a solution in which the solid is dissolved. These two processes are related by the fact that the physical process allowing the formation of the new phase is \textit{nucleation} \citep{Hoose2012,Kalikmanov2012,Sosso2016,Sleutel2014} on which we will have an extensive discussion in the subsequent sections.

One of the most studied systems undergoing freezing is probably water \citep{Zhang2016,espinosa2016interfacial,Montero2023,Espinosa2016,Zhang2018Ice}, not only because of its theoretical importance, but also for its wide spectrum of applications. For example, a fundamental understanding of ice formation is essential, as it plays a crucial role in atmospheric science for the accurate representation of world climate \citep{Knopf2023,Hakimian2021}, and for the design of functional materials, such as materials with anti-icing properties \citep{Huang2022,Shen2019}. Such materials will have a wide spectrum of applications, ranging from increasing safety in aviation \citep{Cao2018}, where ice formation on wings is one of the main safety concerns, increasing the performance of wind turbines in cold climates \citep{Dalili2009,Kraj2010}, to reducing damage on overhead power lines \citep{Ducloux2018,Jeong2018}.

In solidification science, solid-liquid IFE controls the formation of solidification microstructures in metallic materials \citep{Wang2020Metal}, from which, in turn, the quality of the final product in casting \citep{Asta2009} depends. In particular, the dendritic growth velocity depends on the anisotropy in solid-liquid IFE \citep{Amar1993,Langer1980}. This anisotropy will be discussed in more detail in the next sections, and it is important in several systems, \textit{e.g.}, the Al-Cu alloy \citep{Liu2001AlCu,Azizi2022} and the solid-liquid coexistence in Ni, Cu, Al \citep{Asadi2015} and Ti \citep{Kavousi2019}. In fact, when a solid is in contact with a liquid, the value of the IFE depends on the particular orientation of the crystal phase in contact with the liquid. A different arrangement of the atoms in the solid will result in a different arrangement of the layers of the liquid closer to the solid \citep{Spaepen1975}, making the surface energy dependent on the orientation of the crystal structure. This effect highlights one of the difficulties that concern solid interfaces in general (not only in the context of the solidification process), which we will discuss in the next section: when solids are involved, the IFE is not a single quantity but assumes different values depending on the particular structure of the solid-liquid interface.

In the process of formation of a new solid phase from a solution, the role of IFE between the solid and liquid phases in the nucleation process is well established \citep{Vinet2002} in different systems, from polymers \citep{Cole1978,Long1995} to biomineralization \citep{Giuffre2013}. The impact of the surface properties on the formation of the new solid phase through nucleation is related not only to the formation of the new phase but also which polymorph will be formed \citep{Teychene2008,beneduce2024simple}. A given substance exhibits polymorphism if there is more than one solid crystalline arrangement of the molecules of that substance. The prediction of which polymorph can be formed is important for biological processes \citep{Giuffre2013} but also in the crystallisation of pharmaceutical products where the formation of the right polymorphs of drugs is essential for their effectiveness and safety \citep{Thakore2020,Bhatia2018}.

The importance of surface properties in the pharmaceutical industry is not limited to the prediction of the most stable product in drug production but encompasses several key aspects. In fact, the behaviour of such industrial products, (including both excipients and active pharmaceutical ingredients) with respect to binder-drug adhesion \citep{Begat2004}, granulation performance \citep{Thielmann2007,Zhang2002,Thapa2019}, powder flow, and compaction \citep{Li2004,Ahfat1997} can be related to the surface properties \citep{Ho2010,Hadjittofis2018,Klitou2022,Jefferson2011}.

An application that has become extremely important in recent times, in which interface properties play a pivotal role, is the design of next generation graphene-based energy storage devices such as electrochemical double layer (super)capacitors (EDLC)\citep{Simon2008, Wang2021, Elliott2022rev}.  Compared to standard energy storage devices, the new systems (using materials such as graphene \citep{Wang2009capacitors,Liu2010,Yu2010,Zhang2010}, porous activated carbon \citep{Zhu2011}  and carbon nanotubes \citep{An2001,Yang2019}) have higher charge storage capacities, favourable specific energy-to-power ratios (owing to rapid charge-discharge cycling \citep{Liu2010} controlled by changes of an applied potential) and lifetimes that can reach millions of cycles \citep{Zhu2011}. Energy storage in graphene-based supercapacitors is based on a reversible non-Faradaic physisorption of ions in the electrical double layer \citep{DiPasquale2023}. With its high surface area, graphene can in principle guarantee a higher capacitance than amorphous carbon-based electrodes. However, the area offered by the electrode is not the only parameter that enters the quantification of the capacitance. In order to be successful, the electrochemically active surface area of the electrode should be easily accessible by the electrolyte. This, in turn, depends on the ability of the electrolyte to wet the electrode surface. This is another manifestation of the solid-liquid IFE. In addition, the ability of the electrolyte to wet the graphene surface changes as a function of the potential applied to the electrode, a phenomenon known as ``electrowetting'' \citep{Mugele2005}. To capture the behaviour of graphene in contact with electrolytes, a more detailed account of electrostatic interactions is needed \citep{DiPasquale2023}. The development of such improved descriptions of the interactions requires quantum mechanical/molecular dynamics (QMMD) models \citep{Elliott2020} and more accurate force-fields, possibly based on machine learning \citep{DiPasquale2021NN}. The combination of such advanced descriptions with the methodologies presented here will surely be at the forefront of the investigation of such systems.

Another interesting area where knowledge of interface properties in solid-liquid systems is essential is thermal transport across solid-liquid interfaces, in particular when the size of the system considered is microscopic and therefore the interface/volume ratio of the system involved becomes large \citep{Jiang2008}. The control of thermal transport at the nanoscale is important for medical applications, water purification, and microelectronics \citep{Cahill2014} (see \citep{Ramos2024} and the references therein for a more detailed account of the different applications). The interfacial thermal conductance is related to the affinity between the solid and the liquid at the interface. This effect was observed for the first time by \citet{Kapitza1941}, and is now known by its name -``Kapitza resistance'' (although it refers to conductance rather than resistance) and state that the stronger the attraction between the two phases is the lower the thermal resistance becomes \citep{Ramos2024}. A measure of the strength of this attraction is the wettability of the solid interface with the liquid and previous works show that there is a direct relation between wettability and thermal conductance \citep{Shenogina2009,Ge2006}. In turn, as we discussed in the case of electrochemical devices, we can consider the wettability as just another manifestation of the IFE, and the ability to obtain a reliable value of this property in a variety of systems becomes essential. 

Until this moment, we have talked about ``surfaces'' and ``interfaces'' without providing a proper definition, appealing instead to their common meaning. From now on, we will more rigorously define these concepts and put them into the context of thermodynamic theory for quantitative analysis and discussion. Indeed, the analysis of the properties of interfaces is deeply rooted in thermodynamics, as shown in the pioneering work of J. W. Gibbs, one of the earliest contributors to this topic, but also one of the founders of modern thermodynamics. In his work \citep{Gibbs1957}, he defined the interface between two different phases as a zero-width plane (later called  ``Gibbs dividing plane''), to which he ascribed the excess of the thermodynamic quantities which characterise the presence of an interface between two phases. One of these quantities is the IFE, $\gamma$, which represents the reversible work required to create a unit area of interface at the coexistence conditions for the two phases. We introduce here some of the notation that will be used throughout the rest of the work, by explicitly stating what are the two phases in contact through the interface. The liquid can be either the melt corresponding to the solid phase or a solution in which the solid phase is the solute. Where both cases are applicable, we shall refer to the IFE as $\gamma_{sl}$. When we consider only the solid-melt interface, we shall use $\gamma_{sm}$, and if we are considering only the solution-solid interface we shall use $\gamma_{sx}$.

The Gibbs approach was later generalized by Cahn, with a formulation which avoids the need to locate the position of the dividing surface \citep{Cahn1998}. As we shall discuss in the following sections, the definition of interfacial properties, while straightforward for liquid-liquid systems, involves some subtleties when the system contains a solid interface. Such differences between liquid-liquid and solid-liquid systems are the main reasons for the different way to approach the calculation of interfacial properties using Molecular Dynamics simulations in the two kinds of systems. Whereas in liquid-liquid systems it is possible to leverage rather simple relations to calculate the interface properties using MD simulations (such as using the stress within the system as a proxy for the energy required to create the interface) such shortcuts are not possible for solid-liquid interfaces. In the latter case, more complicated calculations are needed, as one often has to resort to using very basic thermodynamic relations. For instance, the energy needed to create an interface is calculated by literally creating a new interface in a simulation box, which is an operation that is much more complicated than just calculating the force acting between atoms in the system.

This review is organized as follows: we will first introduce the surface-specific quantities in solid-liquid systems starting from their thermodynamic definitions. We provide a brief account of the reasons why solid-liquid interfaces are different from liquid-liquid ones. From this we will move on to the presentation of the different methods devised in the literature to access the interface properties for solid-liquid systems. Here, we choose to partition the different techniques into two main groups, which we label as direct and non-direct methods, based on the way the interfacial properties are computed. This is the most extensive part of the review. Although such methodologies have been applied to several different systems, there are some systems which can be considered as ``benchmarks'' against which the results of any new extension of existing methodologies or the development of new ones should be compared: the hard-sphere and Lennard-Jones models. For this reason, we include a detailed description of such systems, along with a detailed comparison of the different approaches against the benchmark. We then discuss in some detail the calculation of interfacial properties for more realistic systems, namely water and hydrates. We selected these systems both for their importance and also because they exemplify the kind of complications one has to face in considering a realistic system for the description of the IFE. We then describe the importance of determining surface properties in the theory of nucleation and devote a section to curved interfaces. These last two sections (\cref{sec:curved,sec:nucleation}) must be considered ``open'', in the sense that, despite being long-known problems (the first appearance of the problem of curved interface is in the work of Young in 1805 \citep{young1805thesis}, and the theory of nucleation is now a century old \citep{Volmer1926,Becker1935}), they have not reached the maturity which we can find, \textit{e.g.}, in the study of liquid-liquid interfaces (a problem which was completely solved by Gibbs well over a century ago \citep{Gibbs1957}). In the last part, we draw some conclusions and give some perspective on future applications and ideas related to the methodologies described here.

\section{Challenges in characterising the physics of solid-liquid interfaces}
Whereas the IFE for liquid-vapour and liquid-liquid interfaces are well known quantities, characterised both theoretically \citep{Ghoufi2016,Andersson2014} and experimentally \citep{Janczuk1993,Drelich2002}, this is not the case for solid-liquid interfaces. One of the most important reasons for this difference comes from the fact that, unlike liquid-vapour and liquid-liquid interfaces, the IFE of interfaces involving solids is anisotropic, as it depends on the orientation of the interface with respect to the solid lattice \citep{Davidchack2003,Asta2002}. In particular, the IFE for a particular solid-liquid system has a different value for $\gamma_{sl}$ for the orientations of the crystal lattice that are not related by symmetry (for example, see the results reported in \citep{Davidchack2003,Asta2002}). This fact has a subtle consequence: in solid-liquid systems, the interfacial stress and the IFE are two distinct concepts. 

\subsection{Experimental Methods to determine the solid-liquid IFE} \label{sec:Exp}
The estimation of the IFE associated to solid-liquid interfaces is extremely challenging not only from a theoretical/numerical point of view but also in experiments, and, in this section, we will give a brief overview of experimental methodologies to determine the solid-liquid IFE. 

Although we refer to the literature for a more detailed account of different techniques (see \citep{Kumar2024} for a full discussion of such methodologies and \citep{Zaera2012} for a discussion of experimental techniques to probe the structure of solid-liquid interfaces), we will highlight some of the major difficulties in the experimental determination of the IFE for solid-liquid interfaces. The reader is warned that in some of the early work (both simulations and experiments), there is some ambiguity about what is being compared to what. The common assumption that the entropic contribution to the interfacial free energy is negligible means that frequent reference was made to ``surface energies'' or ``interfacial energies'', without clarifying whether those quantities estimated the free energy, the enthalpy or the configurational (potential) energy per unit area (see, \textit{e.g.} \citep{Radha2013} Tables 4 \& 5).

One of the most widely used approaches to determine the solid-liquid IFE is based on the Young equation \cite{young1805thesis} and consists of measuring the angle that a liquid droplet makes with respect to a solid surface with which it is in contact \citep{Kwok1999}. Although the idea itself is relatively straightforward, the measurement of the droplet angle is plagued by several issues, either kinetic (evaporation, vapour adsorption, swelling) or thermodynamic (because the surface on which the droplet is located has to be flat and chemically homogeneous down to the molecular-scale, and gravity must not disturb the solid–liquid–vapour system). If these conditions are not met, the departure from ideality generates hysteresis between the direct process, wetting, and its inverse, i.e., surface dewetting \citep{Cha2019}. This hysteresis results in non-unique measurements of the contact angle, making the Young equation inapplicable for the calculation of the solid-liquid IFE \citep{Bruel2019} and special care must be taken to minimise this effect during experiments (see, e.g., \citep{Calvimontes2017}). Moreover, new methodologies must be developed, e.g., to deal with rough surfaces \citep{Sarkar2023}.

Another common way to determine IFEs is through the use of measurements of crystal nucleation rates in supercooled fluids, from which the IFE can be determined by leveraging the Classical Nucleation Theory (CNT) \cite{Turnbull1950,volmer1926keimbildung,becker1935kinetische,Ickes2015} -  see also \cref{sec:nucleation}. As an example of the difficulty of measuring solid-liquid IFEs, we highlight the case of the ice-water interface, for which there is still no consensus on its experimental value \citep{Ickes2015}, despite the importance of such a system. 

Another method of evaluating the IFE involves the measurement of groove morphology at the intersections between a solid-liquid interfacial boundary and grain boundaries in the solid phase \citep{Gunduz1985}. For the archetypal hard-sphere nucleus-melt, the IFE was inferred from nucleation measurements using the CNT framework \citep{sinn2001solidification}, and later compared to the equilibrium crystal-fluid IFE directly obtained from the analysis of the groove morphology under co-existence conditions \citep{rogers2011measurement}. Interestingly, the observed values ranged from 0.51 $k_BT$ to 0.66 $k_BT$ \citep{luck1963kristallisation}, which agrees reasonably well with simulations and theory \citep{sanchez2021fcc,davidchack2010hard,davidchack2003direct,auer2004numerical} and shows a moderate systematic dependence on metastability. Nevertheless, it is worth noting that this technique usually requires additional alloying elements in the liquid phase to mark the interface position, which has a significant influence on the measurement result. Moreover, this technique cannot resolve the crystal anisotropy of the IFE.

The development of new techniques for the determination of solid-liquid IFEs is an active area of research, with newly proposed techniques for the determination of $\gamma_{sl}$, such as the Sessile Drop Accelerometry \citep{Calvimontes2017} 

The challenges and approximations required on the experimental side to determine the solid-liquid IFE increase the value of numerical methods for determining this quantity from molecular simulations, as evidenced by the amount of research that has been dedicated to the development of the methodologies discussed in the rest of this review. 

\section{Thermodynamics of Interfaces}\label{sec:thermo}
Before starting the analysis of the properties of interfaces, we give a brief discussion of terminology. Here we choose to call $\gamma$ the IFE (in the literature the term \textit{Superficial Work} \citep{Linford1978} is also sometimes used). As we have already introduced before if we need to distinguish further between IFEs, we shall use subscripts; \textit{e.g.}, $\gamma_{sl}$ denotes a solid-liquid interface. $\helmA$  denotes the surface free energy, \textit{i.e.} the free energy of formation of a vacuum/condensed matter interface of a phase, and $f_{ij}$ the interfacial stress -  for the same interface - where $i,j$ refers to the two Cartesian coordinates parallel to the interface. As was already noted \citep{DiPasquale2020} it is better to avoid using ``surface tension'' for the quantity $\gamma$: it is harmless for the case of a (single component) liquid, but it can be a source of misunderstanding every time a solid phase is involved.

Although the particular properties of interfaces have been known since the times of Young and Laplace, Gibbs was the first to offer a detailed and quantitative analysis of such systems in his monumental work \textit{On the Equilibrium of Heterogeneous Substances}\citep{Gibbs1957}. When two masses of materials are put in contact, the total energy, the entropy, and other extensive quantities,  cannot be estimated as a simple algebraic sum of terms referring to each coexisting system considered without any interface. More formally, let us consider a system $\alpha$ in contact with a system $\beta$, and call the energy of the new composite system $\energy$.  Let $\energy^\alpha$ be the energy of the (sub-)system $\alpha$ when it is not in contact with $\beta$ (that is, in Gibbs' words, without a \textit{surface of discontinuities}) and let $\energy^\beta$ have an analogous meaning for subsystem $\beta$. Then, Gibbs reasoned:
\begin{equation}\label{eq:endef}
    \energy \neq \energy^\alpha + \energy^\beta 
\end{equation}
The difference is due to the presence of the interface between the two subsystems: if the interface changes, so does the energy of the composite system. This is different from any process involving other extensive quantities (such as volume or mass). From \cref{eq:endef} we obtain the definition of \textit{Surface Excess} quantities as: 
\begin{align}\label{eq:energy}
    \energy^{XS} = \energy - \energy^\alpha - \energy^\beta     
\end{align}
where we use the superscript $XS$ to indicate an excess property of the interface. Similarly, we can define an excess entropy $S^{XS}$ and an excess of the number of atoms at the interface $N^{XS}$ (see, \textit{e.g.}, \citep{Laird2009GC}). The difference between the extensive quantities in a system with and without an interface is therefore taken into account by the excess surface quantities. In the work of Gibbs, excess quantities are reported as relative to an interface of area $\area$, with a different notation from the one used here and which we include for completeness, as it can be useful in the reading of older works: energy $e=\energy^{XS}/\area$, entropy $\eta=S^{XS}/\area$. Gibbs then assigned such excess quantities to a geometrical surface that separates the two subsystems considered as if they were not in contact with a different phase \citep{Gibbs1957}. Another way of assigning these excess quantities was devised by Guggenheim \citep{Guggenheim1940}. In this approach, we identify a small, but non-zero volume in the neighbourhood of the interface, whose boundaries are located sufficiently far away from the interface such that its properties are homogeneous and equal to the subsystems considered without any interface. The equivalence of these two approaches was shown by Cahn \citep{Cahn1998} and will be briefly discussed in the section dedicated to the Cahn thermodynamic model of interfaces (see \cref{sec:Gibbs-Cahn}).

From the definition of excess properties, we can now define the (specific) Surface Free Energy through the so-called \textit{Fundamental Surface Thermodynamic Equation} \citep{Johnson1959,Herring1953,Linford1978}:
\begin{equation}\label{eq:fundSurf}
    \gamma = \helmA^{XS} - \sum_{k} \mu_k \Gamma_k 
\end{equation}
where we indicate with a $\widetilde{\ }$ the thermodynamic quantities per unit area, and $\Gamma_k = N^{XS}_k/\area$ is the excess of component $k$ at the interface per unit of area $\area$. \Cref{eq:fundSurf} shows an important terminology problem with the different quantities used to describe an interface. The quantity $\gamma$ is equal to the Helmholtz surface free energy $\helmA^{XS}$ only when all the quantities $\Gamma_k$ are zero, which is the case for a one-component system. For multicomponent systems, we should make clear the distinction between them \citep{Johnson1959}. The situation will become even more complicated when we introduce the concept of \textit{interfacial stress}, $f_{ij}$ in the following paragraphs. Although the correct identification of these quantities is essential to be consistent in their analysis and description, the debate on the best names for these quantities started long ago \citep{Frumkin1972,Linford1973}, but has not been settled yet. Unfortunately, this problem has been overlooked in the past because, for a liquid-liquid system, $\gamma = f_{ij} = \helmA^{XS}$\citep{Linford1973}\footnote{We are committing an abuse of notation here. For a liquid-liquid interface, the interfacial stress tensor is isotropic and diagonal. Therefore, it can be described by a single term $f$ which is the one to be equated with $\gamma$.}. In the most general case, multi-component solid-liquid interfaces, this is not true anymore, and therefore, one must give the proper name to each quantity.

Using \cref{eq:fundSurf} and the analysis in \citep{Herring1951} we can work out the implications of such definitions for solid interfaces. In \cref{eq:fundSurf}, $\gamma$ is defined as the surface excess of the grand potential $\Psi=\helm - \sum_k \mu_k N_k$ (also known as the Kramer potential \citep{Linford1978} or the Landau potential), i.e. $\area \gamma \coloneqq \Psi^{XS}$, where the meaning of excess quantities is explained in \cref{eq:energy} \footnote{The grand potential contains the terms related to mechanical work acting on a system and for a bulk phase is  $\Psi^\alpha=-PV^\alpha$, whereas for the surface of discontinuities include the IFE}. 

In a system with an interface, for any reversible transformation at constant $T$ and $\mu$, generated by the action of mechanical forces, the work on the system is equal to the work done on the bulk phase plus a surface work $W^{XS}$ \citep{Herring1951}:
\begin{equation}\label{eq:ws}
    W^{XS} = \Delta \int_{\area} \gamma \de s \;\; (T,\,\mu_i\,\,\mbox{const.})
\end{equation}
where $\Delta$ here represents the difference between the integral calculated on a system of area $\area + \Delta \area$ and a system of area $\area$.
The previous equation implies that, in a transformation where the volume of the bulk phases $\alpha$ and $\beta$ is kept constant, $\Delta \Psi = W^{XS}$. For a liquid-liquid interface, the IFE is completely specified by $T$ and $\mu_i$. Therefore, from \cref{eq:ws} we obtain again the result that the surface work is equal to $\gamma \Delta\area$, where $\Delta \area$ is the change in the interfacial area. However, for an interface involving a solid, the IFE may depend on the particular crystallographic orientation of the interface (see, \textit{e.g.}, results reported in \citep{Broughton1986IV} for a Lennard-Jones system), or the state of strain of the crystal (see, \textit{e.g.}, results reported in \citep{DiPasquale2020} for a Lennard-Jones system). In the latter case, the more general expression \cref{eq:ws} must be considered. Gibbs first pointed out that while the state of tension within liquids caused by surface-tension-related forces can be directly related to the work needed to create the interface, no such simple relation exists when solid phases are considered. In that case, the work needed to create a new surface and the work involved in stretching it may be different and must be distinguished. 

From Gibbs' observation we therefore define two quantities related to an interface: the IFE, defined as the reversible work needed to create a new unit of interface in a system without an interface, and the interfacial stress, $f_{ij}$, which describes the excess stresses occurring in a system with an interface with respect to the bulk \citep{Cahn1979}. The interfacial stress is a 2-dimensional second-order tensor (which means that it can be described by a 2x2 matrix). These seemingly unrelated concepts have a straightforward connection in liquid-liquid (or liquid-vapour) systems: here, due to the rotational symmetry, the interfacial stress tensor can be described by a single number $f$, which is related to the IFE by the simple relation $\gamma=f$.

For interfaces involving solids, the relationship between $\gamma$ and $f$ was first established by Shuttleworth \citep{Shuttleworth1950}, who derived the equation (which now bears his name) from thermodynamic considerations:
\begin{equation}\label{eq:Shutt}
    f_{ij} = \delta_{ij} \gamma + \frac{1}{\area} \frac{\partial \gamma}{\partial u_{ij}}
\end{equation}
where $u_{ij}$ is the $i,j^{\rm th}$ element of the strain tensor of the interface, $\delta_{ij}$ is the Kronecker delta, and the indices $i, j = 1, 2$ refer to the two Cartesian coordinates parallel to the interface. Herring \citep{Herring1951,Herring1999} gave a simple derivation of \cref{eq:Shutt} and we reproduce here its main features. Let us consider a region of interface bounded by walls normal to the interface and let us deform the interface by displacing such walls. In general, the interfacial work reported in \cref{eq:ws} will change as:
\begin{equation}\label{eq:derSh}
    W^{XS} = \gamma\delta\area + \area\delta\gamma = \gamma\area\sum_{i=1,2} \strain_{ii} + \area\sum_{i,j=1,2} \frac{\partial \gamma}{\partial \strain_{ij}}\strain_{ij}.
\end{equation}
For a plane that is normal to the interface, the material on one side of that plane exerts a force on the material on the other side. The excess force (with respect to its value in the bulk) is the \textit{interfacial force} acting across this plane. The \textit{interfacial stress} is now defined as the interfacial force per unit length of the line of intersection of the plane with the dividing surface. Because the orientation of the plane normal to the interface is arbitrary, this force can be expressed as $\sum_{j=1,2}{f_{ij} \hat{\bf n}_j}$, where $\hat{\bf n}_j$ are the components of the unit vector perpendicular to the plane defining the interface. Equating now \cref{eq:derSh} with $\sum_{j=1,2}{f_{ij} \strain_{ij}}$ we obtain \cref{eq:Shutt}.

\Cref{eq:Shutt} has been subject to several criticisms since its formulation, and different authors debated its validity (see \citep{Kramer2007} for a critical review). In  \citep{DiPasquale2020}, the authors derived it from first principles, starting from a statistical mechanics description of a solid-liquid system, and tested it against molecular dynamics simulations of a Lennard-Jones system, showing that \cref{eq:Shutt} matches the results.

 When a new interface is created in a liquid, its orientation does not matter and the energy associated with its creation will always be equal to $\gamma$, so the IFE is said to be isotropic. When solids are involved, $\gamma$ is anisotropic and becomes a function of the orientation of the surface, usually indicated as $\gamma\hat{\mathbf{n}}$, where $\hat{\mathbf{n}}$ is the unit normal vector to a particular crystal interface. The effect of the anisotropy of the IFE in solids is macroscopically observable, because the equilibrium shape of a crystal suspended in its melt depends on the relative values of the IFE for each different orientation of the crystal. Roughly speaking, because the free energy should be a minimum, certain crystal planes will be preferred over others with higher values of the IFE. This last effect determines the shape of a crystal and was made more rigorous by Wulff \citep{Wulff1901} in a theorem that now bears his name\footnote{although it is also known as the Gibbs-Curie-Wulff theorem, as the final form of the theorem includes contributions from Gibbs and Curie \citep{Li2016Wulff}}. We use the statement of Wulff's theorem as reported in \citep{Frank1963}: ``\textit{When a crystal is in its equilibrium shape, under negligible gravitational or other body forces or surface constraints, there exists a point whose perpendicular distances from all faces of the crystal are proportional to their specific surface free energies; any other possible face not belonging to the equilibrium shape has such a surface free energy that a plane drawn with the corresponding orientation and distance from this point would lie entirely outside the crystal}''. The Wulff theorem allows one to predict the shape of nanoparticles \citep{Marks2016} (which may not be composed of a single crystal and can have a complicated structure \citep{Ringe2013}). Knowing the IFEs of the different facets, the shape can be predicted by the Wulff construction using the so-called polar plot \citep{Herring1951}, as done by automatic tools (see, \textit{e.g.}, \citep{Rahm2020,sanchez2023direct}).

From the discussion in this section it is clear that the calculation of the IFE includes some complications intrinsic to the solid phase which are completely absent in the case of liquid-liquid systems. These complications, as we will see in the following sections, imply that the way to determine surface properties for solid-liquid interfaces using Molecular Dynamics simulations must be different from the liquid-liquid case. This latter fact is the main reason for the existence of such a large number of approaches for the calculation of IFE.  

\subsection{The Failure of the Mechanical Route}\label{sec:failure}
\noindent The validity of the Shuttleworth equation is the source of the complications in the calculation of the IFE for a solid-liquid system using MD simulations. The standard route to access an IFE was devised by Kirkwood and Buff \citep{Kirkwood1949} in an equation that states that:
\begin{equation}\label{eq:KB}
	\gamma = \int_{-\infty}^{+\infty}{(\press_N - \press_T(z))}\de z
\end{equation}
where $\press_N$ is the uniform pressure (normal to the interface) in the system and $\press_T$ is the pressure in the direction tangential to the interface. By calculating the component of the stress tensor within the system using the virial expression for the pressure \citep{Thompson2009}, and by solving the integral in \cref{eq:KB}, the value of $\gamma$ at the interface can be obtained. 

In the original work of Kirkwood and Buff, the expression for the virial was given only for a pair potential \citep{Kirkwood1949}. Today, the calculation of the pressure using the virial expression is a standard routine in MD simulation codes, with several variants presented in the literature to take into account all possible situations, e.g., including the long-range component of Coulombic interactions calculated with the Ewald summation \citep{Heyes1994} or the PPPM model \citep{Sirk2013}, or polar and charged systems \citep{Hummer1998}.

However, \cref{eq:KB} comes with a catch. As was shown in \citep{DiPasquale2020}, the quantity calculated in \cref{eq:KB} is the interfacial stress which, as we have just discussed, is equivalent to the IFE only in liquid-liquid (or liquid-vapour) systems. The latter means that while \cref{eq:KB} can be safely employed to calculate the IFE in a liquid-liquid (or liquid-vapour) system, it cannot be used to determine the IFE in a solid-liquid system as, in this case, $f\neq \gamma$ \cite{sanchez2024predictions}. 

Unlike fluid-fluid interfaces, there is no mechanical route to $\gamma$ through the pressure tensor for solid-liquid interfaces, which makes it particularly challenging to determine $\gamma$, both experimentally and theoretically. As an example, the experimental value of $\gamma$ at room temperature for the vapour-liquid interface of water is precisely known to be 71.99 $mJ/m^2$, the experimental value of $\gamma$ for the ice Ih-water interface at its melting point is uncertain, with estimates ranging between 25 and 35 $mJ/m^2$\cite{Granasy2002}. 
 
The determination of the IFE through MD simulations must therefore use the definition of the IFE, as the work needed to form a new interface in the system. In terms of MD simulations, that means using one of the methodologies to determine the free energy in a molecular system, using, e.g. thermodynamic integration.  Such difficulties explains why the measurement of $\gamma$ for solid-liquid interfaces has remained elusive until recent advances in simulation techniques over the past two decades have provided significant progress. 
Presenting such simulation techniques is the objective of the rest of this review, starting from the next section.

\section{Molecular Dynamics Free Energy calculations}
\noindent The discussion of the previous paragraph should have made clear that the calculation of the IFE between a solid and a liquid cannot rely on the direct determination of the stresses within the system, because of the validity of the Shuttleworth equation \cref{eq:Shutt}. The only way to determine $\gamma_{sl}$ is to resort to using its thermodynamic definition as the \textit{reversible} work needed to create a new interface of area $\de \area$. At constant system volume, $V$, temperature, $T$, and number of moles of the components, $N_k$, we can write $\gamma_{sl}$  in terms of the variation of the Helmholtz free energy $\helm$
\begin{equation}
	\gamma_{sl} = \left( \frac{\partial \helm}{\partial \area}\right)_{N,V,T}
\end{equation}
which can be written in integral form as
\begin{equation}\label{eq:helmgam}
	\gamma_{sl} = \frac{\helm^{fin}-\helm^{init}}{\area}
\end{equation}
where the superscript $init$ identifies an initial state, i.e. the system without an interface, and the superscript $fin$ a final state, i.e., the system with an interface. \Cref{eq:helmgam} is nothing other than the Fundamental Surface Thermodynamic Equation (\cref{eq:fundSurf}) applied to interfaces where we used the definition of the excess of $\Tilde{A}^{XS}$. In turn, \cref{eq:helmgam} is the starting point for obtaining $\gamma_{sl}$ through MD simulations. There are several ways to determine a difference in free energy using MD simulations which will be outlined in the rest of this work.

\section{Indirect methods to determine the solid-liquid IFE} \label{sec:indirmeth}
\noindent The methods for determining the IFE involving solids can be divided into two broad categories: direct and indirect methods. As the names suggest, to the former group belong all methodologies in which the IFE is determined directly from the measurement of the reversible work required to create a unit area of the interface, while in the latter, it is obtained as a by-product of the calculation of some related quantity. We begin by briefly reviewing some of the most popular indirect methods below.

\subsection{Contact angle methods} 
\noindent This class of methods use simulations that mimic the experimental setup to obtain $\gamma_{sl}$. A liquid droplet is placed on a solid surface and, after equilibration, the contact angle $\theta$, which is the angle formed between the tangent to the liquid surface and the solid surface at the point where they meet, is obtained from the density contour of the droplet \citep{Shi2009,DoHong2009,Zykova2004,Zykova2005,zykova2005physics}. Young's equation \cite{young1805thesis,Johnson1959} gives the relation between $\theta$ and the solid-liquid ($sl$) liquid-vapour ($lv$) and solid-vapour ($sv$) IFEs: 
\begin{equation}\label{eq:Young}
    \cos \theta = \frac{\gamma_{sv}-\gamma_{sl}}{\gamma_{lv}}.
\end{equation}
Obtaining $\gamma_{sl}$ from Young's equation requires independently computing $\gamma_{lv}$ and $\gamma_{sv}$. The former can be readily obtained by simulating the liquid in contact with the vapour and using \cref{eq:KB}. However, computing $\gamma_{sv}$, is not that straightforward, and there are no standard techniques to do it. For instance, in Refs. \citep{Zykova2005,zykova2005physics}, $\gamma_{sv}$ was calculated by thermodynamic integration of the energy difference between a bulk solid and a free solid slab from low temperatures to the melting point.

The contact angle approach has been used to estimate $\gamma_{sl}$ for pure NaCl by simulating a drop of liquid NaCl  on top of the halite solid at its melting temperature \citep{Zykova2005,zykova2005physics}. The value thus obtained, 36 mN/m \citep{Zykova2005,zykova2005physics}, is in sharp contrast with the much higher values (90-100 mN/m) obtained using Classical Nucleation Theory \citep{valeriani2005rate,espinosa2015crystal}, Mold Integration \citep{espinosa2015crystal}, Capillary Fluctuations \citep{benet2015interfacial} or Test area \cite{bahadur2007surface} (see \cref{tab:tablnacl} in \cref{sec:CNT}). More work is needed to understand why the contact angle method is inconsistent with these techniques. For example, a change in contact angle due to the finite drop size \cite{macdowell2002droplets} could be behind the reported discrepancies.
\begin{figure}[!h]
\centering
\includegraphics[clip,scale=0.6,angle=0.0]{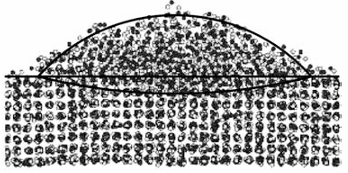}
\caption{NaCl liquid drop on top of an NaCl solid lattice. Dark and light circles correspond to Na$^+$ and Cl$^-$ ions respectively. The solid line contour is an eye guide for the average drop shape. Reproduced from Ref. \citep{Zykova2004}, with permission. Copyright 2004 by Elsevier. }
\label{fig:CAM}
\end{figure}

\subsection{The Test Area (TA) method}
\noindent The TA method \citep{Gloor2005} belongs to a class of free energy perturbation approaches \citep{Zwanzig1954} in which the free energy difference between two states is obtained by varying a parameter throughout the simulation. Using the Zwanzig equation \citep{Zwanzig1954}, the free energy difference can be obtained by an ensemble average of a function of the potential energy difference between the perturbed and the reference system. 

In its original implementation \citep{Gloor2005}, the interfacial area (at constant volume) is varied, thus allowing the calculation of the free energy difference between the two states which correspond to a reference system of interest and a perturbed system, with different interfacial areas. According to the Zwanzig equation \citep{Zwanzig1954}, interfacial free energy can be computed from a ratio of configurational phase-space integrals for isothermal perturbations,
\begin{equation}
\gamma=\left(\dfrac{\partial \helm}{\partial \area}\right)_{NVT}=
\lim_{\Delta\mathcal{A}\rightarrow 0}-\dfrac{k_{B}T}{\Delta\area}\,
\text{ln}\,\Bigg<\text{exp}\biggl(-\dfrac{\Delta U}{k_{B}T}\biggr)\Bigg>_{NVT}
\end{equation}
\noindent where $\helm$ is the Helmholtz free energy, $\area$ is the surface area, $\Delta\area$ is the difference in interfacial area (between the perturbed and reference system) and $\Delta U$ is the difference in configurational energy between the perturbed and reference systems, $T$ is the temperature, and $k_{B}$ is the Boltzmann constant. Angle brackets denote the average in the ensemble associated with the reference system.

Alternative methods to the traditional mechanical route for calculating fluid-fluid IFEs were developed several years ago. The most important are the volume perturbation method of de Miguel and co-workers \citep{deMiguel2006b,deMiguel2006c,Brumby2010a}, the wandering interface method, introduced by MacDowell and Bryk \citep{MacDowell2007a}, and the use of the expanded ensemble, based on the original work of Lyuvartsev \etal \citep{Lyuvartsev1992a}, to calculate the IFE (proposed independently by Errington and Kofke \citep{Errington2007a} and de Miguel \citep{deMiguel2008a}). 

The TA method can be easily implemented to calculate fluid-fluid IFEs in the canonical or $NVT$ ensemble \cite{Gloor2005}, as well as in the isothermal-isobaric ensemble or $NPT$ \citep{Ghoufi2016a} and the grand canonical Monte Carlo or $\mu VT$ ensembles \citep{Miguez2012a}. In addition to that, there is no restriction for considering pure systems or binary mixtures that exhibit vapour-liquid phase separation or liquid-liquid immiscibility. The technique has been used to deal with the vapour-liquid IFEs of simple spherical models, including systems that interact through the Lennard-Jones \cite{Gloor2005} and Mie \citep{Galliero2009a} intermolecular potentials. In addition to that, the methodology has also been implemented to calculate IFEs of simple models of molecular systems, including fully flexible and linear tangent Lennard-Jones chains and mixtures exhibiting vapour-liquid and liquid-liquid phase separation \citep{Blas2008a,MacDowell2009a,Sampayo2010b,Blas2012b,Blas2014a,Martinez-Ruiz2014a,Martinez-Ruiz2015a,Martinez-Ruiz2015b}, as well as Mie chains\citep{Galliero2009a}. The values of vapour-liquid and fluid-fluid IFE for pure and mixtures of realistic molecular systems have also been determined in the last twenty years using the TA technique, including pure water and aqueous solutions, as well as pure and binary mixtures of hydrocarbons, inorganic compounds \citep{Vega2007a,Miguez2010a,Muller2009a,Miqueu2011a,Miguez2013a,Miguez2014a,Ibergay2007a,Ghoufi2008a,Ghoufi2008b,Biscay2008a,Biscay2009a,Biscay2009b,Biscay2011a,Biscay2011b,Biscay2011c}. 

The TA method has been extended to include the calculation of fluid-fluid IFEs in different geometries \citep{Sampayo2010a,Malijevsky2012a,Lau2015a,Bourasseau2015a,Ghoufi2017a}, as well as to determine the solid-liquid IFEs \citep{Nair2012,Blas2013a,dOliveira2017a,Miguez2018a}. It is important to mention that the solid walls are treated at the level of external potential in all cases and cannot be considered as truly solid-fluid IFE calculations at coexistence conditions. For a more detailed account of the work devoted to determining the IFE of fluid-fluid interfaces we recommend to the reader the review of Ghoufi \emph{et al.} \cite{Ghoufi2016a}. In addition, for a more complete recasting of the TA technique to determine the IFE of curved fluid interfaces we recommend the work of Jackson and collaborators \cite{Lau2015a}.

\subsection{ Classical Nucleation Theory (CNT)} \label{sec:CNT}
\noindent CNT \cite{ZPC_1926_119_277_nolotengo,becker-doring,gibbsCNT1} predicts that the {\it thermodynamic} barrier associated with the formation of a crystalline nucleus in a metastable parent liquid or saturated solution is given by \citep{Kelton2010}:
\begin{equation}
    \Delta G = \gamma \area - |\Delta \mu| N
    \label{eq:CNT}
\end{equation}
where $N$ and $\area$ are the number of growth units and the surface area of the nucleus, respectively. If we consider the solidification process, then $\Delta \mu$ is the chemical potential difference between the crystal and the liquid, where the liquid is the melt \citep{Sosso2016,Zhang2018Ice}. Instead, if the liquid is a solution of the solid, then we must use the identity $\Delta\mu = kT\ln{S}$ where $S=a_0/a_{\rm sat}$ is the supersaturation ratio; the ratio of the solute activity in the solution, $a_0$, to the solute activity at saturation, $a_{\rm sat}$ \citep{DeYoreo2003}. The first term on the right hand side of \cref{eq:CNT} increases the nucleation barrier whereas the second decreases it. The competition between these two terms (the first proportional to the square of the size of the nucleus, whereas the second proportional to the cube of this size) gives rise to a  maximum in $\Delta G$, which we denote with $\Delta G_{crit}$, and represents the free energy barrier to the formation of a critical nucleus of the new phase, beyond which growth is spontaneous \citep{Kelton2010}: 
\begin{subequations}\label{eq:CNTgamma}
\begin{align} 
    \Delta G_{crit} &  = \frac{16 \pi \gamma_{sl}^3}{3 \rho_s^2 |\Delta \mu|^2} \label{eq:CNTgammaFreez} \\
    \Delta G_{crit} &  = \frac{16 \pi \gamma_{sl}^3v_c^2}{3 \rho_s^2k^2T^2(\ln{S})^2} \label{eq:CNTgammaSol}
\end{align}
\end{subequations}
where $\rho_s$ is the number density of the solid at the temperature and pressure of interest. Usually, we want to determine $\Delta G_{crit}$ from the parameters on which it depends (supersaturation, IFE, etc.). In turn, $\Delta G_{crit}$ allows us to determine the nucleation rate, essential in different applications, such as (for the crystallization case) precipitation of nanoparticles \citep{DiPasquale2012,DiPasquale2014pcl,DiPasquale2013} and in general in the study of macroscopic models such as the Population Balance Equation \citep{Ramkrishna2000}. The use of $\gamma_{sl}$ to determine the rate of nucleation in the context of the CNT will be discussed in \cref{sec:CNT,sec:curved}.

However, \cref{eq:CNTgamma} can also be used in the other direction: given $\Delta G_{crit}$, and the other factors inside \cref{eq:CNTgamma} determine $\gamma_{sl}$. The problem is now to determine each factor in \cref{eq:CNTgamma} (except $\gamma_{sl}$) and we will give a brief account in this section. 

The numerical factor $16\pi /3$ assumes a spherical nucleus; the equivalent factor for other geometries can be calculated. The chemical potential difference can be obtained through thermodynamic integration for both liquid cases, the melt and the supersaturated solution. For the melt case, we use the fact that $\mu$ is the same for both phases at the melting point \citep{vega2008determination}. For the solution case we use the identity $\Delta\mu = k_BT\ln{S}$.  However, computing $\Delta G_{crit}$ is more challenging as special rare event techniques such as umbrella sampling \citep{torrie1974monte} or metadynamics \citep{laio2002escaping,laio2008metadynamics} are needed to bias the formation of the critical nucleus from the liquid and compute the associated free energy change \citep{van1992computer,auer2004numerical,valeriani2005rate,quigley2008metadynamics,reinhardt2012free,cheng2018theoretical,niu2019temperature}. $\Delta G_{crit}$ can also be obtained with the seeding method \cite{bai2006calculation,knott2012homogeneous,sanz2013homogeneous,espinosa2016seeding}. This identifies the critical nucleus, which, by definition, has equal chances of growing or redissolving, by inserting nuclei of different sizes in the liquid. However, this approach  depends on the mathematical criterion with which the number of particles in the critical nucleus, $N_{crit}$, is determined \citep{espinosa2016seeding,Zimmermann2018,gispen2024variational}, given that $\Delta G_{crit}$ is obtained through the following CNT expression:
\begin{equation}
    \Delta G_{crit}=\frac{N_{crit}|\Delta \mu|}{2}. 
\end{equation}
Although it is true that within this framework $\gamma_{sl}$ depends on the selected order parameter to determine the size of the crystal cluster, it has been shown that, using a judicious choice of order parameter, consistent results are obtained between CNT seeding and direct methods to compute $\gamma_{sl}$ \cite{espinosa2016seeding,lamas2021homogeneous,garaizar2022alternating}.

Regardless of the approach used to compute $\Delta G_{crit}$ (rare event techniques or seeding), \cref{eq:CNTgamma} is typically used to obtain $\gamma_{sl}$ for several state points where the crystal is more stable than the melt (or solution, i.e., $S > 1$). Then $\gamma_{sl}$ at coexistence is obtained through an extrapolation. This extrapolation involves the usual assumption that the IFE, $\gamma_{sl}$, is a scalar, independent of the morphology of the critical nucleus. The $\gamma_{sl}$ thus obtained can be identified with an averaged IFE, $\bar{\gamma}_{sl}$ over the crystal orientations that the nucleus exposes to the liquid. The extrapolation can be avoided by using \cref{eq:CNT} at coexistence conditions, where the second term on the right hand side of the equation is zero ($|\Delta \mu|=0$) \citep{cacciuto2003solid}. However, in such conditions, the free energy does not reach a maximum but increases monotonically as $\gamma \area$. Therefore, the calculation of the IFE depends on the size and shape of the cluster, which are parameters that cannot be unambiguously defined. Moreover, curvature corrections must be included in order to get $\gamma_{sl}$ for a flat interface out of free-energy calculations of finite clusters \cite{cacciuto2003solid}. 

In summary, CNT can be a useful framework to get estimates of $\bar{\gamma}_{sl}$ at coexistence from the free energy of crystal clusters. However one should be aware of the shortcomings of this approach: (i) information about the anisotropy of $\gamma_{sl}$ is lost, (ii) the route to obtain $\gamma_{sl}$ depends on arbitrary criteria to determine the cluster size in the case of seeding and of free energy calculations of clusters (iii) extrapolations to coexistence are required for seeding and for free energy calculations away from coexistence (iv) to extract an IFE we must use the capillary approximation or correct for the finite curvature of the nucleus and (for small critical nuclei) face the problem of drawing a meaningful distinction between the interface and the bulk. We discuss these issues further in \cref{sec:nucleation} and \cref{sec:curved}.

\subsection{Capillary Fluctuations} 
\begin{figure}[!h]
\centering
\includegraphics[clip,scale=0.4,angle=0.0]{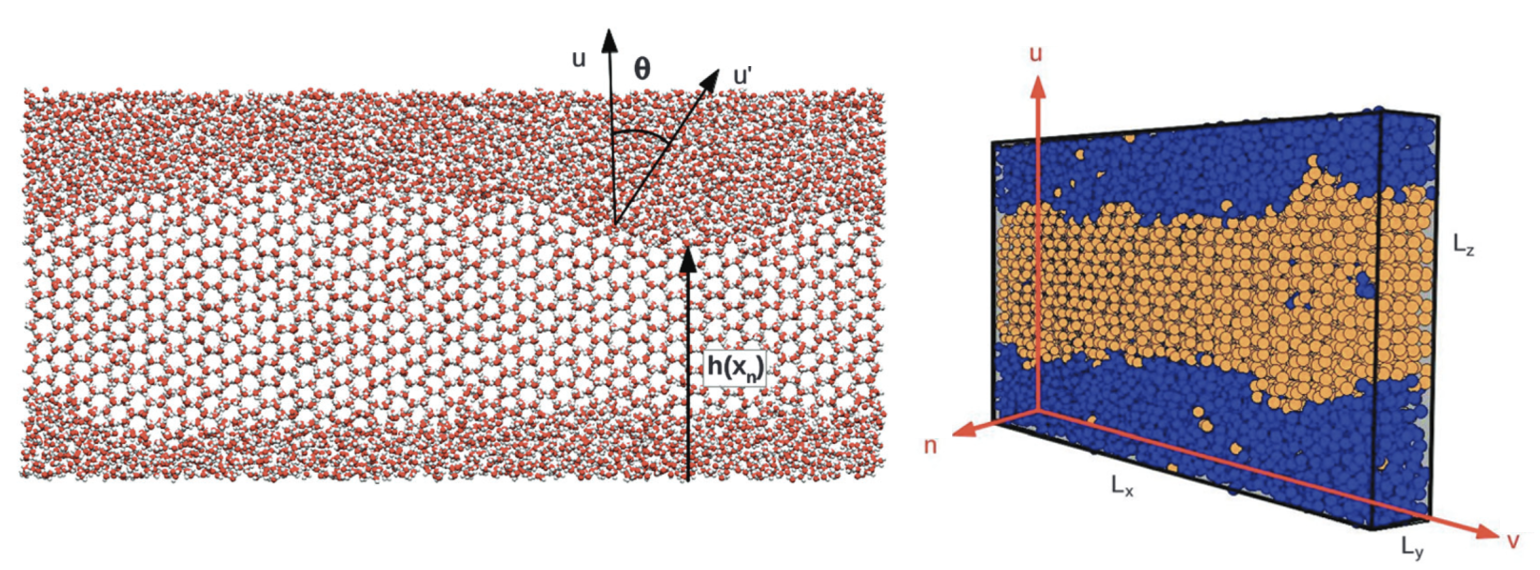}
\caption{Left, front view of a simulation snapshot of hexagonal ice in coexistence with liquid water (water molecules are represented as red and white spheres for oxygen and hydrogen atoms, respectively). The angle $\theta$ that quantifies the deviation with respect to the average interfacial orientation is defined in the figure. In addition, a point of the function $h(x_n)$ is indicated that defines a discretised interfacial profile in the real space. Right,  view of the simulation box showing the elongated strip geometry of the $x-y$ side where the solid interface is exposed to the liquid. Ice and liquid molecules are depicted as orange and blue spheres to enhance the visual contrast between both phases. Figures adapted from Ref. \citep{benet2014study} with permission. Copyright 2014 by the Royal Society of Chemistry.}
\label{fig:CFM}
\end{figure}

\noindent The capillary fluctuations method was proposed by Hoyt, Asta, and Karma \citep{Hoyt2001} and extended by Davidchack, Morris and Laird \citep{Davidchack2006,Morris2002}.  It is one of the most popular {\em indirect} methods for determining IFE and its anisotropy.  This is based on the observation that a diffuse (or rough, i.e., not faceted) solid-liquid interface will fluctuate because of thermal energy.  In \cref{fig:CFM}, an ice-liquid interface with capillary fluctuations is shown \cite{benet2014study}. These capillary fluctuations are enhanced by thermal energy and damped by interfacial stiffness, $\Tilde{\gamma}_{sl}$.

The capillary fluctuations method does not provide a direct calculation of $\gamma_{sl}$ but rather of the stiffness $\Tilde{\gamma}_{sl}$, which is equal to $\gamma_{sl}$ plus the curvature of the dependence of $\gamma_{sl}$ on the orientation of the interface: 
\begin{equation}\label{eq:stiff}
    \Tilde{\gamma}_{sl}(\theta=0)=\left(\gamma_{sl}(\theta) + \frac{\partial^2 \gamma_{sl}(\theta)}{\partial \theta^2}\right)_{\theta=0}, 
\end{equation}
where $\theta$ is the angle formed by the normal to the average interfacial plane, represented by the vector $\mathbf{u}$ in \cref{fig:CFM}, and the normal to a local fluctuation with respect to the average orientation, represented by the vector $\mathbf{u}'$ in \cref{fig:CFM}.

Although capillary waves propagate in 2D \cite{schrader2009methods}, for practical purposes --in order to reduce system size and to control the propagation direction-- the simulation box is built so that the interface is a thin elongated strip (see \cref{fig:CFM}, where $L_x \gg L_y$ for this purpose). In the example of \cref{fig:CFM}, the interface is exposed in the $x-y$ plane, and the capillary waves propagate along the $x$ direction. Then, the stiffness is obtained for a given orientation of the solid with respect to the liquid and for a given direction of propagation of the capillary waves. To ensure a stable interface, the simulation must be run in the NVT ensemble at the melting temperature and with an intermediate density between those of the coexisting solid and molten phases. To avoid stress, special care must be taken to set the edges tangential to the interface ($L_x$ and $L_y$ in \cref{fig:CFM}) to the value corresponding to a solid equilibrated at coexistence conditions. To obtain the stiffness, an interfacial profile, $h(x_n)$, is first computed for $N$ discrete points along the elongated side of the simulation box (see \cref{fig:CFM}). Its Fourier transform is given by:
\begin{equation}
    h(q)=1/N \sum^N_{n=1} h(x_n)e^{iqx_n}
\end{equation}
where the wave vector $q$ is a multiple of $2\pi/L_x$. Capillary Wave Theory \cite{privman1992fluctuating,fisher1983shape,jasnow1984critical} uses the equipartition theorem to provide a relationship between the amplitude of $h(q)$ and the stiffness: 
\begin{equation}
    \langle |h(q)|^2 \rangle=\frac{kT}{\area \bar{\gamma} q^2}
\end{equation}
where $\area$ is the interfacial area  ($L_x \cdot L_y$ in the nomenclature of \cref{fig:CFM}). This expression is valid in the limit of small $q$ vectors, i.e. long wave-length capillary fluctuations, and reveals that the size of capillary fluctuations is inversely proportional to the interfacial stiffness.

Then, knowing the symmetry of the crystal, $\gamma_{sl}(\theta)$ is expanded around $\theta=0$, typically using spherical harmonics (or combinations thereof) \cite{fehlner1976product,Hoyt2001,benet2014study}. With such expansions and \cref{eq:stiff}, a set of equivalent expansions is obtained for the stiffness. These expansions depend on $\gamma_{sl}$ and several coefficients (typically 3-4 coefficients are needed). Therefore, the stiffness has to be obtained for different interface orientations and directions of wave propagation in order to solve a system of equations that provides the coefficients of the expansions. These coefficients, in turn, provide $\gamma_{sl}$ for each crystal orientation studied. Therefore, the capillary fluctuations method is particularly well suited to study the anisotropy of the interfacial free energy with respect to the crystal orientation. 

 The method was first applied to various pure metals, alloys, and other atomic systems with the fcc and bcc crystal structures, such as Ni \citep{Hoyt2001,rozas2011capillary}, Cu \citep{Asta2002}, Al \citep{Morris2002}, Fe \citep{Sun2004}, hard spheres \citep{Davidchack2006,Mu2005,amini2008crystal,hartel2012tension}, or Lennard-Jones \cite{morris2003anisotropic}, and then extended to other systems and solid structures such as Mg with an hcp solid \citep{Sun2023Mg,Wilson2016}, sodium chloride \cite{benet2015interfacial} water with a hexagonal ice solid \cite{benet2014study}, succinonitrile\cite{feng2006}, charged colloids with a bcc solid \cite{heinonen2013bcc}, or the dipolar Stockmayer fluid with an fcc solid \cite{wang2013freezing}.

\section{Direct simulation methods to determine the solid-liquid interfacial free energy} \label{sec:dirmeth}
Direct simulation methods are based on the thermodynamic definition of the IFE as the reversible work required to create a unit area of interface between a solid and a liquid phase under solid-liquid coexistence conditions.  Such methods require the construction of a thermodynamic transformation path from a system without an interface (for example, isolated bulk solid and liquid systems under coexistence conditions) to a system containing the interface.  The reversible work or, equivalently, the free energy difference between the two states can be calculated by a variety of free-energy calculation methods such as (see, e.g., \citep{Frenkel2001}) Thermodynamic Integration (TI), free-energy perturbation, Bennett acceptance ratio and non-equilibrium switching (employing the Jarzynski identity). Because the different methodologies presented here are mostly based on the TI procedure, we included a general description of the methodology in  \cref{sec:TI} leaving only the details of each methodology to be explained here. 

\subsection{Cleaving methods}
\noindent These approaches have been used for the first time by Broughton and Gilmer \citep{Broughton1986} to calculate the solid-liquid IFE for a truncated Lennard-Jones potential, although the idea was first proposed by Miyazaki {\em et al.} for the liquid-vapour interface \citep{Miyazaki1976}.  It is based on the calculation of a free energy change along a reversible path that starts from separate solid and liquid bulk systems under co-existence conditions and ends with the solid-liquid interfacial system under the same conditions.  The free energy change is thus related to the creation of the interface and the IFE is determined as the ratio of this change to the area of the created interface. 

The calculation starts by preparing separate solid and liquid systems under solid-liquid coexistence conditions with periodic boundary conditions in all directions.  The systems should have the same size in the directions parallel to the interfacial plane (typically aligned with the $x$ and $y$ axes, while the $z$ axis is taken to be normal to the interface). The transformation path is then constructed with the help of external {\em cleaving} potentials which play the dual role of $i$) splitting the solid and liquid bulk systems along a plane (chosen to be between two crystal layers in the solid system and arbitrarily in the liquid system), so they can be later combined into the interfacial system, and $ii$) introducing a structure in the liquid phase near the chosen plane that is compatible with the crystal structure of the solid phase.

\begin{figure}[!h]
\centering
\includegraphics[clip,scale=0.5,angle=0.0]{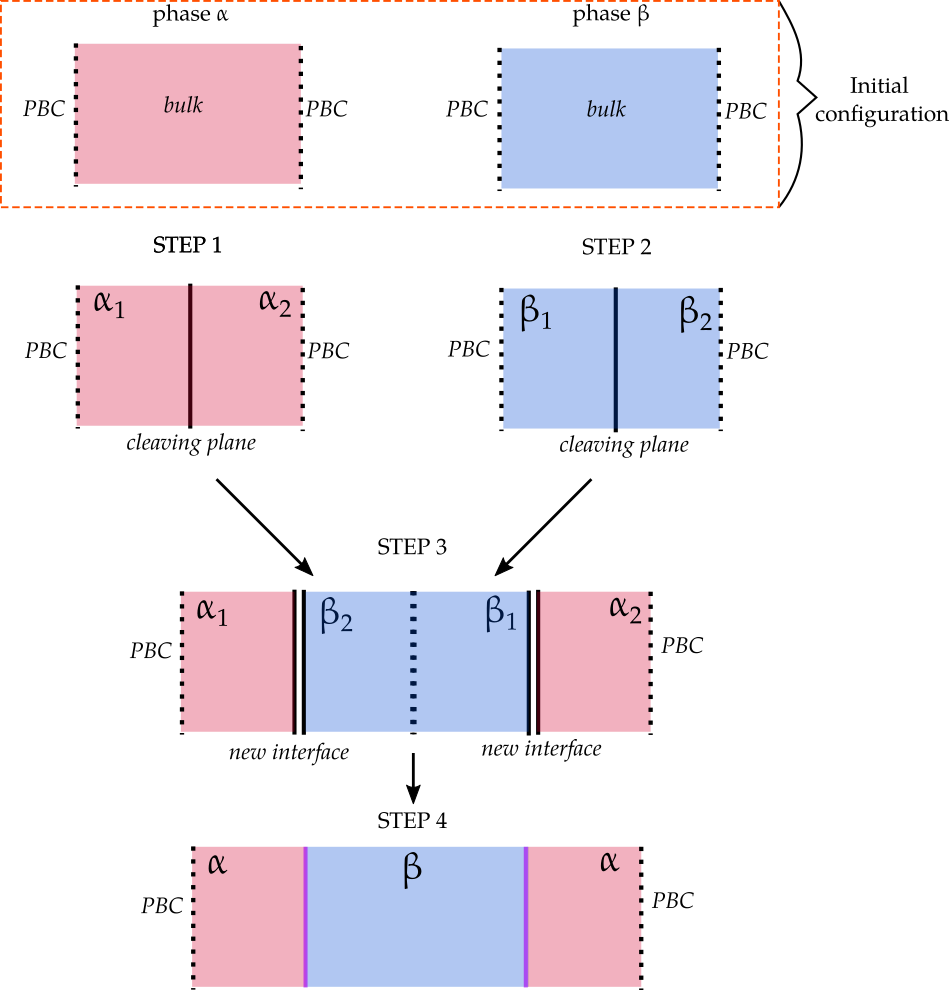}
\caption{Schematic transformation of the cleaving procedure, with highlighted the four steps described in the text. The initial point is represented by two different systems, phase $\alpha$ bulk and phase $\beta$ bulk. The final point is represented by a single system with two new interfaces between phases   $\alpha$  and  $\beta$.  }
\label{fig:cleav}
\end{figure}

The transformation path consists of four basic steps, as shown in \cref{fig:cleav}: 
\begin{enumerate}
\item insert the cleaving potential into the solid system along a plane between crystal layers of a given orientation (called the cleaving plane); 
\item insert the cleaving potential in the liquid system with the same dimensions as the solid system in the directions tangential to the plane;
\item gradually (and reversibly) switch the interactions from solid-solid and liquid-liquid to solid-liquid across the cleaving plane, while maintaining the cleaving potentials;
\item remove the cleaving potentials from the combined solid-liquid system.  
\end{enumerate}

Each transformation step can be implemented using a standard coupling parameter approach \citep{Frenkel2001}, where the total potential energy of the system in step $n$, $V_n(\lambda)$, depends on a coupling parameter $\lambda$ in such a way that changing the parameter from 0 to 1 transforms the system from the thermodynamic state at the start of the step to that at the end. In its simplest implementation, the potential energies $V_n(\lambda)$ take the form
\begin{eqnarray} \label{eq:Vn}
    V_1(\lambda) &=& U_s + \lambda \Phi_s\,,\nonumber\\
    V_2(\lambda) &=& U_l + \lambda \Phi_l\,,\nonumber\\
    V_3(\lambda) &=& (1-\lambda)(U_s+U_l) + \lambda U_{sl} + \Phi_s + \Phi_l\,,\nonumber\\
    V_4(\lambda) &=& U_{sl} + (1-\lambda) (\Phi_s + \Phi_l)\,,
\end{eqnarray}
where $U_s$, $U_l$, $U_{sl}$ are the potential energies of the solid, liquid, and combined systems, respectively, and $\Phi_s$, $\Phi_l$ are the cleaving potentials introduced in the solid and liquid systems.  Note that the simple linear dependencies on $\lambda$ in the above equations can be replaced by any continuous functions $g(\lambda)$ such that $g(0) = 0$ and $g(1) = 1$.  

In the TI formulation of the coupling parameter approach, the reversible work, $W_n$, $n = 1, 2, 3, 4$, required to perform each step is calculated as 
\begin{equation}\label{eq:work}
    W_n = \int_0^1 \left\langle \partial V_n / \partial \lambda \right\rangle_\lambda \de \lambda \,,
\end{equation}
where $\langle \ldots \rangle_\lambda$ denotes an average over the equilibrium state at a fixed value of $\lambda$.  The solid-liquid IFE is then given by
\begin{equation}\label{eq:gamma-cleaving}
    \gamma_{sl} = \area^{-1}\sum_{n=1}^4 W_n\,,
\end{equation}
where $\area$ is the area of the created interface. There is considerable flexibility in the design of the cleaving potentials $\Phi_s$ and $\Phi_l$.  In their calculation of the IFEs of (100), (110), and (111) solid-liquid interfaces in the truncated Lennard-Jones system at the triple point, Broughton and Gilmer \cite{Broughton1986} used the cleaving potentials in the form of a simple repulsive potential for the solid system (in Lennard-Jones units)
\begin{equation}
    \Phi_s = 3\mathrm{e}^{-\alpha z^4}
\end{equation}
and a combination of repulsive and attractive potentials for the liquid system
\begin{equation}
    \Phi_l = 3\mathrm{e}^{-\beta z^4} - [A + B F(x,y)]\mathrm{e}^{-\delta (z - z_{\rm min})^2}
\end{equation}
where $z$ is the distance to the cleaving plane, and $z_{\rm min}$ is the distance from the cleaving plane to the nearest crystal layer (which is equal to half the inter-layer distance for a given crystal orientation). The attractive part of the liquid cleaving potential is modulated by the function $F(x,y)$ to produce local minima at the locations of particles in the crystalline structure in order to induce formation of crystal-like layers in the liquid system that match the crystal layers in the solid system at the cleaving plane. Parameters $\alpha$, $\beta$, $\delta$, $A$, and $B$ are chosen empirically to: $i$) ensure atoms in solid and liquid systems do not mix during Step 3, $ii$) introduce sufficient structure in the first liquid layer to match the corresponding crystal layers in the solid systems, and $iii$) perturb the systems as little as possible in order to minimise the amount of reversible work performed in Steps 1 and 2.  

This last requirement is satisfied by making the cleaving potential relatively short-range (i.e., choosing large values for $\alpha$, $\beta$, and $\delta$).  While this is fine for Step 1, in Step 2 it leads to a large uncertainty in the calculated reversible work due to the presence of a first-order transition associated with the formation of crystalline layers in the liquid system, and thus requires crossing a nucleation barrier, resulting in a hysteresis loop in the process of switching the cleaving potential on and off in Step 2.  Broughton and Gilmer noted that the size of the loop depends on the range of the attractive part of the cleaving potential $\Phi_l$.  To reduce it, they proposed to increase the range of attractive potential away from the cleaving plane.  As such, while $\delta$ has values between 40 and 100 (depending on the orientation of the interface) for $z \leq z_{\rm min}$, $\delta = 4.0$ for $z > z_{\rm min}$. The results obtained by Broughton and Gilmer (see \cref{tab:tablagamms_LJ}) had a precision of about 3-6\%, which was not sufficient to resolve the anisotropy of $\gamma_{sl}$.  Further development of the cleaving method \citep{Davidchack2003} and introduction of the Capillary Fluctuations method \citep{morris2003anisotropic} were necessary to achieve a precision sufficient to resolve the anisotropy.

The cleaving method was then adapted by Davidchack and Laird to obtain the first direct calculation for the IFE of the hard-sphere crystal-melt interfaces with orientations (100), (110), and (111).\citep{Davidchack2000}  Because the event-driven implementation of the time evolution of a hard-sphere system is conceptually very different from the time-stepping numerical solution of the equations of motion for continuous potentials, the cleaving method for calculation of the hard-sphere crystal-melt IFE needs to be adapted to this event-driven paradigm.  To achieve this, Davidchack and Laird proposed to cleave the solid and liquid systems using a pair of moving ``walls'' consisting of ideal crystal layers placed on either side of the cleaving plane and interacting only with the system spheres on the opposite side of the plane.  At the start of Steps 1 and 2, the two walls are placed sufficiently far from the cleaving plane so they do not interact with the system hard-spheres.  During the steps, the two walls move closer to the cleaving planes and start colliding with the system spheres. At the end of the steps the walls reach positions where the system spheres on the opposite sides of the cleaving plane no longer collide with each other.  Thus, the cleaving of the systems in Steps 1 and 2 is achieved, and rearrangement of the boundary conditions in Step 3 does not require additional work (i.e. $W_3 = 0$).  Step 4 is then performed on the combined solid-liquid system by moving the walls back to their initial positions.
\begin{figure}
    \centering
    \includegraphics[width=0.7\linewidth]{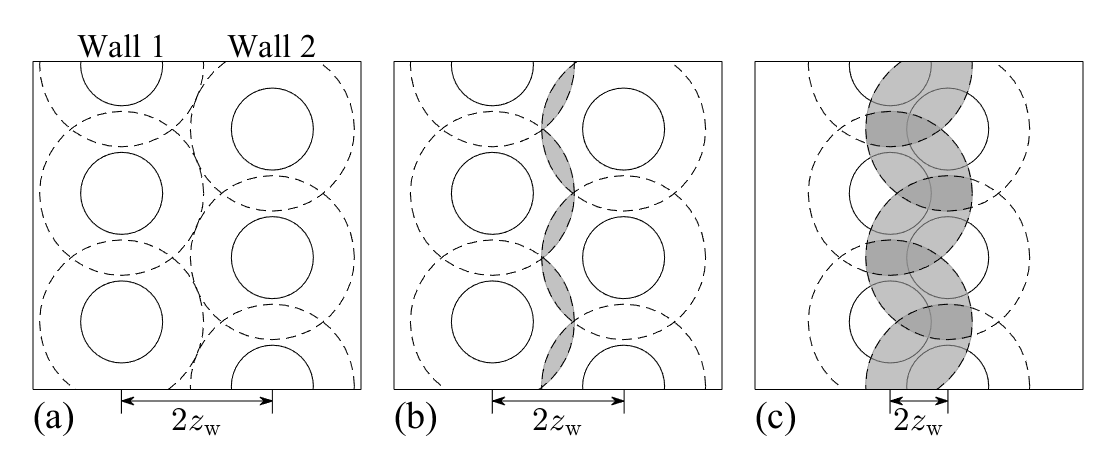}
    \caption{Illustration of the cleaving walls needed to calculate the IFE of the (100) crystal-melt interface in a hard-sphere system. Solid circles outline the wall spheres of diameter $\sigma$.  Dashed circles outline spheres of radius $\sigma$ centred at the wall spheres.  Shaded regions indicate the excluded volume introduced by the cleaving walls, i.e., where the cleaving walls potential is infinite (see \cref{eq:Pcw}). (a) Initial position of the walls, where they do not interact with the system. (b) Intermediate wall position, where the system sphere can no longer pass through the cleaving plane. (c) Final position of the walls, where the system spheres cannot collide across the cleaving plane.}
    \label{fig:hs_cleave}
\end{figure}
Note that the original implementation of the cleaving method for hard-spheres systems \citep{Davidchack2000} contained an error that was later corrected \citep{davidchack2010hard1}, so here we outline the corrected version of the method.  Let the spheres of the system have diameter $\sigma$.  The spheres comprising the walls have the same diameter and are located at positions ${\bf R}_j^{(1,2)} = (X_j^{(1,2)}, Y_j^{(1,2)}, Z_j^{(1,2)})$ in ideal crystal layers with the same orientation as the solid system in Step 1.  Depending on the orientation, each wall consists of one or two layers.  The positions of the walls with respect to the cleaving plane are $-z_w$ for Wall-1 and $+z_w$ for Wall-2, where $z_w$ is half the distance between the nearest layers of the two walls: $z_w = \frac{1}{2}\left(\min_j Z_j^{(2)} - \max_j Z_j^{(1)}\right)$.  The system spheres interact with the wall spheres as follows: a sphere collides with one of the wall spheres only if, at the moment of collision, it overlaps with a sphere belonging to the other wall.  Therefore, during the hard-sphere molecular dynamics simulation, collisions with wall spheres are predicted, and, while processing a sphere collision with one wall, it is checked to see if the sphere overlaps with the other wall. If it does, the collision takes place; otherwise, the sphere continues to move in the same direction as before. This interaction can be described by the following cleaving potential exerted by the moving walls on the system spheres
\begin{equation}\label{eq:Pcw}
  \Phi_s({\bf r}) = \Phi_l({\bf r}) = \left\{ \begin{array}{l} \infty,\;\;
  \mathrm{if}\;\min_j|{\bf r}-{\bf R}_j^{(1)}| < \sigma\;\;\mathrm{and}\\
  \hspace{3.1em}\min_j|{\bf r}-{\bf R}_j^{(2)}| < \sigma,\\[1ex]
  0,\quad\mathrm{otherwise}\,. \end{array}\right.\,.
\end{equation}
The cleaving process in Steps 1 and 2 is illustrated in \cref{fig:hs_cleave}, where the shaded regions are inaccessible to the system spheres.  When the two walls are placed sufficiently far apart, as in \cref{fig:hs_cleave}(a), they do not interact with the system spheres.  As the two walls move closer together, inaccessible regions appear and grow, so that at some point, as in \cref{fig:hs_cleave}(b), they form a fully connected region which the spheres cannot cross.  In order to prevent the system spheres from colliding across the cleaving plane, the walls should be moved to the position shown in \cref{fig:hs_cleave}(c), where the minimal thickness of the inaccessible region is larger than the diameter of the hard sphere.

Within the TI approach, for Step $n$ of the moving wall method, the system is equilibrated at a number of intermediate positions of the walls and the average value of the pressure on the walls is computed as a function of the wall position $\press_n(z_w)$. The reversible work is obtained by numerically evaluating the integral
\begin{equation}\label{eq:MW-HS}
    W_n = \area \int_{z_i}^{z_f}  \press_n(z_w) \de z_w\,,\quad n = 1, 2, 4\,,
\end{equation}
where $\mathcal{A}$ is the are of the created interfaces and $z_i$ and $z_f$ are the initial and final wall positions, respectively.

Another approach to computing the reversible work in the cleaving method is by performing {\em non-equilibrium work measurements}, where the transformation (either by changing $\lambda$ or by moving walls) is performed over a finite time interval.  This approach is based on the Jarzynski equality \citep{Jarzynski97prl,Jarzynski97pre} relating the work $\mathcal{W}$ performed on the system during a nonequilibrium process starting from initial states sampled at equilibrium with inverse temperature $\beta = (k_B T)^{-1}$ and the reversible work $W$ between the same initial and final thermodynamic states:
\begin{equation}\label{eq:noneq1}
    \left\langle e^{-\beta\mathcal{W}}\right\rangle = e^{-\beta W}\,,
\end{equation}
where the angle brackets denote the average over an ensemble of nonequilibrium processes starting from an equilibrium ensemble of initial states.  The nonequilibrium work measurements are typically preferable to TI because they provide more efficient calculation and better error estimates for the reversible work, especially when nonequilibrium work measurements can be performed in both forward and reverse directions.  In this case, as shown by Crooks \citep{Crooks00}, the probability distributions of forward (F) and reverse (R) nonequilibrium work measurements are related by the formula
\begin{equation}\label{eq:crooks}
    P_F(\mathcal{W}) = e^{\beta(\mathcal{W}-W)} P_F(-\mathcal{W})\,,
\end{equation}
where the reverse process starts from equilibrium with inverse temperature $\beta$ and the switching protocol mirrors that of the forward process.  Then the reversible work can be calculated from averages over these distributions
\begin{equation}\label{eq:bar1}
   e^{-\beta W} = \frac{\left\langle h(\mathcal{W})\right\rangle_F} {\left\langle h(-\mathcal{W})e^{-\beta\mathcal{W}}\right\rangle_R}\,,
\end{equation}
where angle brackets with subscripts $F$ and $R$ denote averages over nonequilibrium work measurements in forward and reverse directions, respectively, and 
\begin{equation}\label{eq:bar2}
  h(\mathcal{W}) = \left(1 + {\textstyle \frac{n_F}{n_R}}\mathrm{e}^{\beta(\mathcal{W} - W)} \right)^{-1}
\end{equation}
provides an asymptotically unbiased estimator for $W$ with minimal variance, with $n_F$ and $n_R$ being the numbers of independent forward and reverse measurements, respectively.

After successful application of the moving walls method to the hard-sphere IFE calculation, it was adapted for application to continuous potentials and applied to the truncated Lennard-Jones potential, improving on the precision of the results obtained by Broughton and Gilmer at the triple point, as well as calculating solid-liquid IFE at temperatures $T = 1.0$ and $1.5\,\epsilon/k_B$ \citep{davidchack2003direct1}. The cleaving walls potential was constructed from short-range repulsive potentials $\phi(r)$ centred at ideal crystal positions ${\bf R}_j^{(1,2)}$ and defined as the minimum of the two wall potentials
\begin{equation}\label{eq:phimin}
    \Phi({\bf r},z) = \min(\Phi_1({\bf r},z),\Phi_2({\bf r},z))\,,
\end{equation}
where
\begin{equation}\label{eq:phi12}
    \Phi_{1,2}({\bf r},z) = \sum_j \phi(|{\bf r}-{\bf R}_j^{(1,2)} \pm {\bf n}z|)\,.
\end{equation}
with $z$ denoting the distance of the cleaving walls from the cleaving plane and ${\bf n}$ denoting unit vector normal to the cleaving plane.  Similar approach was used to calculate IFE for inverse-power potentials (aka soft spheres), $u(r) = r^{-n}$, with $n = 6$, 7, 8, 10, 12, 14, 20, 30, 50, and 100 for the fcc-liquid interface and $n = 6$, 7, and 8 for the bcc-liquid interface \citep{Davidchack2005}.

Further extension of the cleaving method was introduced for the calculation of IFE in ice-water interfaces, modelled by rigid-body TIP4P and TIP5P potentials \citep{Handel2008,Davidchack2012ice}.  In order to induce formation of crystal layers with correctly oriented molecules in Step 2, attractive cleaving {\em wells} potential was introduces instead of repulsive walls.  The short-range attractive potential $\phi(r)$ was placed at the ideal crystal positions ${\bf R}_j$, modulated by an orientation-dependent factor $\theta$:
\begin{equation}\label{eq:clwells}
    \Phi({\bf r},{\bf q}) = \sum_j \phi(|{\bf r} - {\bf R}_j|) \theta({\bf q},{\bf Q}_j)\,,
\end{equation}
where ${\bf r}$ and ${\bf q}$ are the translational and rotational coordinates of a molecule, respectively, and ${\bf Q}_j$ is the orientation of a molecule in the ideal crystal at position ${\bf R}_j$.  The attractive potential has the form $\phi(r) = d_w[(r/r_w)^2-1]^3$ for $r < r_w$ and zero otherwise, where $d_w$ and $r_w$ are the well depth and range parameters, respectively.  The orientation dependent factor has the form  of a dot product between unit vectors directed from the oxygen atom to the midpoint between the hydrogen atoms in a water molecule: $\theta({\bf q},{\bf Q}) = {\bf n}({\bf q})\cdot{\bf n}({\bf Q})$.  The results obtained by the cleaving method had sufficient precision to resolve the anisotropy of the IFE and show that the basal interface has the lowest IFE \citep{Davidchack2012ice}.

The attractive wells external potential (without the orienting factor) was also used successfully in the calculation of the IFE by the cleaving method in Fe bcc crystal-liquid interface modelled via the Embedded Atom Method \citep{Liu2013}.  More recently, the cleaving approach was extended to more complex materials, such as Ag-ethylene glycol \citep{Qi2016}, orcinol-chloroform and orcinol-nitromethane \citep{Addula2020} interfaces.  Because these are heterogeneous interfaces (i.e., interfaces between dissimilar materials that do not mix), the cleaving process can be simplified by switching off the interactions across the cleaving plane while introducing cleaving potentials in Steps 1 and 2.  Then, the rearrangement of boundary conditions in Step 3 does not require any work, so can be performed instantaneously.

In addition to the Lennard-Jones system \citep{Broughton1986,davidchack2003direct1,Zhou2024}, the cleaving method has been used to calculate the solid-liquid IFE for the hard-sphere system \citep{Davidchack2000,davidchack2010hard1}, inverse power potentials \citep{Davidchack2005}, TIP4P water \citep{Handel2008,Davidchack2012ice}, iron modelled via the Embedded Atom Method \citep{Liu2013}, Ag-ethylene glycol \citep{Qi2016},  orcinol-chloroform and orcinol-nitromethane \citep{Addula2020} and mannitol molecular crystal \citep{DiPasquale2022}.

\subsection{Mold Integration (MI)}\label{sec:mold_integration}
\begin{figure}
    \centering
    \includegraphics[width=0.35\linewidth]{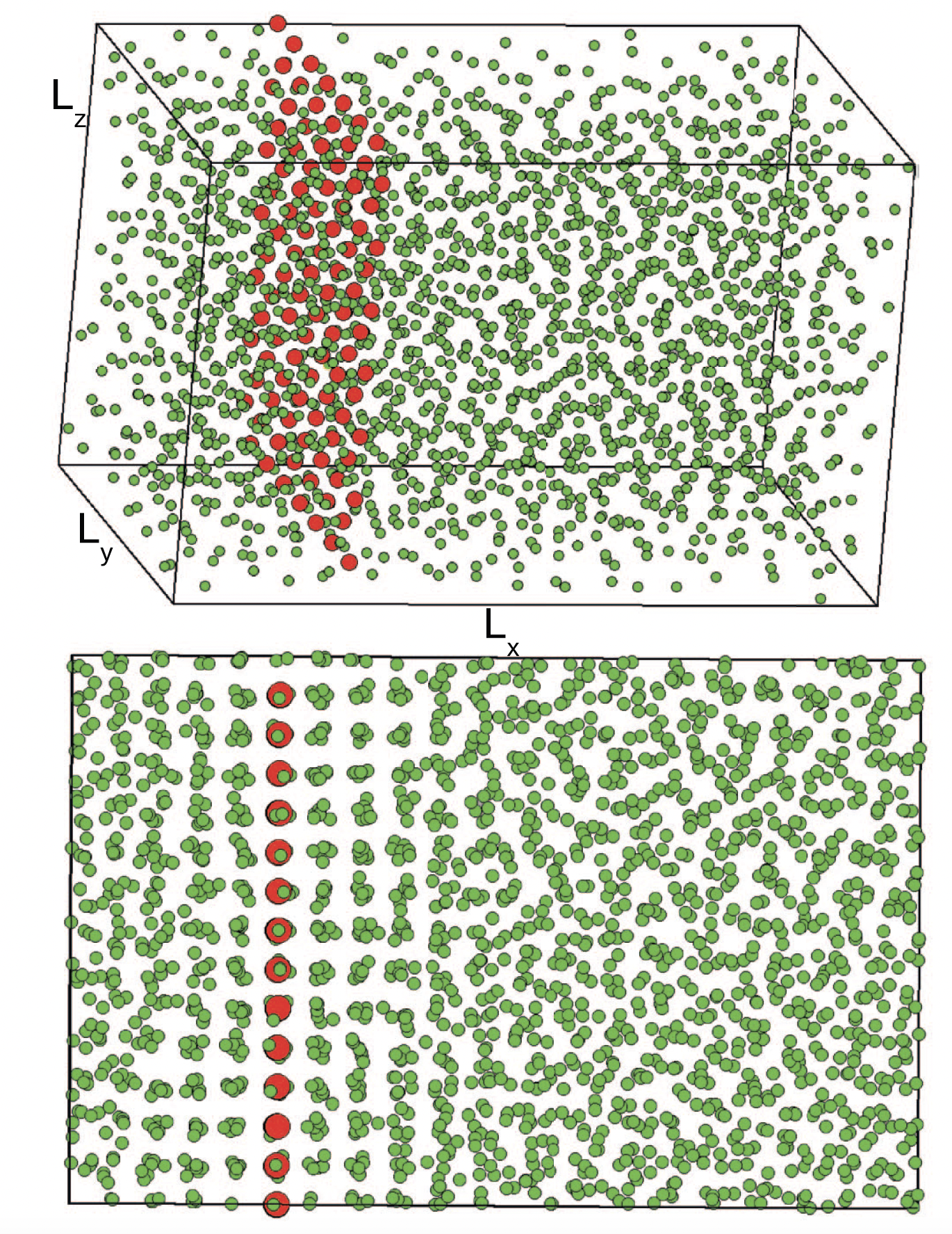}
    \caption{Top: Snapshot of a hard-sphere fluid under coexistence conditions (green particles). Bottom: Snapshot of a fluid with a thin crystal slab under coexistence conditions (a projection on the $x-z$ plane is shown). The mold that induces the formation of the crystal slab consists of a set of potential energy wells (red spheres) whose positions are given by the lattice sites of the selected crystal plane at coexistence conditions. The interaction between the mold and the hard-spheres is switched off in the top configuration and on in the bottom one. The diameter of green particles has been reduced to $1/4$ of its original size. Reproduced from Ref. \citep{espinosa2014mold} with permission. Copyright 2014 by the American Institute of Physics. }
    \label{moldfigure}
\end{figure}

\indent Mold Integration calculates $\gamma_{sl}$ by reversibly inducing the formation of a crystalline slab in the fluid under coexistence conditions \cite{espinosa2014mold}. The free energy needed to form such a crystalline slab, $\Delta G$, is related to $\gamma_{sl}$ by: 
\begin{equation}
\gamma_{sl} = \Delta G / \area
\end{equation}
where $\area$ is the area of the interface between the crystal slab and the coexisting liquid. Because the formation of the slab is performed under coexistence conditions, the fluid and the crystal have the same chemical potential. Hence, $\Delta G$ is just the specific IFE times the area of the interface which corresponds to twice the cross-sectional area of the simulation box because the slab exposes two interfaces to the fluid (see \cref{moldfigure}). 
To induce the formation of the crystal phase, MI uses a mold of potential energy wells located at the equilibrium positions of the perfect crystal lattice under coexistence conditions. \Cref{moldfigure} shows a snapshot of the mold used for the calculation of $\gamma_{sl}$ for the (100) plane of hard spheres (red spheres) in the original reference in which the method was proposed \citep{espinosa2014mold}. Each potential well must be small enough to accommodate no more than one particle. When the mold is turned off, particles freely diffuse in the liquid (see \cref{moldfigure}, top), whereas when the mold is on, every well contains a particle. Moreover, depending on the width and depth chosen for the potential, a crystal slab can be induced. Although for wide or shallow wells the crystal slab cannot be formed, if the potential is sufficiently narrow and deep to confine the particle at the crystal lattice position, the mold can induce a slab (see \cref{moldfigure}, bottom) giving rise to two crystal-fluid interfaces. By gradually switching on the interaction between the mold and the particles, the work of formation of the crystal slab at coexistence conditions can be obtained by TI using the following expression: 
\begin{equation}
\gamma_{sl}(r_w)= \frac{1}{2\area} \left ({\epsilon_m N_w - \int_0^{\epsilon_m} d \epsilon (\langle N_{fw}(\epsilon)\rangle_{N\press_xT}) } \right )   \label{gamma2} 
\end{equation}
where $N_w$ is the total number of wells in the mold, and $\langle N_{fw}(\epsilon) \rangle$ is the average number of filled wells throughout an $N\press_xT$ simulation for potential wells of depth $\epsilon$ (the barostat in the simulation is applied only in the direction perpendicular to the crystal-fluid interface to avoid deforming the perfect equilibrium lattice). To compute $\gamma_{sl}$, TI (using \cref{gamma2}) is performed along the path in which the depth of the potential mold wells is gradually increased to a maximum value of $\epsilon_{m}$. The path of \cref{gamma2} must be reversible, and to ensure its reversibility, the crystal structure induced by the mold must quickly melt when the interaction between the potential wells and the fluid is turned off. To that end, the integration needs to be performed at well radii ($r_w$) that are wider than the optimal one, $r^o_w$, at which the crystal slab is fully formed, and thus, its stability no longer depends on the mold-fluid interactions. In practice, as proposed in Ref. \citep{espinosa2014mold}, $\gamma_{sl}(r_w)$ can be estimated for several values of $r_w > r^o_w$, and then be extrapolated to $r^o_w$, which is the well radius that provides the desired value of $\gamma_{sl}$. The width chosen for $r^o_w$ is based on selecting the intermediate potential well radius between two different regimes: (i) in which there is no induction time for the formation of the crystal slab at maximum potential well depth (i.e., 8-10 $k_BT$, and where the potential well radius is small); and (ii) a regime where using the same potential well depth, the formation of the crystal slab still requires to overcome some activation energy barrier, and thus, the system shows a given induction time before the slab crystal growth (e.g., for wider potential wells). Further technical details, on how to precisely evaluate $r^o_w$, $\langle N_{fw}(\epsilon) \rangle$, and ultimately $\gamma_{ls}$ through the MI method can be found in Refs. \citep{espinosa2014mold,espinosa2016ice,sanchez2021fcc}.     

The MI technique enables the direct measurement of $\gamma_{sl}$ for any arbitrary crystal orientation \cite{espinosa2014mold}. The IFE of different crystal phases in hard-spheres (fcc and hcp) \cite{sanchez2021fcc} and water (hexagonal and cubic ice) \cite{espinosa2016interfacial} as well as different crystal orientations in Lennard-Jones particles \cite{espinosa2014mold} and NaCl with its melt \cite{espinosa2015crystal} have been successfully studied using MI. In addition, the technique has been extended to deal with more complex solid structures and coexisting liquids of different components. Some examples are NaCl-water solutions including ice in contact with salty water \cite{soria2018simulation} and crystalline NaCl in contact with a saturated NaCl aqueous solution at the solubility limit\cite{sanchez2023direct}. It has also recently been used to show the direct relation between the slope of the melting line and the pressure dependence of $\gamma_{sl}$ between hexagonal ice and liquid water \cite{montero2023minimum}.

An important extension of the MI was developed to compute hydrate--water  IFEs containing different guest molecules such as carbon dioxide, methane, nitrogen, hydrogen, and tetrahydrofuran hydrates \cite{Algaba2022b,Zeron2022a,Romero-Guzman2023a}. Following the methodology introduced by Espinosa and collaborators \citep{espinosa2014mold}, Algaba \emph{et al.} \citep{Algaba2022b} and Zer\'on \emph{et al.} \citep{Zeron2022a} proposed two different extensions for the MI method. In the first one, called MI-Host (MI-H), they placed attractive interaction sites in the H$_{2}$O-rich liquid phase at the equilibrium positions of the oxygen atoms of water in one of the principal planes of the sI structure of the CO$_{2}$ hydrate\citep{Algaba2022b}. In the second extension, called MI-Guest (MI-G), they placed attractive interaction sites at crystallographic equilibrium positions of a layer of carbon atoms of CO$_{2}$ molecules in the CO$_2$ hydrate \citep{Zeron2022a}.

\begin{figure}
    \centering
    \includegraphics[width=0.60\linewidth]{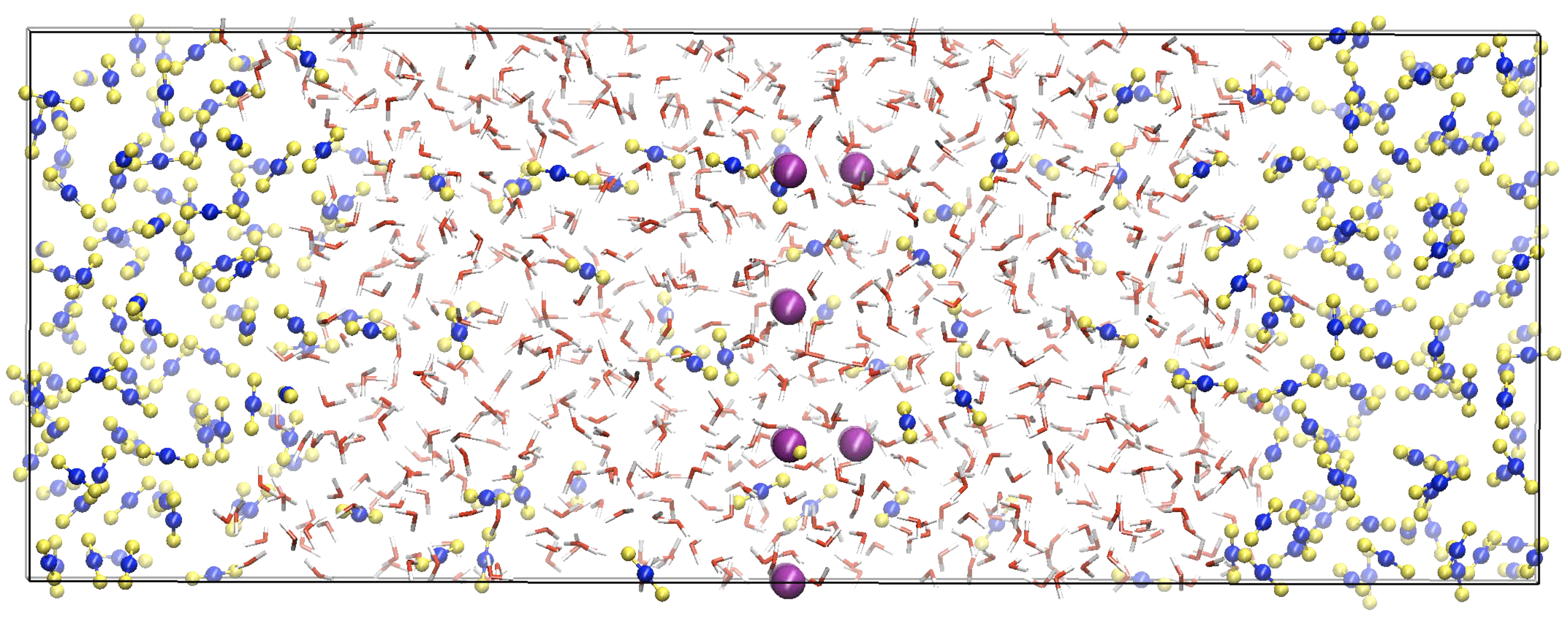}
    \includegraphics[width=0.60\linewidth]{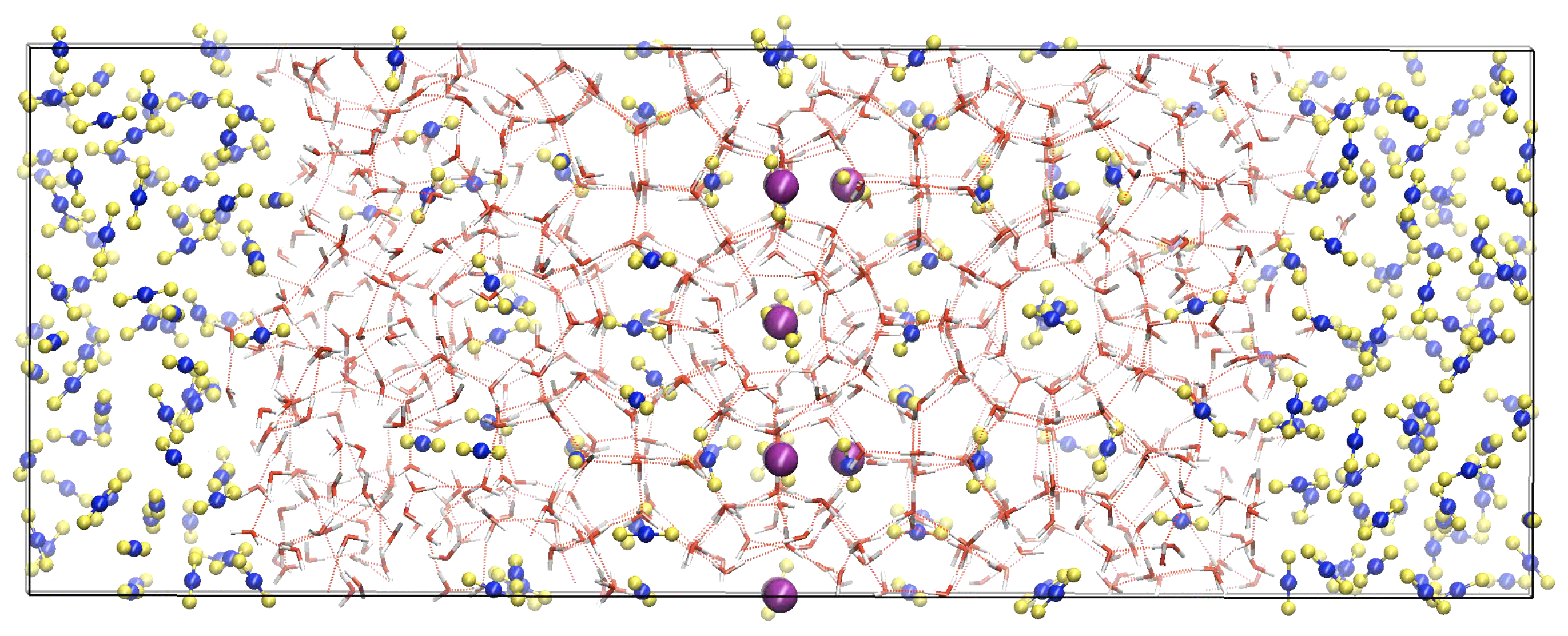}
    \caption{Snapshots representing trajectories extracted from molecular dynamics simulations of the CO$_{2}$-water two-phase coexistence at $400\,\text{bar}$ and $287\,\text{K}$. The interaction between the mold and the carbon atoms of the CO$_{2}$ molecules is switched off in the top configuration and on in the bottom one. The mold that induces the formation of the crystal slab consists of a set of potential energy wells (magenta spheres) located at the crystallographic positions of the carbon atoms of the CO$_{2}$ molecules of the selected crystal planes at coexistence conditions. The red and white licorice representation corresponds to oxygen and hydrogen atoms of water, respectively, blue and yellow spheres (van der Waals representation) correspond to carbon and oxygen atoms of CO$_{2}$, respectively, and magenta spheres (van der Waals representation) correspond to the mold.}
    \label{MIGfigure}
\end{figure}

\Cref{MIGfigure} shows two snapshots corresponding to trajectories obtained from molecular dynamics simulations to determine the CO$_{2}$ hydrate-water interfacial free energy using the MI-G method. Note that the use of a mold in the MI-H technique is similar to the original implementation of MI used for aqueous systems because the associating sites of the mold are located at the crystallographic positions of the oxygen atoms of the water molecules of the selected crystal planes. The use of both extensions of the technique for hydrates requires special attention because the coexistence conditions of the hydrate-water interface involve two different components (H$_{2}$O and CO$_{2}$) and three phases in equilibrium: the CO$_{2}$ hydrate solid, H$_{2}$O-rich liquid, and CO$_{2}$-rich liquid. The presence of the three phases is necessary to ensure that calculations are performed under equilibrium coexistence conditions. It is also necessary to tune the local order parameters to correctly identify hydrate-like and liquid-like water molecules to follow the growth of the thin hydrate layer induced. Zer\'on \emph{et al.} \citep{Zeron2024a} have recently revisited the Lechner and Dellago order parameters \citep{Lechner2008a} and have obtained a new combination of order parameters that can distinguish water molecules in both phases, allowing to correctly characterize the hydrates.

Although both methods (i.e., MI-H and MI-G) are based on the MI technique, the two approaches involve completely different calculations. In the first case \citep{Algaba2022b}, the MI-H technique uses a mold of associating wells positioned at the crystallographic equilibrium locations occupied by the oxygen atoms of water molecules in the primary plane of the sI hydrate. However, the second case \citep{Zeron2022a} employs a mold of associating wells located at the centers of the T and D cages in the sI hydrate structure, corresponding to the equilibrium positions of the carbon atoms of CO$_{2}$ molecules in two different planes. In both calculations, type, number, and well-depth of the associating wells of the molds are different. The results can therefore be regarded as arising from two distinct approaches. They have also extended the study and determined the CO$_{2}$ hydrate-water interfacial energy along the dissociation line of the hydrate at several pressures ($100$, $400$, and $1000\,\text{bar}$) \citep{Romero-Guzman2023a}. The results suggest that there is a weak correlation between interfacial free energy values and pressure indicating that $\gamma_{sx}$ decreases with pressure. Unfortunately, it is not possible to compare with experimental data from the literature because the experimental technique assumes that the interfacial energies are independent of the pressure. We present a more detailed discussion on the hydrates and challenges in obtaining the value of the surface properties in \cref{sec:hydrates}.

\subsection{The Einstein Crystal method} \label{sec:Einstein_crystal}
\indent This is a relatively recent method developed independently by \citeauthor{Addula2020}\citep{Addula2020} and \citeauthor{Yeandel2022}\citep{Yeandel2022}. The key idea is to avoid an explicit real-space transformation of bulk material into an interface by using a reference state to which both bulk and interfacial systems can be easily transformed. The chosen reference state is the Einstein crystal\citep{einstein1907,frenkel1984}, which comprises non-interacting atoms confined to individual harmonic potential wells. The primary benefit of using the Einstein crystal as a reference state is that the real-space position of the harmonic potential does not affect the total free energy of the Einstein crystal, and therefore the thermodynamic work required to rearrange an Einstein crystal in real-space is zero. Exploiting this property allows for the construction of an interfacial system from bulk material without having to identify how the atoms must rearrange to achieve the transformation.

\begin{figure}
    \centering
    \includegraphics[width=0.8\linewidth]{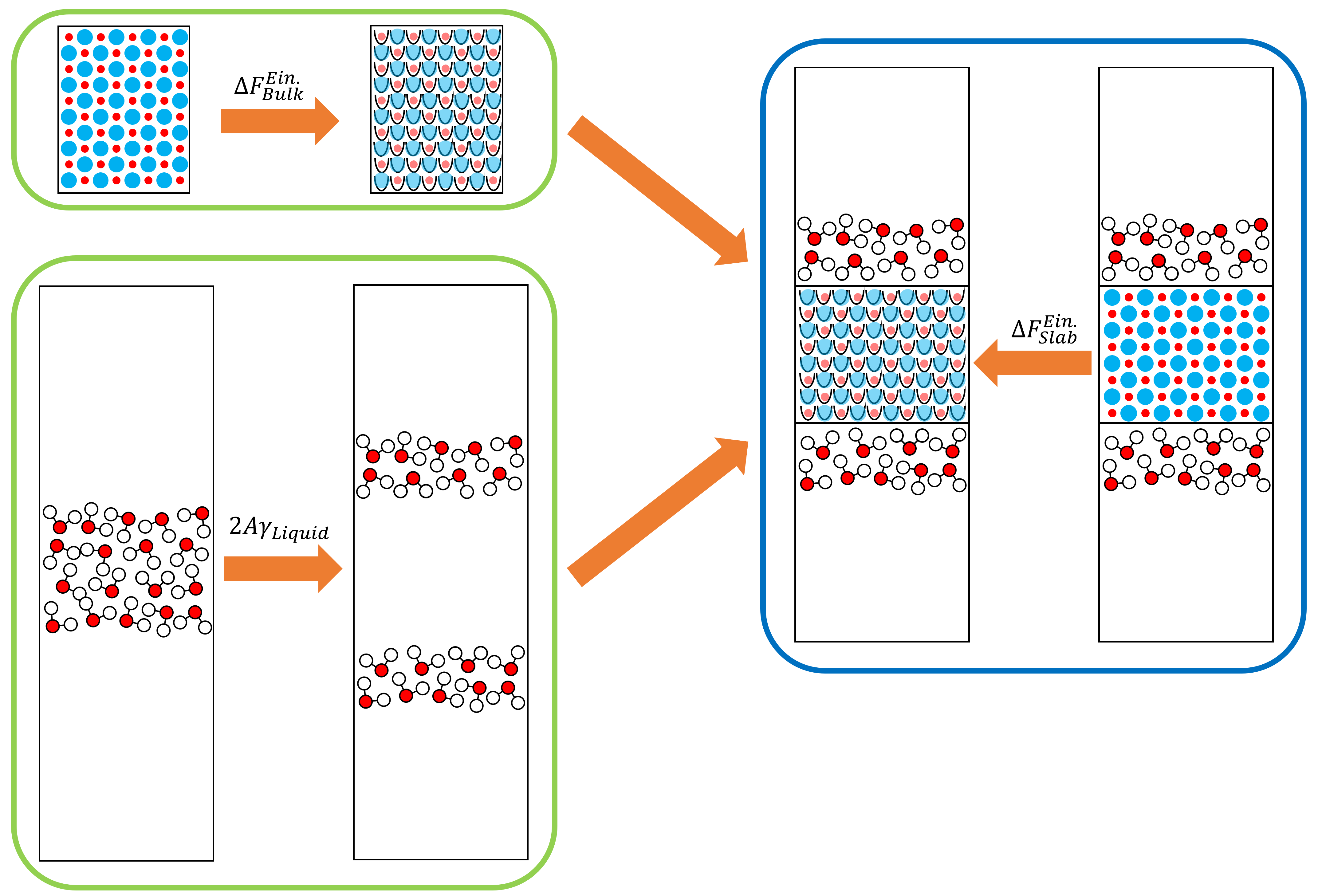}
    \caption{Top Left: Schematic of the transformation of bulk solid material into an Einstein crystal. Bottom Left: Schematic of the creation of a vacuum gap in a liquid. Right: Schematic of the transformation of solid in the slab system into an Einstein crystal. Processes in green boxes need only be performed once and the free energy scaled to match the slab system. The process in the blue box is repeated for each slab system. Reproduced from Ref. \citep{Yeandel2022} under a CC BY 4.0 license.}
    \label{einsteinfigure}
\end{figure}

The usual approach used in the Einstein crystal method is to generate a liquid-vacuum interface, and then replace the vacuum component of the interface with a solid component via an Einstein crystal (see \cref{einsteinfigure}). The stages required for the calculation are:
\begin{enumerate}
    \item Compute the free energy required to generate a vacuum gap in the bulk liquid ($\Delta \helm_{Liquid}^{Liquid+Vacuum}$).
    \item Prepare a bulk solid system and transform it into an Einstein crystal with the work required for the transformation recorded ($\Delta \helm_{Bulk}^{Ein.}$).
    \item Prepare a liquid-solid-liquid ``slab'' system with the desired interfacial configuration (crystal orientation/cutting plane/reconstruction). Transform the solid component of this system into an Einstein crystal and record the work needed ($\Delta \helm_{Slab}^{Ein.}$).
\end{enumerate}
The value of $\Delta \helm_{Liquid}^{Liquid+Vacuum}$ computed in stage 1 is equivalent to creating two liquid/vacuum interfaces, each with an area of $\area$. As this is the free energy of creating a fluid/fluid interface we may use the Shuttleworth equation (see \cref{eq:Shutt}) and identify $\Delta \helm_{Liquid}^{Liquid+Vacuum}$ with the appropriately scaled surface tension ($\Delta \helm_{Liquid}^{Liquid+Vacuum} = 2 \area \gamma_l$). The value of $\gamma_l$ can then be efficiently computed using the method of \citeauthor{Kirkwood1949} \citep{Kirkwood1949} (see \cref{eq:KB}) and re-used for every calculation with the same liquid phase.

The values of $\Delta \helm_{Bulk}^{Ein.}$ and $\Delta \helm_{Slab}^{Ein.}$ required by stages 2 and 3 may be computed using any applicable methods (e.g. TI\citep{Frenkel2001}, Bennett acceptance ratio\citep{bennett1976efficient}). The values of $\Delta \helm_{Bulk}^{Ein.}$ may be computed once for a given bulk solid and scaled for use with multiple different slab systems (surface configurations). The value of $\Delta \helm_{Slab}^{Ein.}$ is computed for each interface of interest.

Although in principle many different approaches may be used to transform the solid material into an Einstein crystal, published studies have thus far opted to use TI\citep{Addula2020,Yeandel2022} (See Appendix~\ref{sec:TI}). There are a number of different ways in which TI can be used to transform the solid into an Einstein crystal. In general a two-stage approach is preferred in which the first stage is used to ``switch on'' the harmonic potential, and then a second stage is used to ``switch off'' all solid-solid and solid-liquid interactions. This choice restricts atoms from approaching too closely as interactions are ``switched off'', which could otherwise lead to instability. Additional TI stages may also be included to further ensure stability of the transformation, such as using a ``cleaving wall'' type approach to first separate the solid and liquid components of the interfacial system before transformation to an Einstein crystal. Throughout the transformation to an Einstein crystal, the positions of the harmonic potentials must be kept stationary to ensure consistency in the TI procedure.

The IFE is then computed using the equation
\begin{equation}\label{einIFE}
    \begin{split}
    \gamma_{sx} & = \frac{\Delta \helm_{Liquid}^{Liquid+Vacuum} + \helm_{Bulk}^{Ein.} - \helm_{Slab}^{Ein.}}{2\area} \\
    & \\
    & = \gamma_l + \frac{\Delta \helm_{Bulk}^{Ein.} - \Delta \helm_{Slab}^{Ein.}}{2\area}
    \end{split}
\end{equation}
where $\Delta \helm_{Bulk}^{Ein.}$ has been scaled to match the stoichiometry of the ``slab'' system.

A key benefit of the Einstein crystal method is the ability to study specific surface configurations, which may not be directly accessible by cleaving or other simple real-space transformations (e.g. stepped surfaces, surface patterning, etc.). The use of the Einstein crystal as a reference state also allows efficient reuse of previous calculations because only a single transformation of the bulk solid needs to be calculated for many different surface configurations and/or liquid phases. The work of adhesion is also accessible by computing the transformation of the dry surface into an Einstein crystal (replacing the transformation of the bulk solid into an Einstein crystal, $\Delta \helm_{Bulk}^{Ein.}$) and discarding the creation of the liquid surface from the bulk liquid ($\gamma_l$) in \cref{einIFE}.

Another advantage of the Einstein crystal approach is that a vacuum gap may be added around the interfacial system. The use of this additional vacuum gap is that dipole corrections \citep{Ballenegger2009,Ballenegger2014} may be added to obtain consistent energies for dipolar surfaces. The additional vacuum gap and corresponding liquid/vacuum interfaces remain in place throughout the entire calculation, and so no additional correction is required to the computed interfacial free energy.

The Einstein crystal method is most appropriate for systems in which the bulk material has low solubility in the liquid phase. In highly soluble systems, or near the coexistence point, difficulties arise in the definition of which atoms belong to the solid and which to the liquid. Constraining an atom in the fluid state to a harmonic potential leads to a divergence in the TI procedure and the free energy is poorly defined. In such cases, other methods described in this section should be preferred. When the solid phase contains species that are miscible in the liquid phase (e.g., water in hydrous clays), corrections can be applied to the Einstein crystal method to obtain a consistent IFE \citep{Yeandel2022}. Although a relatively new approach, the Einstein crystal method has already been applied to a diverse set of interfaces including orcinol/chloroform and orcinol/nitromethane\citep{Addula2020}, $\ce{NaCl}$/water and $\ce{CaSO_4.xH_2O}$/water \citep{Yeandel2022} and \ce{CaCO$_3$}/water \citep{armstrong2024surface}.

\subsection{Phantom-Wall method} 
\noindent Phantom Wall \citep{Leroy2009,Leroy2010} takes its name from the fact that the liquid is separated from the solid by using a wall (for a slab configuration there will be two walls) represented by an external potential which interacts only with a portion of the system. That is to say, the walls interact only with liquid atoms, and it is completely transparent to the solid (hence the specification ``phantom''). $\gamma_{sl}$ is determined by calculating the difference in the Gibbs free energy between a configuration in which the liquid is in contact with the solid and a reference configuration in which the liquid is in contact with the walls acting as an external potential.  The thermodynamic path starts with the walls buried within the solid, sufficiently far away from the liquid to avoid any interactions with it. The walls are then moved in the direction perpendicular to the solid-liquid interface, pushing the liquid away from the solid interface. During this path, the volume of the system changes, and therefore we need to include this contribution in the calculation of the difference in Gibbs free energy.  The phantom wall allows us to determine the Gibbs free energy change per unit area, $\Delta\gamma^{PW}$, by \citep{Leroy2015}:
\begin{equation}\label{eq:PW}
	\Delta \gamma^{PW} = \gamma_{wl} + \gamma_s - \gamma_{sl} + \press_N \frac{\Delta V}{\area}
\end{equation}
where $\press_N$ is the component of the pressure tensor in the direction normal to the interface (as in \cref{eq:KB}),  $\Delta V$ is the change in the system volume after the transformation and $\area$ is the area of the solid-liquid interface. $\gamma_{wl}$ is the wall-liquid interfacial tension, whereas $\gamma_s$ represents the IFE of the solid in contact with vacuum. $\gamma_{wl}$ can be calculated through the mechanical route (\cref{eq:KB}), \citep{Kirkwood1949}) but the term $\gamma_s$ needs to be determined from the thermodynamic definition. If the value of $\gamma_s$ is not available, the methodology can only determine the work of adhesion, $W_{sl}$, between the solid and the liquid  defined as $W_{sl}=\gamma_{lv} + \gamma_s - \gamma_{sl}$ \citep{Tadmor2017,Packham1996} or the heat of immersion, defined as $\gamma_{sl} - \gamma_{s}$, given that we can identify $\gamma_s$, i.e. the IFE of a solid in contact with vacuum, with $\gamma_{sv}$, the IFE of a solid-vapour interface. The latter assumption is approximately correct for surfaces with weak interactions with the fluid \citep{Jiang2019}. The only term remaining in \cref{eq:PW}, $\Delta \gamma^{PW}$, has to be obtained through TI and the thermodynamic path chosen to calculate this term gives the name to this methodology.

The phantom-wall method was applied to the study of the behaviour of a Lennard-Jones liquid in contact with its solid \citep{Leroy2010}, water in contact with rugged graphite \citep{Leroy2011}, and water in contact with $\alpha$-quartz surfaces coated with perfluoro-dimethylsilanes \citep{Mammen2012}. The interest in systems with rough or smooth interfaces stems from the fact that nanoscale roughness can modify the hydrophobicity of an interface \citep{Wong2009}. Other applications involve the determination of the contact angle of a water-graphene system \citep{Taherian2013}.

\subsection{The Dry-Surface Method}
\noindent We also briefly discuss here the dry-surface method developed by \citet{Leroy2015}, even though it was developed primarily to calculate the work of adhesion, $W_{sl}$, \citep{Tadmor2017,Packham1996} between a solid and a liquid phase in contact. In the dry-surface method the quantity $W_{sl}$ is obtained by transforming the solid-liquid interactions into a purely repulsive interaction by turning off its attractive part. The latter effect is achieved by modifying the depth of the well of the solid-liquid interaction potential. The dry-surface method was used to determine the interfacial thermal resistance applied to the evaporation rate of droplets on a heated surface \citep{Han2017}. It was extended to three-phase systems in \citep{Yamaguchi2019, Shintaku2024} for the calculation of the work of adhesion of a droplet to a surface. In this work, a liquid droplet was detached from a solid surface which is also in contact with vapour. In Refs.  \citep{Yamaguchi2019,Yamaguchi2020erratum}, the authors used the work of adhesion determined by this method to determine the contact angle between the droplet and the solid surface predicted by the Young equation and compare it with the one observed in a three-phase system (droplet on a solid surface in contact with vapour). In Ref. \citep{Shintaku2024}, the authors determined the work of adhesion in order to calculate the line tension of a liquid droplet in contact with a solid surface. The line tension is the locus of the intersection of the three phases, the droplet, the surface, and the vapor, and although it is a concept known since the time of Gibbs, there is no satisfactory description of its behaviour in terms of the physical parameters of the system \citep{Bey2020}. As noted in \citep{Shintaku2024}, the use of the dry-surface method to calculate the adhesion work for a droplet system may not give a reversible path, essential if thermodynamic integration is to obtain the equilibrium value. Even if, as noticed in the same article, the error in the final value of the work (calculated by using different initial configurations) is small, one should be aware of these difficulties, as the property of even a small hysteresis should always be noted and carefully checked.

\subsection{Metadynamics simulations}
\noindent Metadynamics is a technique that performs an enhanced sampling of the configuration space in MD simulations and allows to reconstruct the free energy surface of a system in terms of \textit{Collective Variables} \citep{Laio2002}. We do not give any more explanation of the technique for which we refer the interested reader to the many existing reviews and articles published  (see, e.g., \citep{Barducci2011,Bussi2020} and references therein) and its different incarnations \citep{Laio2005,Iannuzzi2003}. In 2016, Angioletti \etal \citep{angioletti2010solid} proposed to use metadynamics to determine $\gamma_{sl}$ for a Lennard-Jones system using the Broughton-Gilmer potential. The idea is to use metadynamics to reconstruct the free energy surface of a system transitioning from a single solid or liquid phase to the space of configurations where two phases coexist. The difference in Gibbs free energy between these two regions at the solid-liquid equilibrium temperature is then proportional to the IFE. The collective variable to control distinguishes between the solid and liquid phases, and it is based on an order parameter depending on the relative position of a particle and its neighbours.

\subsection{Gibbs-Cahn Integration} \label{sec:Gibbs-Cahn}
\noindent We have left this last methodology to the very end of this section, as the Gibbs-Cahn integration is not strictly a technique to calculate $\gamma_{sl}$, but rather a way to determine the law of variation of $\gamma_{sl}$ with respect to thermodynamic conditions, such as pressure, temperature, and composition. All of the methodologies presented above give $\gamma_{sl}$ for a single thermodynamic point along the solid-liquid phase boundary. Finding the IFE at other thermodynamic coexistence conditions requires repeating the calculations (whichever approach is used) for the new thermodynamic point. Gibbs-Cahn integration instead allows one to obtain simple rules and derive a range of values for $\gamma_{sl}$ (for different thermodynamic conditions) by knowing at least one of its values on the thermodynamic coexistence path.

Gibbs-Cahn integration originates from a groundbreaking paper \citep{Cahn1979} (reprinted in \citep{Cahn1998}), where, as already introduced in \cref{sec:thermo}, Cahn generalised the surface thermodynamics by eliminating the artificial construction of the dividing surface proposed by Gibbs. In this formulation, the excess quantities of the interface are now expressed in the form of determinants of matrices whose entries are extensive properties of the interfacial and bulk systems, making it possible to establish a connection between the differential of the IFE and the properties of the system directly measurable in simulations. In turn, MD simulations can calculate these properties from which the IFE  can be obtained by integrating this differential over a parameter of choice (similar in spirit to the well-known Gibbs-Duhem integration \citep{Kofke1993mp,Kofke1993jcp} for the determination of the phase coexistence line).

We will now introduce the most important feature of this methodology, which includes a presentation of the Cahn model of the interface. We believe that the Cahn model is equally important in the treatment of interfaces as the Gibbs model, and we therefore discuss it in detail. Assuming a $r$-component system containing an interface, we can write the total Gibbs energy as  \citep{cahn1978thermodynamics}:
\begin{equation}
G = E - TS + PV
\end{equation}
where $P,T, E, S,$ and $V$ are the pressure, temperature, internal energy, entropy and volume, respectively.  For a bulk system without an interface, the Gibbs energy is given by 
\begin{equation}
G_b = \sum_{k=1}^{r} \mu^k N^k
\end{equation}
Here $\mu^k$ is  the chemical potential of particles of type $k$. The interfacial free  energy, $\gamma$, is given by the difference (per unit area) between the Gibbs energy of the system including the interface and that of the coexisting bulk phases:
\begin{equation}
\gamma \area = G - G_b = E - TS + PV - \sum_{k=1}^{r} \mu^k N^k
\end{equation}
where we are assuming that the solid phase is under hydrostatic stress.
Taking the differential of this quantity gives 
\begin{equation}
d(\gamma \area) = dE- T\,dS - S\,dT  +  P\,dV + V\,dP -  \sum_{i=1}^{r} \mu^k \,dN^k -  \sum_{k=1}^{r} N^k d\,\mu^k.
\label{gamma_diff_1}
\end{equation}
For a system containing a planar interface where one of the coexisting phases is a crystalline solid, the differential for  the energy, still assuming hydrostatic conditions in the crystal, is given by~\citep{Frolov2009}
\begin{equation}
dE = T\,dS - P\,dV + \sum_{i,j = 1,2} (\sigma_{ij} + \delta_{ij}P)V d\epsilon_{ij} +  \sum_{k=1}^{r} \mu^k \,dN^k 
\label{energy_diff}
\end{equation}
where $\sigma_{ij}$ and $\epsilon_{ij}$ are the $ij$ components of the stress and strain tensors, respectively, and $i$ and $j$ are elements of the set $\{1,2\}$, which represent the transverse Cartesian directions. Substituting \cref{energy_diff} into \cref{gamma_diff_1} yields
\begin{align}\label{eq:interface}
d(\gamma \area) = - S\,dT  +   V\,dP  + \sum_{i,j = 1,2} [(\sigma_{ij} +  \delta_{ij}P)V] \, d\epsilon_{ij} -  \sum_{k=1}^{r} N^k d\,\mu^k.
\end{align}
For a solid-liquid interface, in addition to \cref{eq:interface}, we also have the two Gibbs-Duhem equations for the hydrostatic bulk solid and bulk liquid:
\begin{equation}
0 = -S_s  dT + V_s dP  - \sum_{k=1}^{r} \; N^k_s d\mu^k
\label{eq:gd_solid}
\end{equation}
and
\begin{equation}
0 = -S_l  dT + V_l dP - \sum_{k=1}^{r} \; N_l^k d\mu^k,
\label{eq:gd_liquid}
\end{equation}
where the subscripts $s$ and $l$ denote properties of the bulk solid and liquid, respectively. For the set of three simultaneous linear equations (\cref{eq:interface,eq:gd_liquid}), Cahn used Cramer's rule to eliminate any selected pair of differentials $dx$ and $dy$ (e.g. $dP$ and $dN^k$) to give
\begin{eqnarray}
d(\gamma \area)& = & -[S/XY] dT + [V/XY] dP \nonumber \\
&& + \sum_{i,j = 1,2} [(\sigma_{ij} + \delta_{ij}\press)V/XY] \; d\epsilon_{ij} \nonumber \\
&& -  \sum_{k=1}^{r} \; [N^k/XY] d\mu^k,
\label{eq:cahn}
\end{eqnarray}
\noindent
where $X$ and $Y$ are the variables conjugate to the displacements $dx$ and $dy$, and the notation $[A/XY]$ is defined as
\begin{equation}
[A/XY] = \frac{\left | \begin{array}{ccc} A&X&Y\\A_l & X_l & Y_l\\A_s& X_s & Y_s\end{array} \right |}{\left | \begin{array}{cc} X_l & Y_l\\X_s & Y_s\end{array} \right | },
\label{eq:det}
\end{equation}
where quantities without subscripts refer to the entire system (solid + liquid + interface). For a single-component system ($r = 1$),  a common choice is $X = N$ and $Y = V$, which is equivalent to choosing a Gibbs dividing surface in which the excess number of particles is zero ($\Gamma = 0$). With this choice, the $dP$ and $d\mu$ terms in \cref{eq:cahn} are identically zero (because two columns in the determinant in \cref{eq:det}  are identical). Applying this choice gives 
\begin{equation}
d(\gamma \area) = -[S/NV] dT +  \sum_{i,j = 1,2} [(\sigma_{ij} + \delta_{ij}\press)V/NV] d\epsilon_{ij}
\label{eq:cahn_NV}
\end{equation}
The determinant $[S/NV]$ reduces to the total excess entropy $A\eta$, as defined in \cref{eq:fundSurf}. 
Because we assume that the system is hydrostatic, the second term on the right hand side of \cref{eq:cahn_NV} can be obtained as follows:
\begin{eqnarray} 
[(\sigma_{ij} + \delta_{ij}\press)V/NV]  &= &\frac{\left | \begin{array}{ccc} (\sigma_{ij} + \delta_{ij}P)V &N&V\\0 & N_s & V_s\\0& N_l & V_l\end{array} \right |}{\left | \begin{array}{cc} N_s & V_s\\N_l & V_l\end{array} \right | }\\\nonumber
&=& (\sigma_{ij} + \delta_{ij}P)V.
\end{eqnarray}
For simplicity, it is useful to  restrict the discussion to high-symmetry interface orientations where $\sigma_{12} = \sigma_{21}=0$, but application to lower symmetry crystal structures or orientations is straightforward. Mechanical equilibrium at the interface guarantees that
 $\sigma_{33} = -\press_{zz} = -\press $ at the interface, yielding 
\begin{equation}
d(\gamma \area) = -S dT + (\sigma_{11}  + \press)V d\epsilon_{11} + (\sigma_{22} + \press) V d\epsilon_{22} .
\label{eq:dgammaA_1}
\end{equation}
The strain can be related to the change in the interfacial area as the crystal expands as one moves along the coexistence curve:
\begin{equation}
d\epsilon_{11} = d\epsilon_{22} = \frac{d \area}{2\area}.
\end{equation}
Substituting this into \cref{eq:dgammaA_1} and dividing by $\area$  gives
\begin{align} \label{eq:diff1}
\frac{1}{\area} d(\gamma \area)&= -\eta dT + \frac{1}{\area} (\sigma_{11} + \sigma_{22} + 2\press)V \frac{d\area}{2\area}\nonumber\\ 
&= -\eta dT + \tau \frac{d\area}{\area} 
\end{align}
\noindent
where  the excess interfacial stress, $\tau$, is defined as
\begin{equation}
\tau =  \frac{1}{\area}\left (\frac{\sigma_{11} + \sigma_{22}}{2} + \press\right )V = \int^{\infty}_{-\infty} \left [ \press_{zz} - \frac{\press_{xx} + \press_{yy}}{2}\right ] dz
\label{eq:stress}
\end{equation}
where $\press_{zz}$ and $(\press_{xx}+\press_{yy})/2$ are the pressure components normal and transverse to the interface, respectively. The use of \cref{eq:diff1} requires knowledge of the excess interfacial entropy, $\eta$, which is not readily available from the simulations. To remedy this, Frolov and Mishin \citep{Frolov2009} in their work on surface free energy, and Baidakov, \etal \cite{Baidakov2006} in the context of liquid-vapor interfaces, combine the equation $\gamma = e - T\eta$ with the fact that $\eta = -(d\gamma/dT)_\area$ (from \cref{eq:diff1}) to derive 
\begin{equation}
\frac{1}{A} d(\gamma A/T) =  -\frac{e}{T^2}\, dT + \frac{\tau}{T} \, \frac{dA}{A} 
\label{eq:diff2}
\end{equation}
which relates changes in $\gamma$ to the more easily obtainable excess interfacial energy, $e$, by analogy with the familiar the Gibbs-Helmholtz equation in thermodynamics. Dividing both sides of \cref{eq:diff2} by $dT$ along the coexistence curve and using the fact that  the interfacial area, $\area$, is proportional to $\rho_s^{-2/3}$ for high symmetry crystals, where $\rho_s$ is the number density of the solid, we obtain
\begin{equation}
\left [ \frac{d(\gamma_g/T)}{dT}\right ]_\mathrm{coex} =  -\rho_s^{-2/3}\left [\frac{e}{T^2} + \frac{2 \tau}{3\rho_s T} \left (\frac{d\rho_s}{dT}\right )_\mathrm{coex} \right ].
\label{eq:gibbs-cahn}
\end{equation}
Here $\gamma_g = \rho_s^{-2/3}\gamma$ is the ``gram-atomic" IFE per surface atom defined by Turnbull\cite{Turnbull1950}. Given a value of $\gamma$ at a reference point on the coexistence curve determined by one of the methods listed in this section, \cref{eq:gibbs-cahn} can be integrated along the coexistence curve to calculate $\gamma$ at any other point on the curve using the values of $e$ and $\tau$, which are easily calculated from a single simulation. This process is far less computationally expensive than the many simulations required to do a full $\gamma$ calculation at each temperature using direct methods. 

Gibbs-Cahn integration has been successfully applied to solid-vapour and solid-liquid IFEs of metals and metal alloys \citep{Frolov2009,Frolov2009JCP}, to Lennard-Jones systems \citep{Laird2009GC}, as well as to atomistic and coarse-grained water models to investigate the dependence across the coexistence line of the liquid-vapour and liquid-solid IFEs of LJ particles and atomistic and coarse-grained models of water \citep{sanchez2024predictions}. Frolov and Mishin later extended the formalism to include the effect of non-hydrostatic stress on the solid-fluid interfaces \citep{Frolov2010,Frolov2010b}. For systems in which the solid is modelled as a static surface, such as a hard-sphere fluid at a structureless hard wall, the application of the Gibbs-Cahn formalism is simplified by the fact that there is only one Gibbs-Duhem equation and the matrices describing the excess quantities are $2\times2$. This modification has allowed the calculation of the IFE for the hard-sphere (3$D$) and hard-disk (2$D$) fluids at planar hard walls. The method has also been extended to hard-core fluids at curved interfaces in both two and three dimensions \cite{Laird2012,Davidchack2018,Martin2020surface,martin2022inside}. Analysis of the case of a hard-disk fluid inside a circular hard wall (container) \citep{martin2022inside} requires a reformulation of the Gibbs-Cahn formalism within the grand canonical distribution. 

\section{Interfacial solid-liquid free energy for benchmarked systems} 
In the previous section we have provided an account of the different methods available for the calculation of surface properties, specifically for systems involving a solid interface. In the same section we also included a wide spectrum of systems to which such approaches have been applied. However, there are some systems which occupy privileged positions in the development of the methods presented here. These systems are usually characterized by simple interaction potentials, so that they do not feature any of the complications can be often found when dealing with complex molecules and molecular crystals, which simplifies the development of the methodologies (e.g., the determination of a thermodynamic path for thermodynamic integration) and at the same time are general enough to mimic physico-chemical properties of real systems. These benchmark systems, which comprise the hard-sphere and Lennard-Jones models discussed in this section, are usually the first considered in any development of a new methodology, and for this reason, we are including a more detailed description about them. Note that the equations and discussion reported in these section use the reduced Lennard-Jones units. 

\subsection{Hard Spheres}
\subsubsection{Hard-Sphere Crystal-Melt Interface}
The hard-sphere model has been extensively tested to benchmark different computational approaches designed to evaluate melt-solid interfacial free energies. The first calculation for this system was performed in 2000 by Davidchack and Laird \citep{Davidchack2000}. Later, alternative methods such as capillary wave fluctuations \citep{Mu2005,davidchack2006anisotropic}, non-equilibrium capillary simulations \citep{davidchack2010hard1}, tethered Monte Carlo \cite{fernandez2012equilibrium}, thermodynamic integration \citep{benjamin2015crystal}, mold integration \cite{espinosa2014mold,sanchez2021fcc}, and ensemble switch \cite{schmitz2015ensemble} have been used to estimate $\gamma_{sm}$ of different crystal planes of the fcc crystal phase in hard-spheres. Additionally, the analysis of nucleation free energy barriers based on Classical Nucleation Theory \cite{becker-doring} has also provided estimates of $\gamma_{sl}$ as a function of supersaturation and at coexistence conditions by data extrapolation \cite{cacciuto2003solid,espinosa2016seeding,sanchez2021fcc}. In \cref{tablagamms_hs}, we summarise all the published values (as far as we know) of $\gamma_{sm}$ for different crystal orientations of the fcc and hcp phases, as well as the average values of $\gamma_{sl}$ ($\overline{\gamma}_{sm}$) from crystal nucleation studies. 

As can be seen, most of the calculations of $\gamma_{sm}$ for the fcc phase show that the relative IFE anisotropy of the studied planes is: $\gamma_{sm}$(100) $>$ $\gamma_{sm}$(110) $>$ $\gamma_{sm}$(111). Some of the first direct calculations \citep{Davidchack2000,mu2005anisotropic,fernandez2012equilibrium,hartel2012tension} predicted slightly higher values of $\gamma_{sm}$ for these three planes than those from Refs. \citep{davidchack2006anisotropic,davidchack2010hard1,benjamin2015crystal,schmitz2015ensemble,bultmann2020computation,sanchez2021fcc}, with values ranging from 0.60 to 0.64 $k_BT/\sigma^2$ depending on the technique and the crystal plane. More recent calculations have predicted slightly lower values: approximately 0.58-0.59 $k_BT/\sigma^2$ for the (100) plane, 0.56 $k_BT/\sigma^2$ for the (110) plane and 0.54-0.55 $k_BT/\sigma^2$ for the (111) plane, reaching a consensus through different computational techniques for the hard sphere values of the fcc crystal phase \citep{davidchack2006anisotropic,davidchack2010hard1,benjamin2015crystal,schmitz2015ensemble,bultmann2020computation,sanchez2021fcc}. In addition, predictions from nucleation studies using the CNT framework \citep{cacciuto2003solid,espinosa2016seeding,sanchez2021fcc} also agree relatively well with direct estimations of $\gamma_{sm}$ under co-existence conditions with values ranging from 0.57 to 0.61 $k_B T/\sigma^2$, as shown in \cref{tablagamms_hs}. 

Only ref. \citep{sanchez2021fcc} provides values for two additional crystal orientations of the hcp phase (because the (0001) orientation in the hcp and (100) plane of the fcc phase are equivalent). It is unclear which of these two phases would have a lower overall IFE given the small number of crystal orientations studied. However, in \citep{sanchez2021fcc} the authors used seeding calculations to estimate the average $\gamma_{sm}$ for fcc vs. hcp crystal clusters of different sizes (ranging from 300 to 95000 atoms). The values of $\gamma_{sm}$ obtained from these calculations seem to support the idea that the overall $\gamma_{sm}$ for the hcp phase is slightly higher than that of the fcc crystal. Nevertheless, the differences are within the uncertainty of the calculations for most of the clusters. Therefore, if hcp crystals indeed show slightly higher IFEs than fcc ones, the difference is likely to be minimal.    

\begin{table*}[h]
\centering
\begin{tabular}{c|c|c|c|c|c|c}
  $\gamma_{sm}$ fcc  & Technique &  $(100)$ & $(110)$ & $(111)$ & $(120)$ & $\overline{\gamma}_{sm}$ \\
     \hline
    Davidchack and Laird 2000 \cite{Davidchack2000} & CW & 0.62(1)$^*$ & 0.62(1)$^*$ & 0.58(1)$^*$ & - & - \\  
    Mu \etal 2005 \cite{mu2005anisotropic} & CF & 0.64(2) & 0.62(2) & 0.61(2) & - & - \\ 
    Davidchack \etal 2006 \cite{davidchack2006anisotropic} & CF & 0.574(17) & 0.557(17) & 0.546(16) & - & - \\
    Davidchack 2010 \cite{davidchack2010hard1} & CF & 0.582(2) & 0.559(2) & 0.542(3) & 0.567(2) & - \\
    Fernandez \etal 2012 \cite{fernandez2012equilibrium} & TMC & 0.636(11) & - & - & - & - \\
    Hartel \etal 2012 \cite{hartel2012tension} & CF & 0.639(1) & 0.600(1) & 0.600(1) & - & - \\
    Benjamin and Horbach 2015 \cite{benjamin2015crystal} &  TI & 0.596(6) & 0.577(4) & 0.556(3) & - & - \\
    Schmitz and Virnau 2015 \cite{schmitz2015ensemble} & ES & 0.581(3) & 0.559(1) & 0.544(8) & - & - \\
    Bültman and Schilling 2020 \cite{bultmann2020computation} &  TI & 0.591(11) & - & - & - & - \\
    Cacciuto \etal 2003 \cite{cacciuto2003solid} & CNT & - & - & - & - & 0.616(3) \\
    Espinosa \etal 2016 \cite{espinosa2016seeding} & CNT & - & - & - & - & 0.58(3) \\
    Sanchez-Burgos \etal 2021 \cite{sanchez2021fcc} & MI/CNT & 0.586(6) & 0.572(7) & 0.554(6) & - & 0.57(1) \\
     \hline
    $\gamma_{sl}$ hcp & & $(11\overline{2}0)$ & $(10\overline{1}0)$ & (0001) & & $\overline{\gamma}_{sm}$ \\
     \hline
    Sanchez-Burgos \etal 2021 \cite{sanchez2021fcc} & MI/CNT & 0.597(6) & 0.586(6) & 0.554(6) & - & 0.59(1) \\
\end{tabular}
\caption{Melt-solid interfacial free energy ($\gamma_{sm}$ in $kT/\sigma^2$) of the hard-sphere fcc and hcp phases for different crystal orientations as indicated by the Miller indexes. Averaged values of $\gamma_{sm}$ ($\overline{\gamma}_{sm}$) obtained from nucleation studies are also included. Various computational approaches have been employed for the determination of $\gamma_{sm}$: cleaving wall [CW], capillary wave fluctuations [CF], mold integration [MI], tethered Monte Carlo [TMC], ensemble switch [ES], thermodynamic integration [TI], and Classical Nucleation Theory analysis of nucleation free energy barriers [CNT]. $^*$Results contain systematic error, which was corrected in \cite{davidchack2010hard1}.}
    \label{tablagamms_hs}
\end{table*}

\subsubsection{Hard-sphere Fluid at Structureless Hard Walls}
The hard-sphere (HS) fluid at static structureless hard walls (HW)  is a benchmark system for the generic understanding of inhomogeneous fluids and solid-liquid interfaces and for the testing of related theoretical techniques, such as classical density-functional theory (cDFT). A ``static'' wall is one that is rigid and non-elastic and can be either patterned (such as with fixed hard spheres arranged in a regular pattern) or structureless - in other words, the wall acts as a unchanging external field, as opposed to being a dynamic coexisting solid.  The first calculation of the IFE for a HS fluid at a planar HW from atomistic simulation was due to Henderson and van Swol~\citep{Henderson1984}, who used the mechanical route (see \cref{eq:KB} discussed in \cref{sec:failure}).\footnote{The usual failure of the mechanical route for solid-liquid interfaces does not apply in this system because the solid phase is static and not an atomistic elastic solid} This method is numerically challenging because the calculation of the difference between two pressures has a high statistical error and for the highest density studied, $0.901\, \sigma^{-3}$, the IFE value is measured to be 1.8(6) $k_BT/\epsilon$.\footnote{The value actually reported was negative because they used a definition of the wall position that ignores the contribution of the external field to the value of $\gamma$, which is given by $\press\sigma/2$, where $\press$ is the pressure. In general, when comparing calculations for interfaces with static solids, it is important to note the definition of wall position used} Later application of the mechanical KB equation \cref{eq:KB} by de Miguel and Jackson (MJ)~\cite{Miguel2006} still exhibited a significant statistical error, although much improved over the earlier calculation.   
Heni and L\"{o}wen (HL)\citep{Heni1999}, followed later by Fortini and Dijkstra (FD)~\citep{Fortini2006}, calculated $\gamma$ for this system using thermodynamic integration. These calculations also included evaluations of the HS crystal/HW IFE, which together with the HS crystal-melt IFE determined earlier, can be used to evaluate if the HS crystal will wet the HW, to validate simulation evidence for HS surface prefreezing. The results showed clear evidence of partial wetting at the (100) and (110) surfaces, but were not sufficently precise to determine if prefreezing (complete wetting) can occur for (111). 

\begin{figure}[h]
    \centering
    \includegraphics[width=0.8\linewidth]{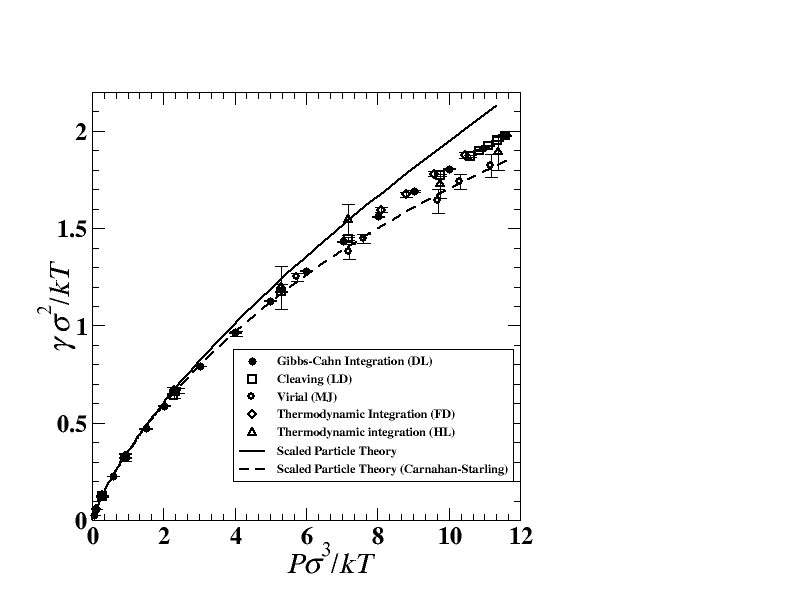}
    \caption{Summary of IFE results for the hard-sphere/hard-wall system calculated  using the Kirkwood-Buff equation (\cref{eq:KB}) (MJ)~\citep{Miguel2006}, thermodynamic integration (HL~\citep{Heni1999}, FD~\citep{Fortini2006}), cleaving walls (LD)~\citep{Laird2007} and Gibbs-Cahn integration~\citep{Laird2010}. The solid and dashed lines are theoretical results from Scaled Particle theory (SPT)~\citep{Reiss1960} and SPT using the Carnahan-Starling equation of state pressure to correct for the position of the wall, respectively.}
    \label{hswallfigure}
\end{figure}

To increase the precision of the HS/HW calculation, Laird and Davidchack (LD)~\citep{Laird2007} adapted the cleaving method to determine the wall-fluid and wall-crystal IFE for the full range of fluid pressures and demonstrated that the (111) crystal does indeed exhibit complete wetting at the surface in the presence of the fluid at densities at and just below the freezing transition. They later repeated the calculation of the wall-fluid IFE using Gibbs-Cahn integration~\citep{Laird2010} to show that this method can be used to determine the IFE over the entire fluid range with a computational effort lower than that required to determine $\gamma$ for a single density by other methods. The results for $\gamma$ for the HS fluid at a planar HW for several methods are shown in \cref{hswallfigure}. The Gibbs-Cahn integration method has also been applied to the binary hard-sphere fluid at a planar hard wall~\citep{Kern2014}.

The Gibbs-Cahn formalism can also be used to calculate $\gamma$ for the hard-sphere (and hard-disk) fluids at curved hard walls to test theories of the curvature dependence, such as the so-called Morphometric Thermodynamics (MT) \citep{Konig2004}, which states, for a 3-d system, that the curvature dependence of $\gamma$ can be determined as a linear combination of the average mean and Gaussian curvatures. Evaluation of the dependence of the IFE for the HS fluid at spherical and cylindrical walls on the wall radius shows that MT is valid except at high densities near the freezing transition\citep{Laird2012,Davidchack2018}. Similar conclusions are seen for the hard-disk fluid at a circular wall.~\citep{Martin2018, Martin2020} MT was also shown to hold for surfaces of negative curvature (a hard-disk fluid inside a circular container) except at high density and very high curvature (small radius $\rad$)~\citep{martin2022inside}. The results for these systems using classical density functional theory were also shown to be in very good agreement with simulations. 

\subsection{Lennard-Jones particles}
The Lennard-Jones potential has also been extensively used to calculate liquid-solid interfacial free energies, and in particular, the Broughton and Gilmer (BG) version of the Lennard-Jones potential is the first model for which $\gamma_{sm}$ was estimated using computer simulations in 1986 \citep{broughton1986molecular}. The methodology they used was the cleaving walls method, presented as a direct TI-based approach to compute IFEs with anisotropic resolution. In 2003 both the cleaving walls \citep{davidchack2003direct1} and the capillary fluctuations \citep{morris2003anisotropic} methods were used with the BG potential to obtain results consistent with those of Ref. \citep{broughton1986molecular} for the crystal orientations of fcc summarised in \cref{tab:tablagamms_LJ}. Later, in 2009, Laird \etal \citep{Laird2009GC} performed cleaving walls calculations to obtain $\gamma_{sm}$ for the BG Lennard-Jones potential at higher temperature and pressure than those used in Ref. \citep{broughton1986molecular}. Interestingly, they also provided a successful integration framework for estimating $\gamma_{sm}$ along the coexistence line based on a single state direct calculation (Gibbs-Cahn integration \citep{Laird2010}). This framework, which relies on the calculation of the IFE and interfacial stress of the crystal-melt interface ($f_{ij}$) has been also tested for the Lennard-Jones potential, producing consistent results in Refs. \citep{baidakov2014surface,sanchez2024predictions}. Alternative techniques, based on thermodynamic integration methods such as the Phantom Wall \citep{Leroy2010} and MI \citep{espinosa2014mold}  have also been employed to compute the interfacial free energy of fcc crystal phases of the standard Lennard-Jones potential and the BG version, respectively. In addition, values from nucleation studies using the seeding technique and a CNT analysis have also predicted consistent average values of $\gamma_{sm}$ with previous independent direct estimates for the BG potential at different pressures \citep{montero2019interfacial}. In \cref{tab:tablagamms_LJ}, we provide the reported values of $\gamma_{sm}$ for both the standard Lennard-Jones potential and the BG version from different direct and indirect calculations.  

\begin{table*}[h]
\centering
\begin{tabular}{c|c|c|c|c|c|c}
  $\gamma_{sl}$ fcc  & Technique & $T^*$ & $(100)$ & $(110)$ & $(111)$ & $\overline{\gamma}_{sm}$ \\
 \hline
 Broughton and Gilmer 1986 \citep{broughton1986molecular} & Cleaving wall & 0.617 & 0.34(2) & 0.36(2) & 0.35(1) & - \\ 
 Morris and Song 2003 \citep{mu2005anisotropic} & CF & 0.617 & 0.369(8) & 0.361(8) & 0.355(8) & - \\ 
 Davidchack and Laird 2003 \citep{davidchack2003direct1} & Cleaving wall & 0.617 & 0.371(3) & 0.360(3) & 0.347(3) & - \\
 Angioletti-Uberti \etal 2010 \cite{angioletti2010solid} & Metadynamics & 0.617 & 0.370(2) & - & - & - \\
 Espinosa \etal 2014 \cite{espinosa2014mold} & MI & 0.617 & 0.372(8) & - & 0.350(8) & - \\
 Sanchez-Burgos \etal 2024 \cite{sanchez2024predictions} & MI & 0.617 & 0.372(8) & - & 0.347(8) & - \\
*Baidakov\etal 2013 \cite{baidakov2014surface} & Cleaving & 0.617 & 0.430(4) & 0.422(4) & 0.408(4) & - \\ 
 Montero de Hijes \etal 2019 \cite{montero2019interfacial} & CNT & 0.617 & - & - & - & 0.358(3) \\
 Davidchack and Laird 2003 \cite{davidchack2003direct1} & Cleaving & 1.0 & 0.562(6) & 0.543(6) & 0.508(8) & - \\
 Sanchez-Burgos \etal 2024 \cite{sanchez2024predictions} & MI & 1.0 & 0.562(8) & - & 0.510(8) & - \\
 Montero de Hijes \etal 2019 \cite{montero2019interfacial} & CNT & 1.0 & - & - & - & 0.543(6) \\
 Davidchack and Laird 2003 \cite{davidchack2003direct1} & Cleaving & 1.5 & 0.84(2) & 0.82(2) & 0.75(3) & - \\
Sanchez-Burgos \etal 2024 \cite{sanchez2024predictions} & MI & 1.5 & 0.845(9) & - & 0.815(9) & - \\
\end{tabular}
\caption{Melt-solid interfacial free energy ($\gamma_{sm}$ in $\epsilon/\sigma^2$) of the BG Lennard-Jones potential (* for a standard Lennard Jones-potential; see Ref. \cite{baidakov2014surface}) for different crystal orientations of an fcc crystal in contact with its melt at coexistence conditions of the temperatures indicated. Averaged values of $\gamma_{sm}$ ($\overline{\gamma}_{sm}$) obtained from nucleation studies are also included. The same abbreviations for the employed techniques as in \cref{tablagamms_hs} have been used.}
    \label{tab:tablagamms_LJ}
\end{table*}

All calculations of $\gamma_{sm}$ for the different crystal planes (100), (110), and (111)) at the triple point temperature ($T^*$=0.617) of the BG Lennard-Jones potential agree independently of the technique employed within the uncertainty. Only the first value provided by Broughton and Gilmer for the (100) plane slightly underestimates the most recent calculated IFEs by different groups. Furthermore, nucleation studies using the seeding technique \citep{montero2019interfacial} also report an average value of $\gamma_{sm}$ that perfectly matches direct estimates under coexistence conditions of different groups (\cref{tab:tablagamms_LJ}). On the other hand, for the standard Lennard-Jones potential (shifted and truncated at 2.5 $\sigma$), the values of $\gamma_{sm}$ at the same temperature are approximately 10$\%$ higher \citep{baidakov2014surface}. This is expected because the potential shape is not equivalent to the BG expression. However, the fact that the relative interfacial anisotropy of the crystal orientations matches that found for the same planes using the BG Lennard-Jones potential gives credibility to these independent calculations. At higher temperatures and pressures, cleaving techniques \citep{davidchack2003direct1}, Mold Integration \citep{espinosa2014mold}, and seeding techniques \citep{montero2019interfacial} have been also used to obtain the value of $\gamma_{sm}$. The agreement at $T^*$=1 between direct estimations for different orientations is excellent, as well as the extrapolation of values from critical fcc clusters under supercooling conditions \citep{montero2019interfacial}. For $T^*$=1.5, the reported IFEs for the (100) plane match within the uncertainty, while those for the (111) orientation differ somewhat \citep{sanchez2024predictions}. More work is probably required to clarify the origin of this small discrepancy. However, the Lennard-Jones model, and in particular the BG version, have excellent potential to validate novel techniques to estimate $\gamma_{sm}$ alongside the hard-sphere model.  

\section{Interfacial free energy for realistic systems} \label{sec:real}
In the previous section, we have shown how some indirect and direct simulation techniques have been used in the literature to deal with two of the most standard reference systems characterized by simple intermolecular interaction potentials, i.e., the hard-sphere and Lennard-Jones examples. Although this strategy is appropriate for developing new methodologies in computer simulation, sooner or later it is necessary to extend the applicability of new techniques to more complex systems. Increasing complexity, in the case of solid-fluid interfaces, can be due to more complex intermolecular interactions between the molecules forming the system and/or more complex solid crystalline structures.

In this section, we concentrate on the determination of ice-aqueous solution IFEs, with particular emphasis on pure water systems and a particular class of aqueous solutions that are able to form clathrate hydrates \citep{Sloan2007a,Ripmeester2022a}.

\subsection{Water}
Water is probably ``the simplest'' molecule that can be found in a solid, liquid, and gas phase in nature under ordinary temperature and pressure conditions. In addition to its significance in our daily lives, water is also a fascinating subject of study because of its remarkable properties. Both condensed phases, liquid and solid, present a series of anomalies compared to other compounds \citep{Debenedetti1996a,Gallo2016a}.  The complexity of the liquid-solid water phase diagram, the thirteen solid ice structures \citep{Petrenko2002a,Salzmann2006a,Salzmann2006b,Falenty2014a,delRosso2016a}, and the existence of a liquid-liquid phase transition are of particular interest \citep{Poole1992a,Debenedetti2020a}. Taking this into account, it is easy to understand why obtaining a deep understanding of the factors that control the homogeneous nucleation of ice in water, including the solid-liquid IFE, is still a formidable challenge. 

Several authors have reported in the literature their findings obtained from different simulation techniques, thermodynamic conditions, and water models. The first time $\gamma_{sm}$ was computed entirely from molecular simulations was in 2005 by Haymet \emph{et al.} \citep{Haymet2005}. They determined the $\gamma_{sm}$ predicted by the SPC/E water model for the basal ice Ih--water interface obtaining a value of $39(4)\,\text{mJ/m}^2$. However, this result was of limited value because it was obtained using the mechanical route, without considering any corrections. In 2007,  Wang \emph{et al.} \citep{Wang2007} determined $\gamma_{sm}$ for the ice Ih-water interface through molecular dynamics simulations using TIP4P-Ew and TIP5P-Ew water models and the indirect superheating-undercooling hysteresis method \citep{Luo2005}. In this work, the values of $\gamma_{sm}$ obtained for both TIP4P-Ew and TIP5P-Ew models, at $1\,\text{bar}$ and the corresponding melting temperature, were $37$ and $42\,\text{mJ/m}^2$, respectively. Although the values obtained in these two papers agree with each other, they are higher than the experimental data reported in the literature. As Wang \emph{et al.} \citep{Wang2007} claimed in their work, more accurate values of $\gamma_{sm}$ can be computed by using more rigorous techniques such as the Cleaving or Capillary Fluctuation Methods. 

In 2008, Handel \emph{et al.} \citep{Handel2008} determined $\gamma_{sm}$ for the first time using a direct simulation technique, the Cleaving Method. In this pioneering work, Handel \emph{et al.} determined $\gamma_{sm}$ for the three principal crystal ice Ih plane, basal, primary prismatic (pI), and secondary prismatic (pII), using molecular dynamics simulations and the TIP4P water model. The $\gamma_{sm}$ values obtained in this work for the same three principal planes of ice Ih in contact via a planar interface with pure water at ambient pressure and coexistence temperature were $23.8(8)$, $23.6(10)$, and $24.7(8)\,\text{mJ/m}^2$, respectively. Later, the same authors extended their original work \citep{Davidchack2012ice} to two other water models: TIP4P-Ew and TIP5P-Ew. In addition, they revisited the $\gamma_{sm}$ results obtained by the TIP4P water model to go beyond the truncated the electrostatic interactions used in the original work  by using Ewald sums to account for the full electrostatic interaction. The new $\gamma_{sm}$ values obtained by using the Cleaving Method and the TIP4P model were $24.5(6)$, $27.6(7)$, and $27.5(7)\,\text{mJ/m}^2$ for the basal, pI, pII Ih ice planes, respectively. They obtained similar results by using the TIP4P-Ew model, $25.5(7)$, $28.9(8)$, and $28.3(7)\,\text{mJ/m}^{2}$,  and the TIP5P-Ew model, $27.8(9)$, $27.4(8)$, and $31.6(7)\,\text{mJ/m}^{2}$, for the three principal planes of ice Ih at $1\,\text{bar}$ and the coexistence temperature. In all cases, the agreement between simulation and experimental data taken from the literature was very good.

Benet \emph{et al.} \citep{benet2014study} determined $\gamma_{sm}$ using the Capillary Fluctuation method for the basal, pI, and pII ice Ih--water interface obtaining $27(2)$, $28(2)$, and $28(2)\,\text{mJ/m}^2$, respectively, using the TIP4P / 2005 water model at $1\,\text{bar}$ and coexistence temperature. Some years later, Ambler \emph{et al.} \citep{ambler2017solid} applied the same technique to determine the average Ih, Ic, and 0 ice--water  $\gamma_{sm}$ value using the coarse-grained monatomic water (mW) model. In all cases, they obtained a $\gamma_{sm}$ value around $35\,\text{mJ/m}^2$ at $1\,\text{bar}$ and the melting temperature. It is interesting to note that the results obtained by Ambler \emph{et al.}\citep{ambler2017solid} are $\approx 20\%$ higher than those reported by Benet \emph{et al.} \citep{benet2014study} even when the technique employed in both works is the same. However, this can be explained by noting that $\gamma_{sm}$ is extremely sensitive to molecular model details, and the water models used in both works, as well as their respective ice-water coexistence temperatures.

As explained in \cref{sec:CNT}, $\gamma_{sm}$ can be related to the free energy barrier, $\Delta G_{crit}$, required for the formation of a critical solid nucleus in the middle of a homogeneous liquid. Although this is an indirect method for determining $\gamma_{sm}$ through the calculation of $\Delta G_{crit}$ and has some shortcomings (see \cref{sec:CNT,sec:curved,sec:nucleation} for more details), this approach has great versatility because $\Delta G_{crit}$ can be determined from different simulation techniques such as Forward Flux Sampling, Umbrella Sampling, and Seeding. In 2011, Li \emph{et al.} \citep{li2011homogeneous} studied homogeneous ice nucleation at $1\,\text{bar}$ from supercooled water using the Forward Flux Sampling Method, molecular dynamics simulations, and the mW water model. The nucleating ice embryo contains ice Ic and Ih structures in a 50/50\% fraction. Combining their findings with CNT they estimated a $\gamma_{sm}$ value of $31.01(21)\,\text{mJ/m}^2$. A year later, Reinhardt and Doye \citep{reinhardt2012free} studied the homogeneous nucleation of ice from supercooled liquid water with Monte Carlo simulations using the Umbrella Sampling Method and the mW water model. By combining their findings with CNT they obtained the value of $\gamma_{sm}$ at $1\,\text{bar}$ and supercooling conditions ($23.0$ and $24.0\,\text{mJ/m}^2$ at $220$ and $240\,\text{K}$ respectively). Later, the same authors \citep{Reinhardt2013} extended their results to the TIP4P/2005 water model, obtaining a $\gamma_{sm}$ value of $24.0$ and $26.1\,\text{mJ/m}^2$ at 240 (supercooled conditions) and $252\,\text{K}$ (melting temperature).  Interestingly, the two models yielded the same value of $\gamma_{sm}$ at $240\,\text{K}$ even though the water model and the degree of supercooling were different; however, their results seem to be lower than those reported by Li \emph{et al.} \citep{li2011homogeneous}. As Reinhardt and Doye claimed in their work \citep{reinhardt2012free}, these discrepancies arise because the two groups used different order parameters to monitor the number of water molecules in the solid critical cluster.

In 2013, Sanz \emph{et al.} \citep{sanz2013homogeneous} combined for the first time the Seeding Method and CNT to determine the ice Ih-water $\gamma_{sm}$ at 1 bar and the melting temperature using the TIP4P/2005 and TIP4P/Ice water models. They obtained, for both models, a value for $\gamma_{sm}$ of 28.9 mJ/m$^2$, in very good agreement with the results obtained by Reinhardt and Doye for the TIP4P/2005 water model \citep{Reinhardt2013}. A year later, some of the authors of the original work of Sanz \emph{et al.} \citep{sanz2013homogeneous} employed the same methodology and determined the ice Ih-water $\gamma_{sm}$ predicted by the TIP4P, TIP4P/Ice, TIP4P/2005, and mW water models at 1 bar and in a broad range of supercooled temperature conditions \citep{Espinosa2014}. By extrapolating $\gamma_{sm}$ to the melting temperature for each model, they reported $\gamma_{sm}$ values of 25.6, 30.8, 29.0, and 29.6 mJ/m$^2$ for the TIP4P, TIP4P/Ice, TIP4P/2005, and mW water models, respectively. The same authors extended this study to determine the ice Ic--water $\gamma_{sm}$ at 1 bar and the corresponding melting temperature predicted by the TIP4P/Ice model \citep{Zaragoza2015}. They obtained a $\gamma_{sm}$ value of 31(3) mJ/m$^2$, in very good agreement with their previous results. The similar results obtained for both Ih and Ic ice--water interfaces are in good agreement with the fact that both ice structures present the same nucleation rate, which means that under these conditions the formation of the two ice I polymorphs is equally favoured. Very recently, Zanotto \emph{et al.} \citep{tipeev2024exploring} have employed the same methodology to determine the Ic--water $\gamma_{sm}$ at 1 bar and supercooled conditions (215-240 K) using the mW water model. They obtained an average value of $27.5(11)\,\text{mJ/m}^{2}$ for the crystal nuclei seeded in the supercooled water. 

Although the seeding + CNT combination have demonstrated to provide very reliable results, it is still an indirect way to evaluate solid--fluid IFEs. On the other hand, the MI methodology (see \cref{sec:mold_integration}) proposed by Espinosa \emph{et al.} \citep{espinosa2014mold} provides a direct and relatively simple method that predicts solid--fluid IFEs from a fundamental point of view. The same authors of the original work where the MI method was proposed, determined the ice Ih-water $\gamma_{sm}$ value using the TIP4P/Ice, TIP4P/2005, TIP4P, and mW water models with the MI method. They calculated the ice Ih-water $\gamma_{sm}$ at $1\,\text{bar}$ and the melting temperature of each water model, for the three main planes of the ice Ih (basal, prism I, and II) obtaining an average value of $29.8(8)$, $28.9(8)$, $27.2(8)$, and $34.9(8)\,\text{mJ/m}^{2}$  for the TIP4P/Ice, TIP4P/2005, TIP4P, and mW water models, respectively \citep{espinosa2016ice}. They also calculated the ice Ic-water $\gamma_{sm}$ value for three different ice Ic planes in contact with the water phase ((100), (110), and (111)), obtaining an average value of $30.1(8)\,\text{mJ/m}^{2}$. As the authors claimed in previous work \citep{Espinosa2014}, there are no significant differences in the ice-water IFE between the ice I polymorphs. These results are in very good agreement with those reported previously in the literature using Seeding + CNT \citep{sanz2013homogeneous,Espinosa2014,Zaragoza2015}, Umbrella Sampling + CNT \citep{Reinhardt2013}, Capillarity Fluctuation method \citep{benet2014study}, and the cleaving method \citep{Handel2008,Davidchack2012ice}. Shortly after the determination of ice Ih-water $\gamma_{sm}$ value at $1\,\text{bar}$ and the melting temperature, some of the authors of the original work extended that previous study and determined $\gamma_{sm}$ at $2000\,\text{bar}$ to analyze the effect of the pressure on the interfacial free energy using the MI methodology and the TIP4P/Ice water model \citep{espinosa2016interfacial}. They obtained an increase of $\gamma_{sm}$ with pressure of $\sim10\,\text{mJ/m}^{2}$. In the same work, they determined the ice Ih basal plane--water $\gamma_{sm}$ at $2000$ and $5000\,\text{bar}$ using the mW water model and the MI methodology. As for the case of the TIP4P/Ice water model, they observed an increase of $\gamma_{sm}$ when the pressure is increased. In the same work \citep{espinosa2016interfacial}, the authors determined the ice 0--water $\gamma_{sm}$ at $1\,\text{bar}$ using the mW water model and the MI technique, obtaining a value of $35.4\,\text{mJ/m}^{2}$. All of the results obtained in this work were also obtained by the Seeding + CNT combination, obtaining excellent agreement with those obtained by the MI methodology. Finally, it is worth mentioning that recently the MI methodology has been employed to determine the basal ice Ih--water $\gamma_{sm}$ at coexistence temperatures from $-2600$ to $500\,\text{bar}$ using the TIP4P/Ice water model \citep{Montero2023}. This study was carried out by some of the authors of the original work of Espinosa \emph{et al.} \citep{espinosa2014mold} and they have reported a $\gamma_{sm}$ minimum of $26(1)\,\text{mJ/m}^{2}$ around $\sim-2000\,\text{bar}$ and a poorly reproduced pressure dependence of $\gamma_{sm}$ below $500\,\text{bar}$.

Sanchez-Burgos \emph{et al.} \citep{sanchez2024predictions} have determined $\gamma_{sm}$ along the ice Ih-water coexistence line from single-state calculations utilizing the Gibbs-Cahn integration method \citep{Cahn1998}. They used the mW water model and the results previously obtained using the MI methodology by some of them \citep{espinosa2016interfacial} as the initial single-state $\gamma_{sm}$ value. As the authors claimed, they find an excellent agreement between the results obtained following the MI methodology  \citep{espinosa2016interfacial} and those obtained from the  Gibbs-Cahn integration approach, proving the power of this approach to quantify the dependence of the IFE along a coexistence line.

\subsection{Hydrates}\label{sec:hydrates}
Clathrates are non-stoichiometric inclusion compounds where guest molecules, such as methane (CH$_{4}$), carbon dioxide (CO$_{2}$), hydrogen (H$_{2}$), nitrogen (N$_{2}$), and tetrahydrofuran (THF), are trapped within cavities formed by a periodic network of associating molecules, or host \citep{Sloan2007a,Ripmeester2022a}. These associating molecules interact not only by Lennard-Jones forces, but also through specific, short-range, and highly directional interactions, that cause the network arrangement of the system. When the associating system is formed by water molecules, the association is mediated through hydrogen bonding, and clathrates are also known as hydrates. Hydrates crystallize into several distinct structures \citep{Sloan2007a,Ripmeester2022a} and also exhibit proton-disorder, satisfying the Bernal-Foller rules \citep{Bernal1933a},  as do various phases of ice, including Ih ice.

However, hydrates are much more complex than ice. The nature and concentration of guest molecules in a hydrate greatly affect the stability conditions of these compounds as well as the crystalline structure adopted by the hydrate. Hydrates of small molecules, such as CO$_{2}$ or CH$_{4}$, crystallize in the so-called sI structure. The unit cell of this structure, which exhibits cubic symmetry, is formed from 46 water molecules distributed in 6 T (tetrakaidecahedron or 5$^{12}$6$^{2}$) cages and 2 D (pentagonal dodecahedron or 5$^{12}$) cages, usually denoted as ``small'' and ``large'' hydrate cages \citep{Sloan2007a,Ripmeester2022a}. The hydrates of medium-size molecules, such as iso-butane, propane, cyclopentane, and THF crystallize in the sII structure, which also shows cubic symmetry. The sII unit cell is more complex than the sI structure. It is composed by 136 water molecules distributed in 16 D (pentagonal dodecahedron or 5$^{12}$) cages and 8 H (hexakaidecahedron or $^{51}$6$^{4}$) cages \citep{Sloan2007a,Ripmeester2022a}. The D or “small cages” are the same in both structures, but the “large cages” (H) are larger in the sII structure, allowing them to accommodate larger molecules. The sII structure has the peculiarity that it can be stabilized by medium or small molecules, such as H$_{2}$ or N$_{2}$ through multiple occupancy of the H cages \citep{Alavi2005a,Alavi2006a,Barnes2013a}.

According to the literature, the CO$_{2}$ and CH$_{4}$ hydrates exhibit mainly a single occupancy in each cage but in such a way that each unit cell can accommodate eight CO$_{2}$ or CH$_{4}$  molecules \citep{Sloan2007a,Brumby2016a,Henning2000a,Udachin2001a,Ikeda1999a,Ripmeester1998a}. However, THF only occupies the H cages (5$^{12}$6$^{4}$) of the sII hydrate structure \citep{Makino2005a,Manakov2003a}. The T cages (5$^{12}$) remain empty and can be occupied by other guest molecules of small size and low molecular weight. Finally, the formation of sII structures of hydrates with small molecules such as N$_{2}$ and H$_{2}$ is unusual. However, the non-stoichiometric nature of hydrates offers the possibility of multiple cage occupancy. The explanation of the preference to form sII hydrates instead of sI hydrates is that the N$_{2}$ and H$_{2}$ molecules better stabilize small hydrate cages, which are in a greater number in the sII crystallographic structure \citep{Michalis2022a,Tsimpanogiannis2018a,Kuhs1997a,Chazallon2002a,Sasaki2003a}. Multiple occupancy of these molecular cages is another complexity that makes hydrates fascinating and very complex substances to model and understand from a molecular perspective. Obviously, the prediction of hydrate-water IFEs is not an exception.

The first calculation of the CH$_{4}$ hydrate-water IFE was performed by Jacobson and Molinero more than ten years ago \citep{Jacobson2011a}. Water molecules were modelled using the mW water model \citep{molinero2009water}. The guest molecule, which the authors call M \citep{Jacobson2010d,Jacobson2010c}, is also represented by a single particle with properties intermediate between CH$_{4}$ and CO$_{2}$. They performed seeding simulations at $50\,\text{MPa}$ of a slab of M liquid in contact with a saturated water solution with M, containing clusters of M hydrates of different sizes to determine the melting temperatures of the crystalline nuclei. Combining these results with the well-known Gibbs-Thomson relationship \citep{Handa1992a,Clennell1999a,Henry1999a}, it is possible to estimate the M hydrate-water IFE, $\gamma_{sx}$. They obtained a value of $\gamma_{sx}\approx 36(2)\,\text{mJ/m}^{2}$. Note that this value corresponds to the case of a hydrate in which the guest molecule has intermediate properties between CH$_{4}$ and CO$_{2}$. The value obtained agrees well with the experimental data obtained by Uchida \emph{et al.} \citep{Uchida1999a,Uchida2002a} and Anderson \emph{et al.} \citep{Anderson2003a,Anderson2003b} for the real CH$_{4}$ hydrate-water interface, $\gamma_{sx}=34(6)$ and $32(3)\,\text{mJ/m}^{2}$, respectively. It is also in good agreement with the experimental CO$_{2}$ hydrate-water interfacial energy values independently obtained by the same authors, $\gamma_{sx}=28(6)$ and $30(3)\,\text{mJ/m}^{2}$. One year later, Knott \emph{et al.} \citep{Knott2012a} used the mW model for water and a single-site Lennard-Jones potential for methane molecules to predict the IFE of the CH$_{4}$ hydrate using seeding simulations in combination with CNT \citep{ZPC_1926_119_277_nolotengo,becker-doring,gibbsCNT1} (see \cref{sec:CNT} for more details). They obtained a value for the IFE, $\gamma_{sx}=31\,\text{mJ/m}^{2}$, also in good agreement with experimental data taken from the literature. 

More recently, Grabowska \emph{et al.} \citep{Grabowska2022a,Grabowska2023a} have estimated homogeneous nucleation rates for the CH$_{4}$ hydrate from seeding simulations at $400\,\text{bar}$ for a supercooling of $35\,\text{K}$ ($260\,\text{K}$) using the TIP4P/ice model \citep{Abascal2005b} and a Lennard-Jones centre to model methane \citep{Guillot1993a,Paschek2004a}. Using simulations and CNT, they compared $\gamma_{sx}$ values for two critical clusters found at $400\,\text{bar}$ and $260\,\text{K}$ as a function of their radius and extrapolated to the planar limit (see Fig.~14 in the work of Grabowska \etal \citep{Grabowska2023a}). Their calculations suggest a value of around $38\,\text{mJ/m}^{2}$ for the CH$_{4}$ hydrate-water planar interface. The coexistence temperature of this hydrate at $400\,\text{bar}$ is $290\,\text{K}$, approximately. Interfacial free energy values under super-cooling conditions \emph{increase} as the temperature increases (at constant pressure) \citep{Sanz2013a}. Thus, the results of this work seem to suggest a higher value of $\gamma_{sx}$ for the planar CH$_{4}$  hydrate–water interface than for the CO$_{2}$ hydrate-water IFE of a planar interface. The value found by Grabowska \emph{et al.} \citep{Grabowska2023a}  for the CH$_{4}$ hydrate-water planar interface from simulation is higher than the experimental value found by Anderson \emph{et al.} \citep{Anderson2003a,Anderson2003b}, which is equal to $32\,\text{mJ/m}^{2}$. However, the value found from seeding simulations seems to be consistent with the preliminary results obtained by Zer\'on \emph{et al.} \citep{Zeron2024b} using the two extensions of the MI technique to estimate the hydrate-water IFEs. In particular, these authors have obtained a value of $43(2)-44(1)\,\text{mJ/m}^{2}$ for CH$_{4}$ hydrate-water IFE using the same molecular models for water and CH$_{4}$ (see below for further details).


All of the work just presented uses indirect methods to determine the CH$_{4}$ and CO$_{2}$ hydrate-water IFE, including the combination of seeding simulations with CNT or the use of the Gibbs-Thompson relationship. However, as discussed in \cref{sec:mold_integration}, Algaba and collaborators have obtained the interfacial free energy of CO$_{2}$ hydrate-water using the MI-H and MI-G methodologies \citep{Algaba2022b,Zeron2022a,Romero-Guzman2023a}. In both cases, water molecules are modelled using the well-known TIP4P/Ice \citep{Abascal2005b} and the TraPPE-UA force field for CO$_{2}$ molecules \cite{Potoff2001a}. In the first case, they obtained a value of $\gamma_{sx}=29(2)\,\text{mJ/m}^{2}$ and in the second case $\gamma_{sx}=30(2)\,\text{mJ/m}^{2}$. Both values are in excellent agreement with the experimental data of Uchida \emph{et al.} \citep{Uchida1999a,Uchida2002a}, $28(6)\,\text{mJ/m}^{2}$, and Anderson \emph{et al.} \citep{Anderson2003a,Anderson2003b}, $30(3)\,\text{mJ/m}^{2}$, discussed above. 


While CH$_{4}$ and CO$_{2}$ hydrates crystallize in the sI structure, many other aqueous solutions form hydrates in the more complex sII structure. Some authors have used the MI technique to estimate IFEs of two different hydrates. Torrej\'on \emph{et al.} \citep{Torrejon2024a} have predicted the THF hydrate-water IFE using the MI-H technique at $500\,\text{bar}$, under the coexistence conditions of the univariant two-phase coexistence line of the hydrate. This hydrate exhibits a sII crystallographic structure, more complex than the sI structure of the CH$_{4}$ and CO$_{2}$ hydrates. The IFE obtained, $27(2)\,\text{mJ/m}^{2}$, is in excellent agreement with the experimental data taken from the literature, $24(8)\,\text{mJ/m}^{2}$ \citep{Lee2007a,Zakrzewski1993a}. 

\section{The role of interfacial free energy in crystal nucleation} \label{sec:nucleation}
\noindent In \cref{sec:CNT} we discuss how CNT could be used to extract the values of IFE. In this section, we detail, instead, how the IFE can be used to explore nucleation and some of its challenges in light of our discussion in previous sections. However, it is not our intention to add another full-scale review on the subject of nucleation; we refer readers who want to be introduced to the vast literature on nucleation to the relevant chapters in \citep{Myerson2019}, and then to reviews such as  \citep{Wu2004,Kashchiev2003,Kashchiev2003b,Karthika2016,Gebauer2018,Wang2022,sosso2016crystal}.

The Classical Nucleation Theory developed by Volmer-Weber and Becker-Döring \cite{volmer_weber, becker-doring, kelton_book} aims at describing homogeneous nucleation, although it has been extended to the heterogeneous case (we refer the interested reader to the classical works of Turnbull \cite{Turnbull1950het}, Fletcher \citep{Fletcher1958het}, and \citep{Kelton2010het,Sear2007het,Kashchiev2000} for a modern account). Nucleation is defined in terms of the nucleation rate $J_{CNT}$, which represents the number of critical nuclei, $N_c$, that appear per unit of time and volume.
The nucleation rate is defined in terms of a product of a kinetic factor, $J_{kin}$, describing the rate of attachment of particles to the growing cluster, and a thermodynamic factor, $J_{thd}$, related to the free energy barrier of nucleation. To date, the thermodynamic contribution has received much more attention in the literature. In general, the nucleation rate can be factorized as follows
\begin{equation}
J_{CNT} =J_{kin}J_{thd} = J_{kin}\exp{\left(-\frac{\Delta G_{crit}}{k_B T} \right)}
\label{nucl_rate}
\end{equation}
where $\Delta G_{crit}$ is the thermodynamic barrier to nucleation. This is strongly dependent on the IFE as can be seen from \cref{eq:CNTgamma}, where $\Delta G_{crit}\propto\gamma_{sl}^3$. The nature of this term will depend on the rate-determining step of the nucleation mechanism. 

In studying systems with solid-melt and solid-semi-solid interfaces such as metals, modelling strategies utilising CNT are now employed in phase prediction and precipitation studies. These can be collectively termed classical nucleation growth theories (CNGTs). These models have been built largely by exploring the kinetic pre-factor in CNT \cite{Becker1935,Zeldovich1943}. The kinetic pre-factor in this case is given by
\begin{equation}
 J_{kin} = \rho_l Z f^{+} 
 \label{kinetic}
 \end{equation}
 where $f^{+}$ is the rate of attachment of formula units to the growing cluster, $Z$ is the so-called Zeldovich factor, and $\rho_l$ is the number density of formula units in the fluid phase. In crystallization from solutions, the latter term represents the number density of solute molecules in solution, whereas in freezing from melt, it represents the number density of molecules in the melt. The rate of attachment is a feature of atoms hopping from the matrix phase into the nucleus and also of the flow of matter into the matrix surrounding the nucleus. There have been many different interpretations of this term, which is often estimated using forms of jump frequency and diffusion coefficients \cite{Russell1968,Turnbull1949,Turnbull1956}.
 The Zeldovich factor, instead, can be expressed as \cite{Becker1935,Zeldovich1943}:
\begin{equation}\label{eq:Zldf}
Z = \frac{1}{2 \pi \rad_{C}^2\rho_s}\sqrt{\frac{\gamma_{sl}}{k_B T}},
\end{equation}
where $\rho_s$ is the number density of the solid and $\rad_{C}$ is the critical nucleus radius \citep{Perez2008,perez2009corrigendum}. We will have a more precise discussion of the Zeldovich factor and the meaning of $\rad_{C}$ in the next section, in which the problem of curved interfaces is introduced (see \cref{sec:curved} and \cref{zeldovich_pedagogical}). As can be seen from \cref{eq:Zldf}, the evaluation of the Zeldovich factor requires knowledge of the IFE. In the first developments of nucleation theory, such a quantity was not readily available (either from experiments or calculations). Moreover, as we will discuss in more detail in \cref{sec:curved}, even if $\gamma_{sl}$ is available from experiments for a planar interface, the value that enters in CNT should be the value for the curved interface, which is in general different from the one for planar interfaces. 


The CNT has been extended in several directions to include different effects. One of these extensions is related to the inclusion of an incubation time, $\tau$, to account for the time required for the clusters to reach a steady state with their environment. In fact, in some applications, such as the solution deposition of organic thin films \citep{deBrujin2024}, the process is far from equilibrium. Therefore, the transient concentration of critical nuclei in such systems differs from its steady-state value.
One model that accounts for transient nucleation was derived by Kampmann and Wagner \citep{Kampmann1984} and describes the variation of the number of nuclei over time, $N_c$, as \cite{Perez2008,perez2009corrigendum}:
\begin{equation}
    \frac{d N_c}{dt} = \rho_l Zf^+ \textrm{exp}\left({-\frac{\Delta G_{crit}}{k_B T}}\right)\textrm{exp}\left({-\frac{\tau}{t}}\right).
\end{equation}
A range of numerical methods have been built around this model, with the Kampmann-Wagner numerical model becoming one of the most popular due to its few basic assumptions and ability to work with grain coarsening\cite{Perez2008,Du2021}. This model has also been extended and implemented in various ways to explore precipitation dynamics e.g. \cite{Perez2008, Wang2024}. The expression of the incubation time $\tau$ depends on the IFE (see \citep{Perez2008}), further increasing the dependence of the theory on this parameter.

It should be clear by now the important role $\gamma_{sl}$ plays in the theory of nucleation. Therefore, in order to make the theory usable for calculations, different approximations were proposed for it. One of the first rather crude approximation was given by Turnbull \citep{Turnbull1950,Bai2008} for crystal-melt systems, based on experimental results, which equates the IFE to the latent heat of fusion $\Delta H_f$:
\begin{equation}\label{eq:Turnbullapp}
\gamma_{sl} = \alpha\rho_s^{2/3}(\Delta H_f/N_A)
\end{equation}
where $\alpha$ is an empirical coefficient (about 0.45 for most metals; about 0.32 for most non-metals \citep{Laird2001}, although the precise value seems to be system-dependent, see e.g., value reported for hcp metals \citep{Sun2006IFEhcp}) and $N_A$ is the Avogadro number. 
 The approximation shown in \cref{eq:Turnbullapp} can give reasonable results when dealing with systems including one or two dominant species where the values of solid-melt IFE is generally thought to be of the order of $\sim$10 mJ/m$^2$, and little distinction is made between enthalpy and free energy. Moving to multi-component systems such as high entropy alloys where there is often much more heterogeneity at the interface, using \cref{eq:Turnbullapp} may become more problematic. Moving beyond metals to ionic melts, a greater directionality appears around atom positioning, and therefore a much greater degree of anisotropy can be found \citep{sanchez-burgos2023nacl,espinosa2015crystal}, which further increases the uncertainty in the calculation of the IFE with such approximated empirical laws.

More challenges emerge when we consider nucleation from solution. The use of atomistic simulations to study nucleation from solution have recently been reviewed by Finney and Salvalaglio \citep{Finney2024}, building on the reviews of Agarwal and Peters \citep{Agarwal2014} and Sosso\etal \citep{Sosso2016}. We therefore limit our discussion to issues concerning the IFE. There are several compilations of IFEs extracted from crystallisation data; the most extensive is probably  that of S\"{o}hnel (1982) \citep{sohnel1982electrolyte}. As with Turnbull's analysis of metal solid-melt interfaces (i.e., \cref{eq:Turnbullapp} reported above), other empirical correlations have been proposed to estimate the IFEs for solutes-solvent systems. These fit reasonably well to expressions of the form \citep{sohnel1982electrolyte}:
\begin{equation}
    \gamma_{sl} = a \log_{10} C_{\rm eq} + b
\end{equation}
where $C_{\rm eq}$ is the solubility of solute in water and $a$, $b$ are fitted parameters\cite{sohnel1982electrolyte}. 

For crystallization from solution, the kinetic prefactor (the term $J_{kin}$ in \cref{nucl_rate,kinetic}) is again a measure of the attachment frequency of the formula units but now the rate-determining step in the mechanism may be the desolvation of the ions in solution. This latter effect can be large and dominate the overall activation energy for nucleation, and an example is given by the nucleation of LiF in water \citep{lanaro2018influence}. Zimmermann \etal \citep{Zimmermann2015} have argued that the kinetic term in the nucleation of NaCl is also dominated by a dehydration mechanism, although in this case it is still possible to estimate the IFE from the thermodynamic term. Typical values of the kinetic prefactor, $J_{kin}$ are of the order of $10^{37\pm 3}$ s$^{-1}$m$^{-3}$ \citep{lamas2021homogeneous,Wang2022}.
Sometimes, in the literature predictions are reported of the $J_{CNT}$ obtained from CNT deviating from experimental values  by 20-30 orders of magnitude\citep{Zanotto1985},  or more\citep{Fokin2000}). However it should be pointed out that since the experimental value of $\gamma$ for most of the systems are unknown, the failure of CNT is most likely due to the use of an incorrect value of $\gamma$. 
Simulations have proved particularly useful to determine the accuracy of CNT as one can compare values of $J_{CNT}$ obtained from simulation with estimates obtained from  CNT for the same potential model. The case of the nucleation rate for the precipitation of NaCl from a supersaturated aqueous solution has been extremely useful. When the Joung-Cheatham potential \citep{Joung2008} for NaCl and SPC/E for water \citep{Berendsen1987} is used, the values of $J_{CNT}$ obtained from CNT where in quite good agreement \citep{lamas2021homogeneous} (i.e., with deviations of about 3-4 orders of magnitude only) with the exact values obtained from the path sampling technique known as forward flux sampling \citep{thanos2018a,thanos2018b}. This latter fact illustrates that CNT works quite well if the correct value of $\gamma$ is used \citep{lamas2021homogeneous,Zimmermann2015,Zimmermann2018}. However, if compared to experiments, the nucleation rates obtained for the same force-field is about 10 orders lower than those observed \citep{thanos2018a,lamas2021homogeneous}. The fact that the comparison between calculated quantities (with the same force-field but different approaches) gives consistent results, but when compared with experiments the agreement deteriorates illustrates the deficiencies of the force field rather than  the nucleation theory itself. In fact, using a polarizable force field for $J_{CNT}$ improves the predictions significantly \cite{thanos2018b}.

 In \cref{tab:tablnacl}, we report the values (both experimental and calculated) for the IFE of the solid-liquid interface between crystalline NaCl and molten NaCl (top) and brine (bottom). For the solid-liquid interface of molten NaCl the values of $\gamma$ are located around 90 $mJ/m^2$ (except for the value reported by Zykova-Timan \etal \citep{zykova2005physics} which is much lower). For the solid-liquid interface with the aqueous solution there is some scatter. Notice that simulations values could be affected by deficiencies of the force field in describing the real system. Also as will be discussed in the next section, values of $\gamma$ for a planar interface are not necessarily identical to those of spherical clusters due to the presence of curvature effects in $\gamma$. For the planar NaCl-aqueous solution interface values seems to be located around 125$mJ/m^2$. For clusters (nucleation) values seems to be lower and around 80 $mJ/m^2$ (notice that the value of Sohnel was obtained using very old experimental data\citep{sohnel1982electrolyte} and the value of Cedeno et al.\citep{cedeno2023microdroplets}  used for the kinetic prefactor J$_{kin}$  a value of $10^{22}$ s$^{-1}$m$^{-3}$) . With the recommended value J$_{kin}$ $10^{37}$ s$^{-1}$m$^{-3}$ their values of $\gamma$ would be in the range 65-85 $mJ/m^2$.

\begin{table*}[h]
\centering
\begin{tabular}{l|c|c|l}
                                                          & $\gamma_{sm}$ & face     &  Technique \\
\hline
Buckle and Ubbelohde 1960 \citep{Buckle1960}              & 84            & clusters  & expt: homogeneous nucleation \\ 
\\
Zykova-Timan \etal   2005 \citep{zykova2005physics}       & 37            & (100)    & calc: contact angle$^a$ \\
Espinosa \etal       2015 \citep{espinosa2015crystal}     & $100\pm 10$, $114\pm 10$ & (100), (111) & calc: mold integration$^a$  \\ 
Benet \etal          2015 \citep{benet2015interfacial}    & $89\pm 6$, $88\pm 6$     & (100), (114) & calc: capillary fluctuations$^a$ \\
\hline
                                                          & $\gamma_{sx}$ & face     & Technique                       \\
\hline
S\"{o}hnel           1982 \citep{sohnel1982electrolyte}   & 38            & clusters  & expt: nucleation  \\
Na \etal             1994 \citep{Na1994}                  & 87            & clusters  & expt: nucleation   \\
Cedeno \etal         2023 \citep{cedeno2023microdroplets} & 47.5--61.9    & clusters & expt: nucleation \\
Bulutoghu \etal       2022 \citep{Bulutoglu2022}          & 60-98         &  clusters   & calc: nucleation Two-step model$^{bc}$ \\
                      \\
 Yeandel \etal         2022 \citep{Yeandel2022}            & 128           & (100)        & calc: Einstein crystal$^{bd}$ \\
 
Sanchez-Burgos \etal  2023 \citep{sanchez-burgos2023nacl} & $104\pm 18$, $153\pm 11$ & (100), (111) & calc: MI$^{bc}$ \\
Sanchez-Burgos \etal  2023 \citep{sanchez-burgos2023nacl} & 137 & Planar average & calc: MI$^{bc}$ \\
Lamas \etal      2021 \citep{lamas2021homogeneous,sanchez-burgos2023nacl}         & 150        & Planar average  & calc: fit to seeding simulations$^{bc}$ \\

\end{tabular}
\caption{Melt-solid ($\gamma_{sm}$) and solution-solid ($\gamma_{sx}$) interfacial free energies in mJ/m$^2$ for NaCl. Force-fields used in calculations:
$^a$NaCl BMHFT  model\citep{Tosi1964,Fumi1964}, $^b$NaCl Joung-Cheatham model\citep{Joung2008}, $^c$water: SPC/E\citep{Berendsen1987}, $^d$water: SPC/Fw\citep{Wu2006flexible}. The empty lines separate results obtained for spherical clusters obtained from nucleation studies (experimental or simulation) from those obtained for a planar interface.} 
\label{tab:tablnacl}
\end{table*}

The CNT theory as discussed in this section includes several simplifying approximations which lead to short-comings of the theory. These have frequently been discussed in the literature (see e.g. \cite{Karthika2016,Reiss1997}) and we also present in \cref{sec:curved} a development of the formulation of the CNT theory in which some of these approximations are removed. Two of the most important approximations usually considered in CNT theory (or at least the most related to the current review) are
\begin{itemize}
\item The clusters grow by one unit (often a molecule or a formula unit) at a time to form a spherical cluster with a sharp interface and a crystal structure that is identical to that of the bulk \cite{Kashchiev2000}. The assumption of a single scalar value for $\gamma$ is reasonable if the nucleus is amorphous (as it will be for a small cluster). However, if the nucleus is faceted the correction is simple. In that case, the IFE is a weighted average over all the facets, given by
\begin{equation}
    \area_{\rm total}\bar{\gamma} = \sum_{\rm\{hkl\}} \area_{\rm\{hkl\}}\gamma_{\rm\{hkl\}}
\end{equation}
where $\area_{\rm total}$ is the total surface area of the nucleus, $\bar\gamma$ is the effective (scalar) IFE and the sum is over all the planes with Miller indices \{hkl\} exhibited by the nucleus. Note the constraint $\sum_{\rm\{hkl\}} \area_{\rm\{hkl\}} = \area_{\rm total}$. The areas $\area_{\rm\{hkl\}}$ can be found using the Wulff construction.
\item The interfacial free energy between the solid and liquid (whether it is a melt or a solution) is constant (independent of the temperature) and equal to the value for an infinite plane (i.e., the curvature of the cluster can be ignored). This is usually part of the \emph{capillary approximation}. Note that sometimes the failures of CNT reported in the literature correspond to failures of CNT within the capillary approximations. 
\end{itemize}
Of these two approximations, the second is more relevant for the present discussion.
For CNT to work within the capillary approximation, it is sufficient that the value of  $\gamma$ does not change much with curvature, but this is not a priori guaranteed.
As discussed in \cref{sec:curved}, we cannot ignore the effects of the curvature of the interface for clusters which certainly much smaller that macroscopic size and microscopic size effects become non-negligible (see \citep{Merikanto2007}, although it focuses on nucleation in liquid-vapour systems).
Simulations by Montero~de~Hijes\etal \citep{deHijes2020} using a hard-sphere model and a spherical solid cluster have shown that the interfacial free energy is a function of the size of the cluster (see discussion of \cref{eq:Tolman} in \cref{sec:curved}). The same conclusions were found for other systems (LJ, water) for values of $\gamma$ obtained from seeding simulations \citep{espinosa2016seeding,sanchez2021fcc,sanz2013homogeneous}). 
These findings also suggest that when using CNT expressions to fit experimental results on the nucleation rate, the value of $\gamma$ obtained from the fit includes curvature effects and does not correspond to the value of $\gamma$ of a planar interface. In recent years, simulations using seeding methods have become increasingly popular because they offer a way to avoid the problems associated with the capillary approximation \citep{garaizar2022alternating,lamas2021homogeneous}. 

\begin{table*}[h]
\centering
\begin{tabular}{l|c|c|c|l}
                                  & $\gamma_{sx}$ & face    &  Technique        \\
 \hline
 S\"{o}hnel and Mullin 1983 \citep{sohnel1982precipitation}  & 98            & average        & expt: homogeneous nucleation    \\
 Liu and Lim            2003 \citep{liu2003microgravity}      & 170           & average        & expt: homogeneous nucleation \\  
 R{\o}yne \etal         2011 \citep{royne2011experimental}    & 150           & $(10\bar{1}4)$ & expt: sub-critical cracking \\
 Ja{\'n}czuk \etal      1986 \citep{janczuk1986spreading}     & 98            & $(10\bar{1}4)$ & expt: contact angle       \\
 Okayama \etal          1997 \citep{Okayama1997}              & 72            & $(10\bar{1}4)$ & expt: contact angle       \\
 Hadjittofis \etal      2021 \citep{hadjittofis2021tauhe}     & 55            & $(10\bar{1}4)$ & expt: inverse gas chromatography \\
 Forbes \etal           2011 \citep{forbes2011nanocalcite}         & $1480\pm 210$ & average        & expt: calorimetry         \\
\hline
DeLeeuw and Parker      1998 \citep{deLeeuw1998}              & 160           & $(10\bar{1}4)$ & calc: internal energy     \\
Duffy and Harding       2004 \citep{Duffy2004}                & 140           & $(10\bar{1}4)$ & calc: internal energy     \\
Kvamme \etal   2009 \citep{kvamme2009modelling}               & 288           & $(10\bar{1}4)$ & calc: internal energy     \\
Bruno \etal    2013 \citep{bruno2013estimate}                 & $412\pm 20$   & $(10\bar{1}4)$ & calc: free energy (estimate) \\
Armstrong \etal 2024 \citep{armstrong2024surface}              & 205           & $(10\bar{1}4)$ & calc: free energy         \\
\end{tabular}
\caption{Solution-solid ($\gamma_{sx}$) interfacial free energies in mJ/m$^2$ for CaCO$_3$ (calcite)}
\label{tab:tabcaco3}
\end{table*}

An illustration of the problems encountered in the comparison of experimental and calculated values for the solid-liquid IFE is shown in \cref{tab:tabcaco3} which gives a (not comprehensive) set of experimental and calculated values for the case of calcium carbonate. Two measurements of the heat of immersion ($q_{imm}=\gamma_{sx}-\gamma_{sv}$) have been omitted (\citep{wade1959heats,goujon1976crystallinity}) because they imply a negative value for $\gamma_{sx}$ for any reasonable estimate of the surface-vapour free energy, $\gamma_{sv}$. In addition, we omitted the value reported in \citep{costa2018experimental} because it is based on the dubious estimate of the free energy given by \citep{bruno2013estimate}. Finally, the value of \citep{forbes2011nanocalcite} seems unreasonably large; in fact, Wang \etal \citep{Wang2022} have argued that it is so large that, if correct, the nucleation of calcite would never be seen. The spread of the values in experimental numbers (55-170 mJ/m$^2$) is similar to that seen in the NaCl values, once the unreliable values are removed. The set of calculated values also have some issues: The three values labeled "internal energy" assume that the configurational energy is a reasonable proxy for the enthalpy and further that the enthalpy is a reasonable proxy for the free energy (i.e. that the entropic contribution is negligible). Bruno \etal \citep{bruno2013estimate} do attempt to estimate the entropic contribution, but their whole calculation assumes that a continuum approximation can be used to describe the water. The calculated free energy of \citep{armstrong2024surface} does not suffer from these problems, but there is the inevitable question of the accuracy of the force field, since the value they obtain is at the high end of the range of experimental values. However, this comparison assumes that the capillary approximation holds, which, as discussed above, is questionable. This suggestion is reinforced by the recent work of Darkins \etal \citep{Darkins2024Nucleation} who used the first nucleation theorem and experimental nucleation rates under a wide range of conditions to determine the number of formula units in the critical cluster. The low values obtained (an average of about 10 formula units) are small enough to rule out prenucleation cluster pathways, and the lack of dependence of the values on the saturation index rules out the capillary approximation.

Despite its shortcomings, CNT continues to provide the framework within which much experimental work continues to be analysed. No other theory combines its simplicity and practical utility. However, even if corrections are made to account for curvature effects, there is an additional issue: Central to CNT is the assumption that the nucleation pathway is characterised only by the size of the cluster, with its ordering being identical to that of the final bulk phase. This discounts the increasing volume of evidence for the importance of clusters (of varying density and composition including dense amorphous phases) in many systems (see \citep{Gebauer2011} for a discussion of pre-nucleation clusters; \citep{Addadi2003,DeYoreo2003} for amorphous phases). 

Many authors have therefore concluded that it is time to look for an alternative approach. We cannot do justice here to the considerable literature on this topic in recent years. Gebauer \etal \cite{gebauer2022nonclass} have produced an interesting map of the territory. Authors such as Kashchiev\citep{kashchiev2020classical} and Jia \etal \citep{jia2024nonclas} continue to argue for CNT-based approaches. A group of theories, the so-called ``mesoscopic nucleation theories'' \citep{Lauer2022,Lutsko2018NJP,Lutsko2018PRE,Lutsko2019} have been advanced to remedy the most fundamental deficiencies of CNT.  The simplest versions of these theories add a second-order parameter to represent the mean inner density of the cluster, thus permitting density fluctuations to evolve independently of the cluster size. However, such theories do not produce the simple analytic connection between the thermodynamic nucleation barrier and the interfacial free energy found in the classical theory. 

Here, our main contribution to this topic is the revision of the concepts of the CNT where the assumption of planarity of the interface between the solid and liquid phases is removed. The latter discussion is one of the main topics of the next section.

\section{Thermodynamics of curved interfaces: an approach to nucleation}\label{sec:curved}
\begin{figure}
    \centering
    \includegraphics[width=0.4\linewidth]{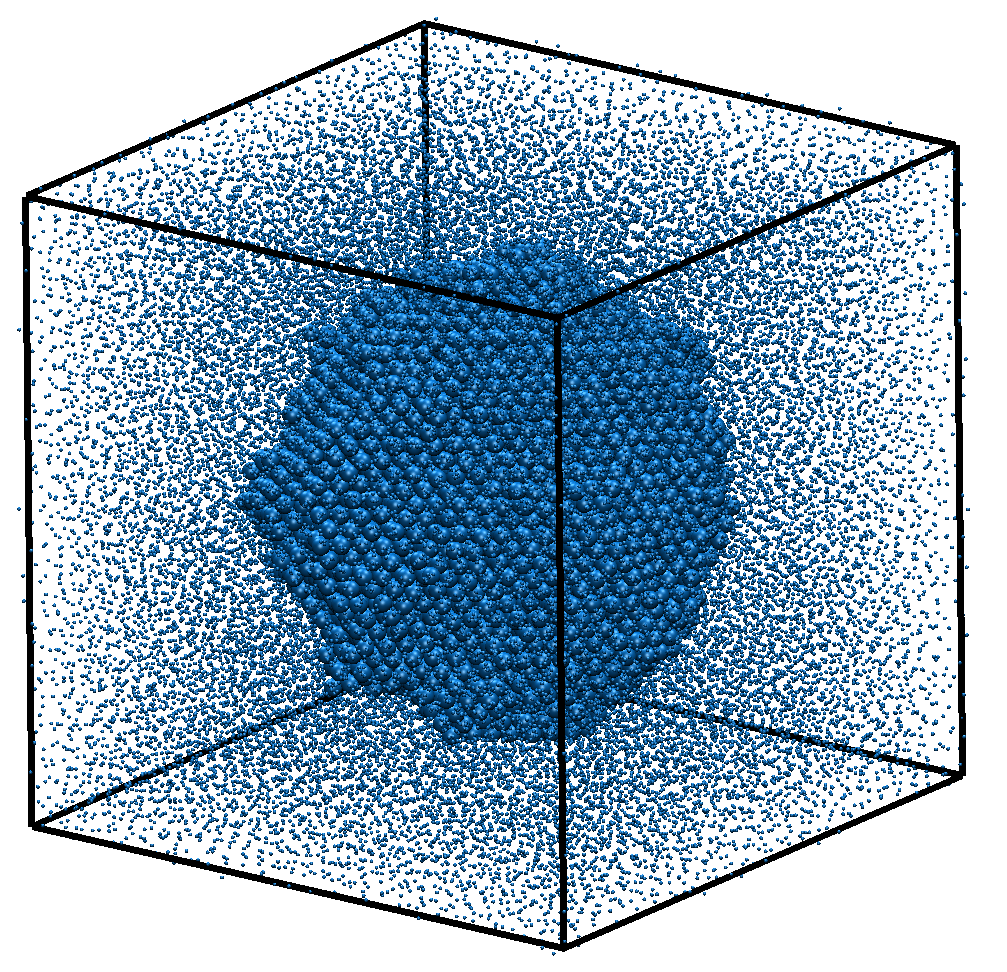}
    \caption{ A solid cluster of hard spheres is shown in equilibrium with the surrounding melt in the NVT ensemble. For clarity, liquid particles are depicted at a reduced size. The system is in thermodynamic equilibrium, meaning that molecular motion ensures both temperature and chemical potential are uniform throughout the system, although, the pressure is not uniform in this case. Reproduced with permission from Ref. \citep{montero2022thermodynamics}. Copyright 2022  by American Institute of Physics.  }
    \label{snaphot}
\end{figure}

\noindent As discussed above, the capillarity approximation, which is often applied within CNT, frequently fails. Therefore, any attempt of applying CNT to probe interfacial properties should consider the effect of curved surfaces. Curved solid-liquid interfaces play a crucial role in various processes such as crystal nucleation from melts or solutions \cite{sosso2016crystal, de2003principles}, nanoparticle sintering \cite{campbell2013energetics, castro2016sintering}, and liquid storage in porous media \cite{lazarenko2022melting, morrow1970physics}.  Technical difficulties may explain the limited amount of development in this part of the field. Experimentally, setting up a system where a curved interface is stable is difficult, if not impossible. Therefore, direct measurements are not practicable. However, over the past two decades computer simulations have significantly contributed to clarifying key thermodynamic aspects of solid-liquid curved interfaces. Here, we focus on the spherical interface formed by a solid nucleus (s) surrounded by a molten phase (l) in a single-component system. \cref{snaphot} shows a snapshot of a system where such configuration is stable in the NVT ensemble \footnote{Under certain conditions, a solid nucleus can remain stable within its melt in various ensembles (NVT, NVE, or NPH) \cite{rusanov1964thermodynamics, yang1985thermodynamical}, as demonstrated in computer simulations \cite{statt2015finite,binder2017,zepeda2017extraction, montero2020interfacial, montero2020young, gunawardana2018theoretical}}. 

Since the system is in equilibrium, the IFE, $\gamma_{sl}$, can be defined from \citep{kondo1956thermodynamical, rowlinson2013molecular}:
\begin{equation}     
           \helm  =  N \mu - \press_s \vol_s - \press_l \vol_l + \gamma_{sl} \area   ,
        \label{rowlinson_book_1}         
\end{equation}
where $\helm$ represents the Helmholtz free energy, $N$ is the total number of particles, $\mu$ is the chemical potential, $\area$ is the interfacial area, $\press_s$ and $\press_l$ are the pressures of the solid and liquid phases, respectively, and $\vol_s$ and $\vol_l$ are their respective volumes (with the total volume given by $\vol = \vol_s + \vol_l$).
It is important to note that \cref{rowlinson_book_1} applies to both planar and curved interfaces. For further extensions to other curved interfaces and multi-component systems, see Refs. \citep{sekerka2015thermal, koenig1950thermodynamic}.

According to Gibbs, we assume that the system consists of a solid up to a certain Gibbs dividing surface, with liquid outside. Is the value of $\gamma_{sl}$ affected by the choice of the location of the dividing surface? For a planar interface, the answer is no—since in this case $\press_s = \press_l$, moving the interface does not change the area $\area$, and thus $\gamma_{sl}$ remains invariant to the choice of the dividing surface. However, for curved interfaces, the value of $\gamma_{sl}$ \emph{does} depends on the arbitrary choice of the dividing surface. In this scenario, $P_s$ and $\press_l$ differ, and changing the dividing surface alters $\vol_s$, $\vol_l$, and $\area$. Therefore, we must express the IFE of a given thermodynamic state with a curved interface as $\gamma[\rad]$, where the brackets imply that $\gamma_{sl}$ changes with the arbitrary choice of the dividing surface and we drop the subscript $sl$ for notational simplicity.

Since $\helm$ (and similarly $\mu$) must remain invariant to the choice of the dividing surface, we can take the notational derivative of \cref{rowlinson_book_1} with respect to the radius of the cluster denoting the dividing surface, $\rad$, and set it to zero, yielding:
 \begin{equation}    
   \Delta \press = \press_s - \press_l =  \frac{2\gamma}{\rad} + \left[ \frac{d\gamma}{d\rad} \right].   
   \label{eq:YLgen}   
\end{equation}
The derivative in square brackets represents a notational derivative, describing how $\gamma$ changes with the \emph{chosen} position of the dividing surface for a given system. Gibbs suggested a particular choice, known as the surface of tension, where $\gamma[\rad]$ reaches its minimum. The radius at which this minimum occurs is denoted $\rad_C$, and the corresponding value of $\gamma$ at this minimum is $\gamma_C$. Therefore, we have:
\begin{equation}         
   \Delta \press = \frac{2 \gamma_C}{\rad_C}.
   \label{eq:YL_naive}     
\end{equation}
The Young-Laplace equation only holds when using the radius at the surface of tension and the corresponding value of $\gamma$ at that surface, which highlights the fact that when reporting values of $\gamma$ for curved interfaces, it is essential to specify the choice of the dividing surface. For example, another commonly used dividing surface is the equimolar surface, $\rad_E$, defined by $N = \rho_s \vol_s[\rad_E] + \rho_l \vol_l[\rad_E]$, where the number of excess surface molecules is zero. However, the Young-Laplace equation does \emph{not} apply to this surface, and the value of $\gamma = \gamma_E$ for this surface is higher than $\gamma_C$ \cite{rowlinson1993thermodynamics, rowlinson2013molecular, buff1951spherical}.
A general expression can be used to describe $\gamma$ at any $\rad$ if $\gamma_C$ and $\rad_C$ are known:
\begin{equation}     
   \gamma [\rad] = \gamma_C \frac{2\rad^3 + \rad_C^3}{3\rad^2\rad_C},     
\end{equation}
which has been extended to account for cylindrical interfaces in Ref. \citep{montero2022thermodynamics}.

\begin{figure}
    \centering
    \includegraphics[width=0.4\linewidth]{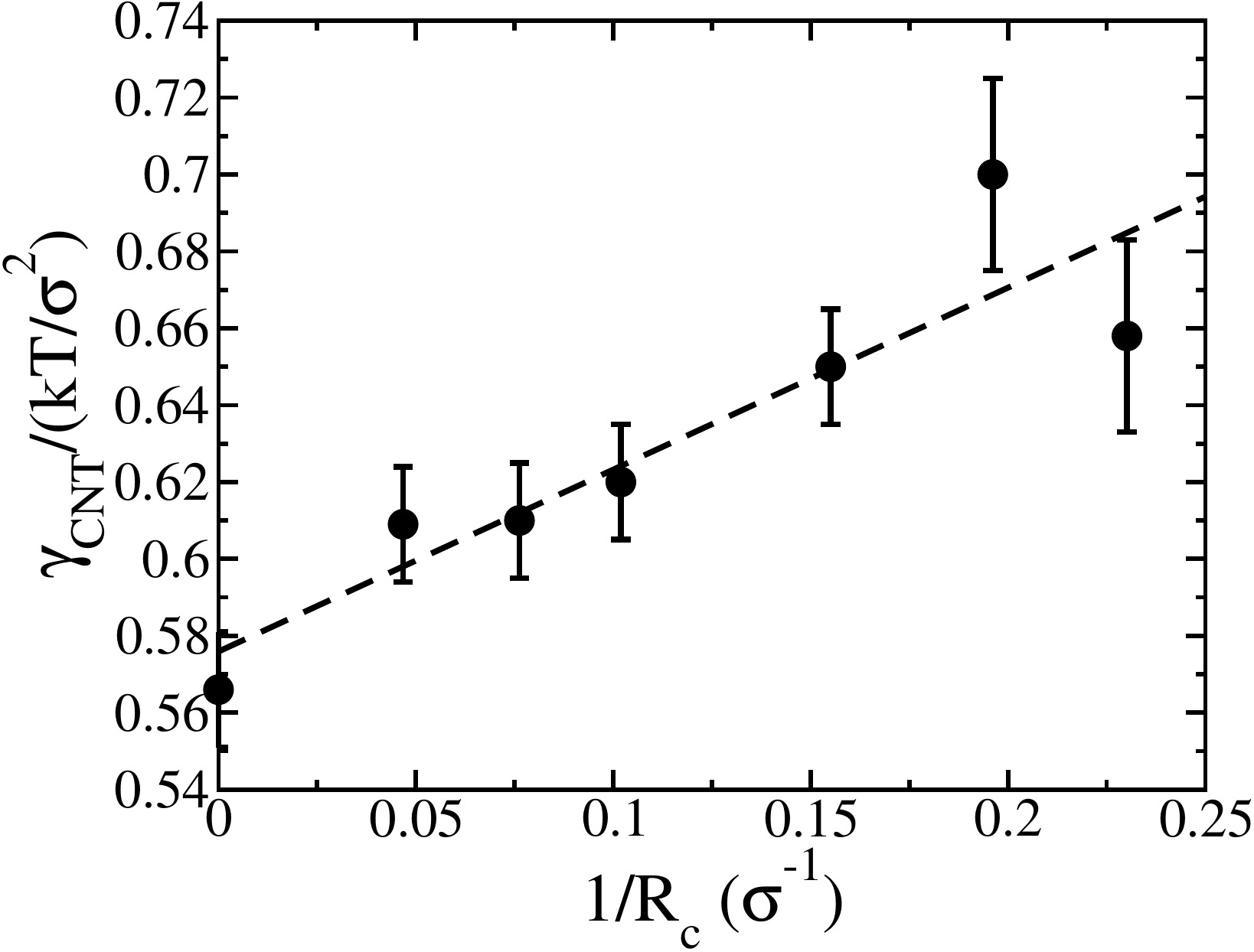} a) 
    \includegraphics[width=0.4\linewidth]{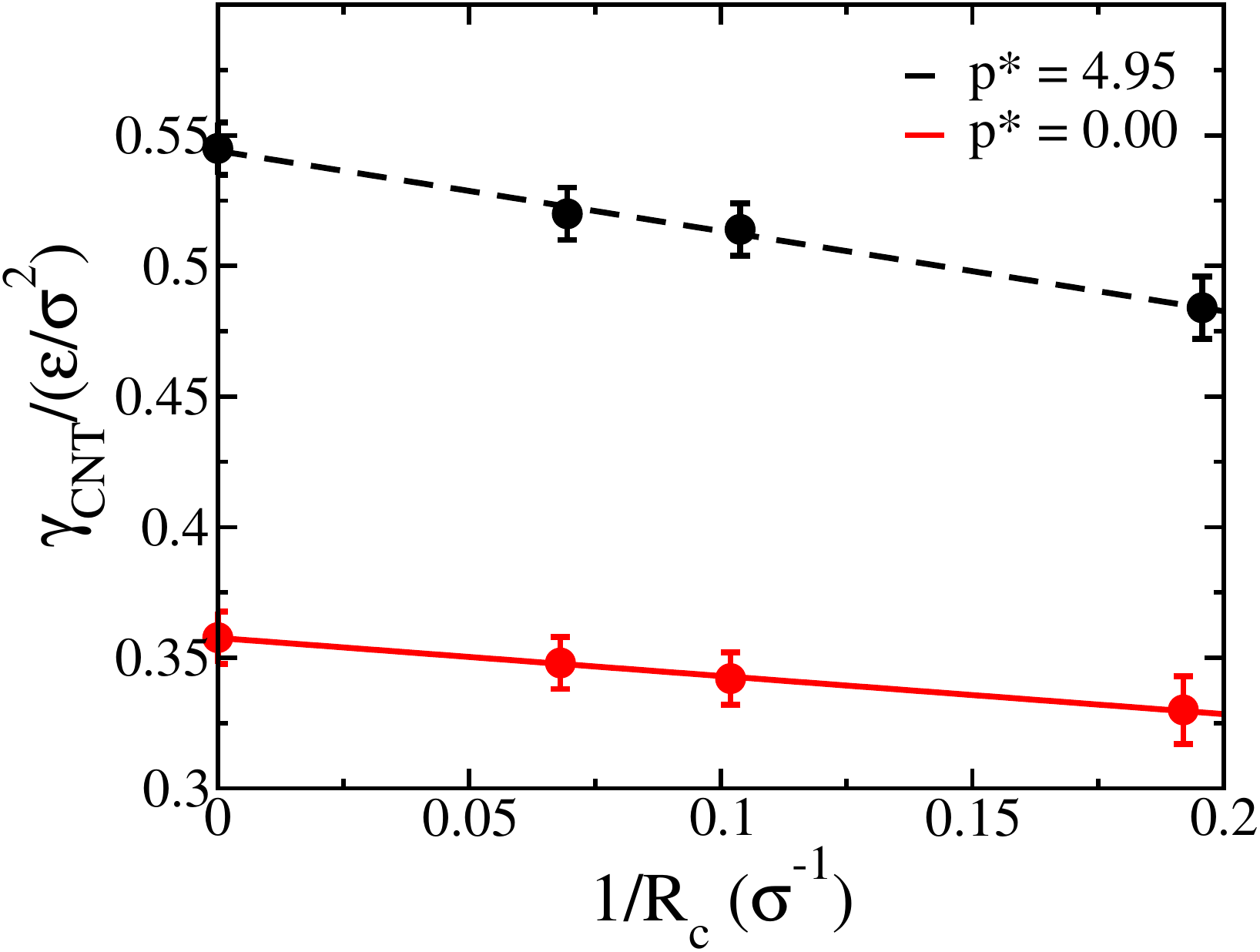} b)
    \includegraphics[width=0.4\linewidth]{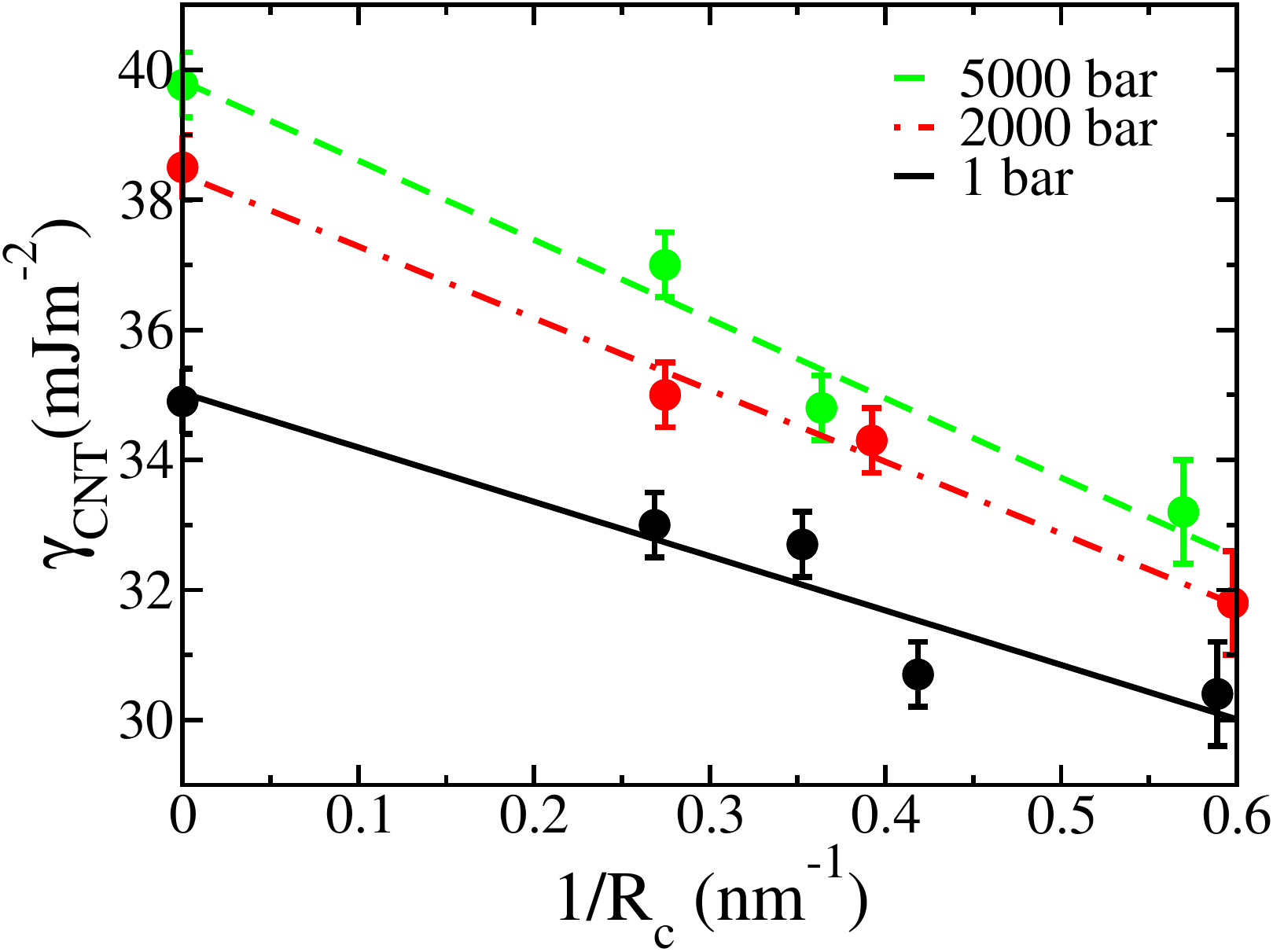} c) 
    \includegraphics[width=0.4\linewidth]{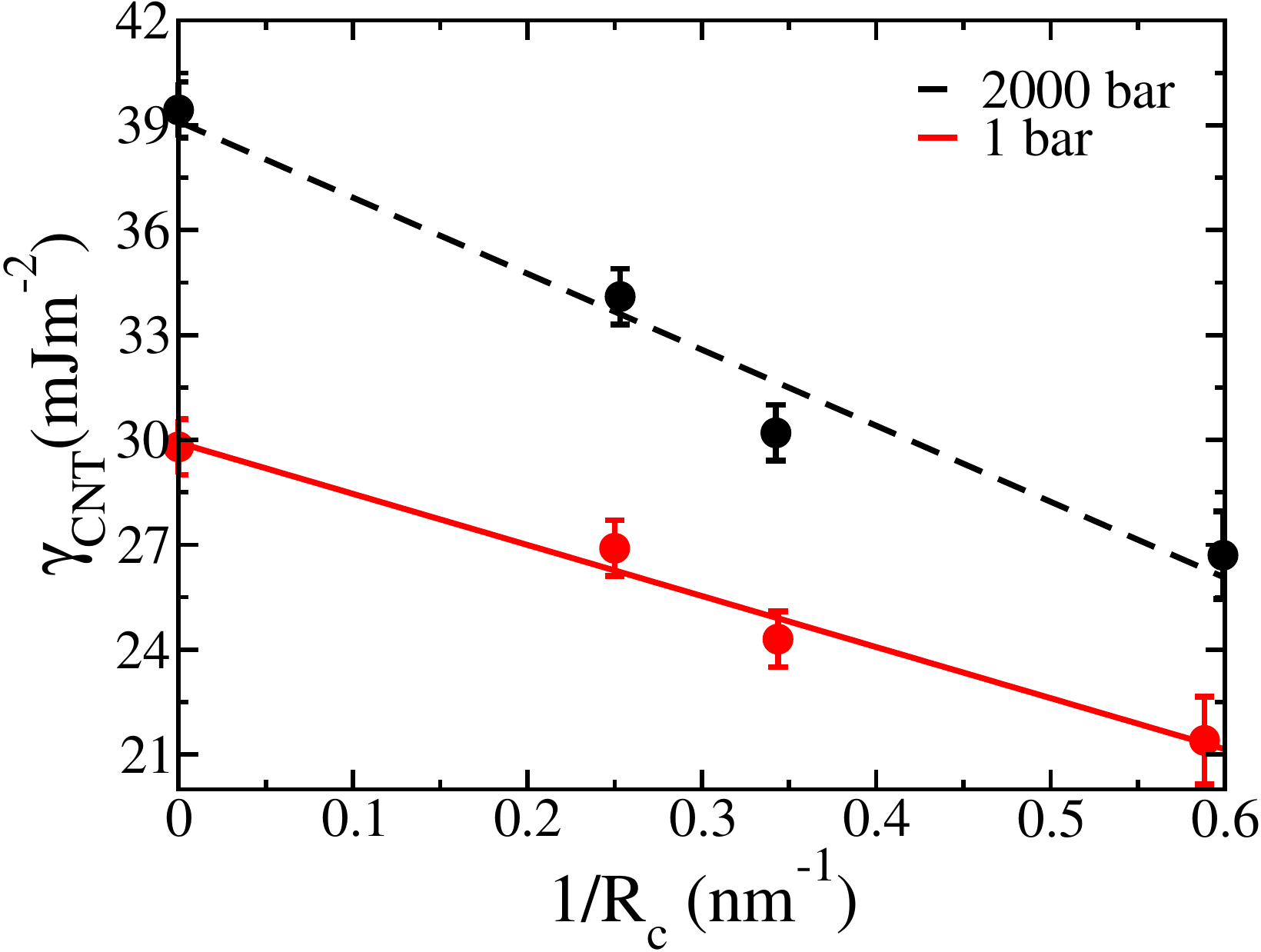} d)
    \caption{ Variation of $\gamma_C$ with the radius of the cluster $1/R_c$ for
     (a) HS, (b) LJ (two isobars), (c) mW (three isobars), and (d) TIP4P/ICE (two isobars). Reproduced with permission from Ref.\cite{montero2019interfacial}. Copyright 2019  by American Institute of Physics. }
    \label{variacion_gamma_radio}
\end{figure}

$\gamma$ can \emph{only} be defined when the system is at equilibrium (i.e., when $T$ and $\mu$ are homogeneous and the divergence of the pressure tensor is zero). Note that $\rad_C$ is not an independent variable, since for each value of $T$ and $\mu$ there is a unique value of $\rad_C$ at which the cluster is in equilibrium with the liquid\footnote{For hard spheres only $\mu$ is required, but this is not the case for Lennard-Jones or other systems}. In \cref{variacion_gamma_radio}, we illustrate the variation of $\gamma_C$ with radius $\rad_C$. As shown, $\gamma_C$ changes with the actual size of the spherical solid cluster (this reflects a physical change, not merely a notational one). The capillarity approximation should be reconsidered in the thermodynamic description of curved interfaces, not only for liquid-solid interfaces but also for liquid-liquid interfaces. Recent simulation evidence overwhelmingly supports this perspective \cite{gamma_hs_frenkel,filion_hs,lau2015surface,bindertolman,montero2019interfacial,kashchiev_2020}

The change in $\gamma_C$ with curvature can be effectively described by an expression first proposed by Tolman \cite{tolman1949effect}:
\begin{equation}    
    \gamma_C = \gamma_0 \left( 1 - \frac{2\delta}{\rad_C} \right),   
\label{eq:Tolman}   
\end{equation}
where $\gamma_0$ represents the value at the planar interface, while $\delta$ has units of length. When relating $\gamma$ at a specific $T$ and $P$ to the value at a planar interface, one can maintain constant $P$ (traversing along an isobar) or keep constant $T$ (traversing along an isotherm). The isotherm path was the choice made by Tolman, which is why $\delta$ is referred to as the Tolman length. Although originally proposed for liquid-liquid interfaces, the Tolman equation has also been useful for crystalline nuclei \cite{montero2019interfacial, montero2020interfacial, granasy2000cahn}. The inclusion of higher-order (quadratic) terms in the expansion has been discussed \cite{baidakov2019spontaneous}, along with its application to isobaric paths \cite{granasy1999cahn, schmelzer2019entropy}. The physical interpretation of $\gamma_0$ in \cref{eq:Tolman} for the solid-liquid interface remains somewhat ambiguous. As $\rad_C$ approaches infinity, the system converges to a planar interface; yet, as we discussed earlier, the value of $\gamma$ for a planar interface varies depending on the specific crystallographic plane considered \citep{baidakov2012crystal}. Furthermore, as a solid cluster grows, it tends to form facets rather than a smooth spherical interface \cite{binder2017}. In practice, the value of $\gamma_0$ is generally close to the average of $\gamma$ for planes with lower Miller indices, but more research is necessary to fully understand this phenomenon.

\begin{figure}
    \centering
    \includegraphics[width=0.6\linewidth]{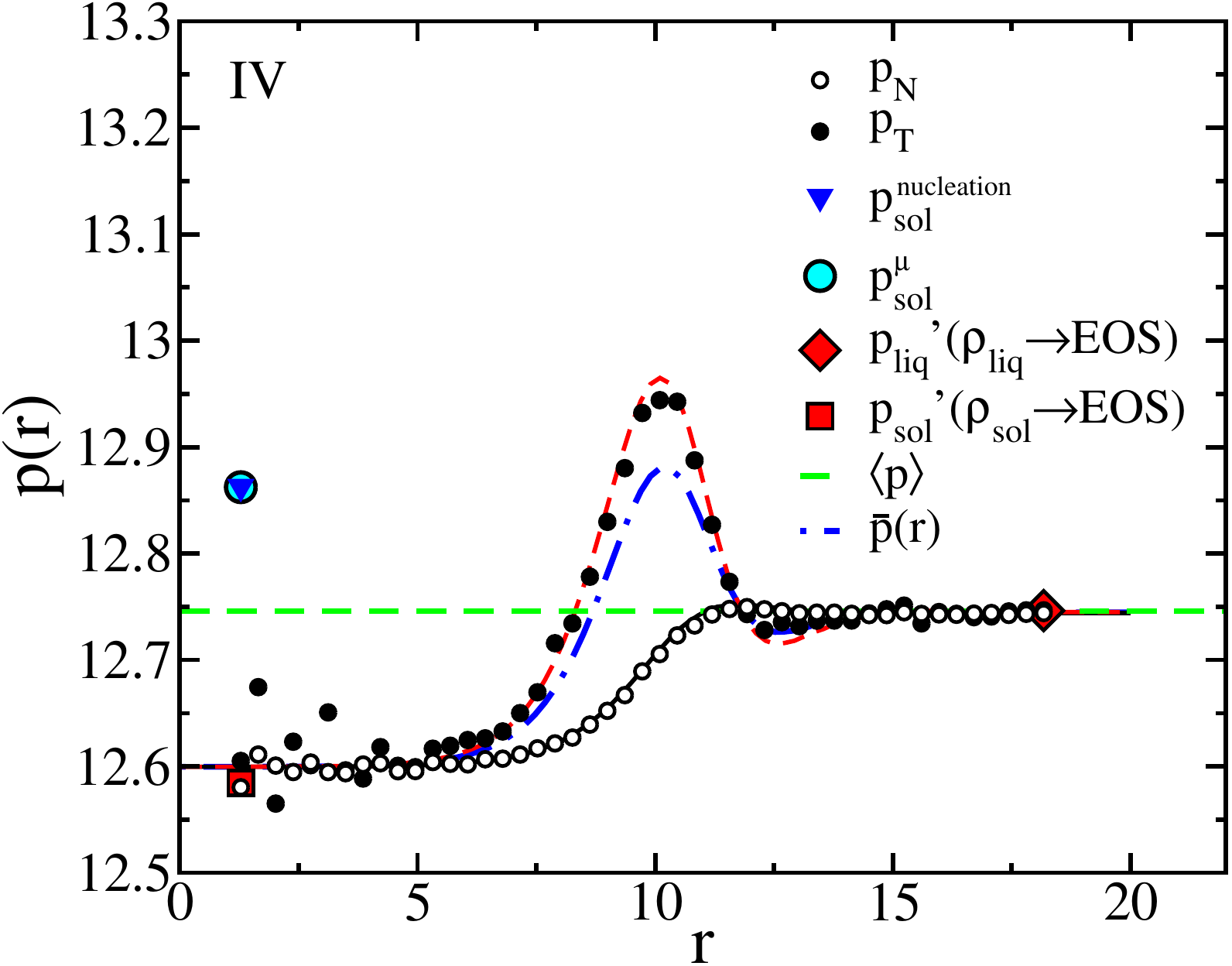}
    \caption{Normal $P_N$ and tangential $P_T$ components of the pressure tensor for a spherical solid cluster of pseudo hard spheres in equilibrium with the liquid at constant $N$, $V$, and $T$. The pressure components are shown as a function of the distance (r) to the center of mass of the cluster.  We refer for the exact definition of all the other symbols to the original reference \citep{montero2020young}. Reproduced with permission from Ref.\citep{montero2020young}. Copyright 2020  by American Institute of Physics.}
    \label{pressure_tensor}
\end{figure}

In all previous equations, $P_l$ refers to the pressure of the external liquid phase. However, what value should be used for $P_s$ (the pressure of the solid phase) in \cref{rowlinson_book_1}? In \cref{pressure_tensor}, the tangential and normal components of the pressure tensor (for a pseudo hard sphere system) are displayed as functions of the distance from the center of the solid cluster illustrated in \cref{snaphot}. As shown, the pressure inside the solid cluster is lower than in the external liquid phase. This result has been further corroborated for pure hard spheres \cite{de2024statistical} and was initially observed in Lennard-Jones solid clusters \cite{gunawardana2018theoretical}. In addition, it was implicitly described via the density of the nuclei of the hard spheres \cite{cacciuto2004breakdown,cacciuto2005stresses,richard2018crystallization}. This anomalous behaviour is not present in liquid-liquid interfaces, where the pressure of the internal spherical liquid phase is always greater than that of the external liquid phase. The lower pressure of the solid, as indicated in \cref{pressure_tensor}, leads, according to \cref{eq:YL_naive}, to a negative value of $\gamma$, which is a non-physical result.

Tolman suggested the solution to this issue following Gibbs' original work, although initially only in the context of small liquid droplets. Following Tolman and Gibbs, we can define $P_s$ as the pressure of a perfect bulk solid (without defects or strain) that has the same chemical potential as the external liquid phase. This definition of $P_s$ is referred to as the thermodynamic pressure $P_s^{\mu}$ (i.e., the pressure of a perfect bulk solid with the same chemical potential $\mu$ as the external liquid phase). Recent findings indicate that this definition should be applied not only to small clusters but also to any spherical solid cluster, regardless of its size. It is important to realise that, in the previous formalism, the properties of a reference solid are utilized rather than those of the actual solid. The use of reference systems is common in thermodynamics; for example, the reference state of the solute in the thermodynamics of mixtures follows similar principles. These properties of this ``reference solid" are necessary for defining $\gamma$ in a curved solid-liquid interface. 

How is it possible to have two solids with the same value of $\mu$, one with $\press_s^{\mu}$ and the other with the actual mechanical pressure of the cluster, $\press_s^{mech}$? The first represents a reference bulk solid without defects at $\press_s^{\mu}$, while the second is the actual solid, which contains vacancies and / or strain, yet maintains the same chemical potential at $\press_s^{mech}$ as the external liquid phase. These internal degrees of freedom could be incorporated into a thermodynamic description of the solid \citep{mullins}, but the solution of using a reference bulk solid is both simple and elegant. The extension of the Gibbsian formalism to explicitly account for the additional state variables arising from the possibility of strained states and defects within the spherical interface was addressed by Mullins \citep{mullins1984thermodynamic}, who already noted that the actual nucleus is not bulk in nature. This approach was later applied in the context of simulations of hard-sphere systems \citep{cacciuto2005stresses,cacciuto2004breakdown}. Mullins expanded the solid variables in terms of unit cell volume, number of unit cells, and number of components per unit cell. Surprisingly, the approach of Mullins has received little attention since it was suggested in 1984. However, it has inspired recent work on the statistical mechanics of a crystalline nucleus of hard spheres in liquid \citep{de2024statistical}, leading to conclusions in agreement with Ref. \citep{montero2020young} while providing further insight into the interfacial stress of the system and the role of vacancies.

Some authors have used a bulk solid without defects at $\press_s^{mech}$ as the reference state for the solid; however, this choice should be avoided \citep{cahn1980surface, cammarata2008generalized} because this reference solid will have a chemical potential different from that of the liquid. This represents a conceptual mistake, as the system is at equilibrium and the chemical potential must be homogeneous \citep{rowlinson1993thermodynamics, perego2018chemical}. Although we have focused on the Young-Laplace equation, other equations commonly used in the literature are influenced by similar reasoning; specifically, the value of $\gamma$ varies with $\rad$ and also depends on the arbitrary choice of the dividing surface. For example, this is true for the Gibbs-Thomson equation, which describes the freezing point depression under confinement and also incorporates $\gamma$ for a curved interface \citep{jackson1990melting}. A source of confusion may arise from the dividing surface proposed by Gibbs, which makes the definition of the interfacial free energy arbitrary. However, an approach that does not come with this limitation and has previously been applied to solids is that proposed by Cahn \citep{cahn1978thermodynamics}, and thoroughly discussed in \cref{sec:Gibbs-Cahn}. However, the Cahn approach has barely been applied to the curved solid-liquid interface \citep{gunawardana2018theoretical,Martin2020surface,martin2022inside}. 

\begin{figure}
    \centering
    \includegraphics[width=0.9\linewidth]{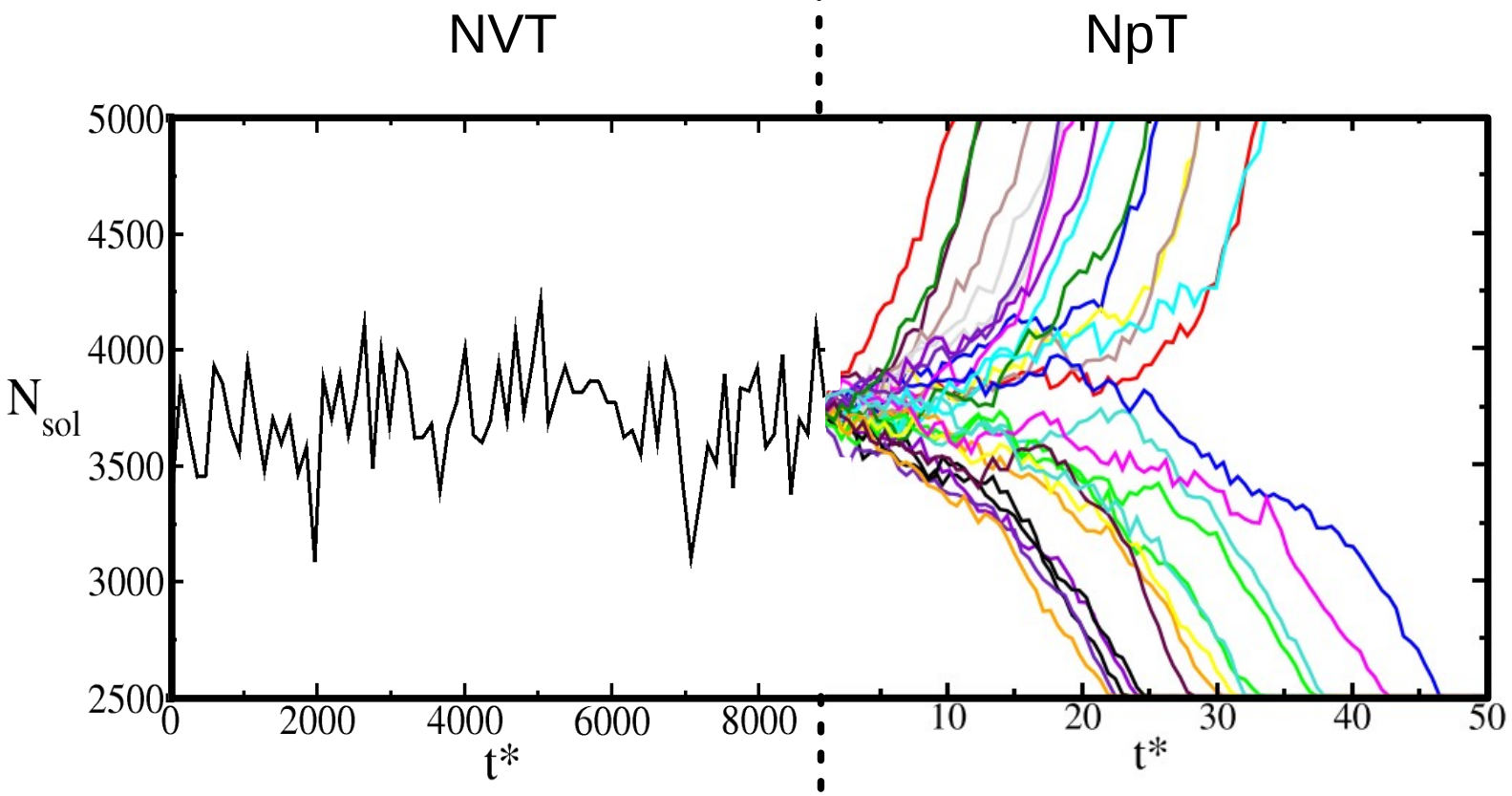}    
    \caption{Trajectories in the NPT ensemble from a configuration of the stable solid cluster in the NVT shown in \cref{snaphot}. Results shown were obtained for the hard sphere potential introducing a spherical solid cluster of size $N_{sol}$ in the fluid phase.  Adapted reproduction with permission from Ref.\citep{montero2020interfacial}. Copyright 2020 by American Chemical Society.} 
    \label{cluster_de_equilibrio_es_critico_a}
\end{figure}

\begin{figure}
    \centering
    \includegraphics[width=0.6\linewidth]{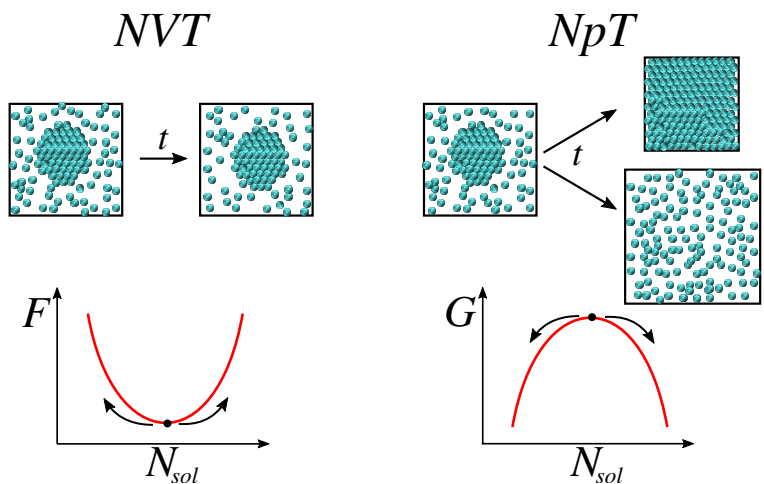}
    \caption{ Sketch showing a stable solid cluster in the NVT ensemble (minimum in $\helm$, called $F$ in the figure) corresponding to a saddle point in the NPT ensemble corresponding to a critical cluster.
    Reproduced with permission from Ref.\citep{montero2020interfacial}. Copyright 2020 by American Chemical Society.}
    \label{cluster_de_equilibrio_es_critico_b}
\end{figure}

Can these stable spherical clusters (in the NVT ensemble) provide insights into nucleation? What happens when we switch to the NPT ensemble using the average pressure obtained during the NVT simulation? In \cref{cluster_de_equilibrio_es_critico_a}, we illustrate that upon changing the ensemble, these stable clusters either melt or grow, each process occurring approximately half of the time. In other words, they behave as critical clusters. It is important to note that in both ensembles, they are at equilibrium (i.e., both temperature and chemical potentials are homogeneous); however, they exist in stable equilibrium in the NVT ensemble and unstable equilibrium in the NPT ensemble. This is represented in \cref{cluster_de_equilibrio_es_critico_b} where the same system is at a minimum in $\helm$ in the NVT ensemble and at a maximum in $G$ in the NPT ensemble or $\Omega$ in the grand-canonical ensemble \citep{oxtoby_evans_1988}. \Cref{cluster_de_equilibrio_es_critico_a,cluster_de_equilibrio_es_critico_b} have significant implications because they connect the thermodynamics of curved interfaces with the nucleation realm. 
 
We will now summarize some of the notation and ideas reported in \cref{sec:CNT} for nucleation, as we want to discuss CNT in light of relaxing some of the approximations considered in the previous section. In particular, we will highlight the role of the curved interface, which was neglected in \cref{sec:CNT}. 
We start by reporting the expression of the nucleation rate $J_{CNT}$ in a slightly different form than that discussed in \cref{sec:CNT}, where the main difference is the expression of the Zeldovich factor (see \cref{sec:CNT}, \cref{eq:Zldf}):
%
%
%
\begin{equation}
        J_{CNT}  =  \rho_l \sqrt{ \frac { |\Delta G_{crit}''|} {(2 \pi k_BT)} }  f^{+} \exp \left(- \frac{\Delta G_{crit}}{k_BT}\right)
        = \rho_l Z f^{+}  \exp \left(- \frac{\Delta G_{crit}}{k_BT} \right),
        \label{eqt:J}
\end{equation}
\noindent where again $\rho_l$ represents the density of molecules in the liquid phase, $Z$ denotes the Zeldovich factor, $f^{+}$ is the attachment rate, and $\Delta G_c''$ indicates the curvature (that is, the second derivative) of the free energy profile at the critical maximum, $\Delta G_{crit}$.

Computer simulations enable the testing of CNT, since the quantities in \cref{eqt:J} can be evaluated numerically. Pioneering studies by Frenkel \etal \cite{Nature_2001_409_1020, ARPC_2004_55_333, N_2004_428_00404, PRL_2004_93_068303} demonstrated that the free energy profile as a function of the size of the largest solid cluster could be determined using the umbrella sampling technique. From this energy profile, one can determine $\Delta G_{crit}$ and $Z$, while $f^{+}$ can be obtained from additional simulations that estimate the diffusive behaviour of the cluster at the top of the barrier. This technique has proven to be highly successful in accurately estimating the values of $J$ for various systems, including hard spheres \cite{auer_frenkel, filion_hs}, Lennard-Jones \cite{seeding_2016}, water (mW) \cite{reinhardt:054501, seeding_2016}, silicon \cite{frenkel_si}, and sodium chloride \cite{nacl_valeriani}, among many others. Estimates of $J$ have been found to be generally quite accurate (except in the cases of two-step nucleation processes) and are not sensitive to the choice of the order parameter used to classify molecules as liquid or solid. It should be noted that the umbrella sampling technique does not require the prediction or definition of any value for the interfacial free energy between the liquid and the solid.

 However, since equilibrium clusters in NVT are critical in NPT, it is possible to connect the thermodynamics of curved interfaces at equilibrium with $\Delta G_{crit}$, that is, with the Gibbs free energy difference (now at constant $N$, $P$, and $T$) between a system with a critical cluster (given by $\helm + \press_l \vol$ of the inhomogeneous systems) and that of a homogeneous liquid (given by $N \mu$). By subtracting both terms one obtains (using the equations of the thermodynamics of curved interfaces of this section):
\begin{equation}
        \Delta G_{crit} = \gamma \area - \vol_s  (\press_{s}^{\mu} - \press_{l} ).
        \label{eqt:final_1}
\end{equation}
This equation is exact. As described above, the values of $\gamma$, $\area$, and $\vol_s$ for the critical cluster depend on the arbitrary choice of the dividing surface, but $\Delta G_{crit}$ does not depend on this choice. By setting the notational derivative of $\Delta G_{crit}$ to zero, one recovers \cref{eq:YLgen}. By selecting the value of $\rad$ (i.e., $\rad_C$ at which $\gamma$ is the minimum $\gamma_C$), one recovers the Young-Laplace equation (\cref{eq:YL_naive}). By using the Young-Laplace equation, one can rewrite \cref{eqt:final_1} as:
\begin{equation}
\Delta G_{crit} =  \frac{1}{3}  \gamma_C \area_C = \frac{1}{2}  \vol_{C} ( \press_{s}^{\mu} - \press_{l} ) =  \frac{16  \pi (\gamma_C)^3} {3  (\press_{s}^{\mu}-\press_{l})^2  }
      \label{freeenergybarrier}
\end{equation}
which is an exact result (being $\area_C$ and $\vol_C$ the area and volume of the solid critical cluster evaluated at the surface of tension), already known to Gibbs. The free energy barrier for nucleation (which is needed to determine $J$ within CNT) is therefore one third of the interfacial free energy of the critical solid cluster (when choosing the radius at the surface of tension, which is always the recommended choice). Now, the connection between nucleation and $\gamma$ has been established (through $\Delta G_{crit}$). We shall omit the $C$ subscript in what follows, but it should be understood that the default choice is the surface of tension for $\gamma$, $\area$, and $\vol_s$. Notice that \cref{freeenergybarrier} holds only when equilibrium holds (i.e., for the critical cluster) and not for a cluster of arbitrary size.

Some confusion about \cref{eqt:final_1} should be clarified. Both \cref{eqt:final_1} and \cref{freeenergybarrier} are exact (that is, they contain no approximations), but they are exact only for the critical cluster (which is at equilibrium and where thermodynamics holds). They are obtained by using a rigorous thermodynamic formalism. It is tempting to assume that \cref{eqt:final_1} is valid for solid clusters of any size ($\rad$) to write:
\begin{equation}
        \Delta G = \gamma \area(\rad)  - \vol_s(\rad)   ( \press_{s}^{\mu} - \press_{l} )
        \label{eqt:final_pedagogical}
\end{equation}
but this is not exact, as clusters of size different from the critical cluster are not at equilibrium and therefore thermodynamics does not hold. However, if $\gamma$ and $\press_s^{\mu} - \press_l$ are taken as constants and set to zero the first derivative of $\Delta G$ with respect to $\rad$, then the correct \cref{freeenergybarrier} is recovered. This implies avoiding the formalism of the thermodynamics of curved interfaces, but it is not rigorous, and in fact leads to wrong values if not applied to the critical cluster, so that \cref{eqt:final_pedagogical} should be used for pedagogical purposes only.

The curvature at the top of the free energy profile, which is not available from thermodynamic reasoning and is needed to evaluate $J$ within the CNT, can be nevertheless estimated by leveraging \cref{eqt:final_pedagogical}. Indeed, the latter equation can be used to describe the free energy profile at the top of the barrier, making it possible to obtain an estimate of the Zeldovich factor $Z$ (which we introduced in \cref{sec:CNT}, see \cref{eq:Zldf}):
\begin{equation}
        Z = \sqrt {\frac{ ( \press_{s}^{\mu} - \press_{l} ) } { 8 \pi^2 k_BT \rho_s^2 \rad_C^3 }  } =  \sqrt {  \frac{ ( \gamma ) } { 4 \pi^2 k_B T \rho_s^2 \rad_C^4 }  }
        \label{zeldovich_pedagogical}
\end{equation}
where $\rho_s$ is the number density of the solid. As shown, $\gamma$ also contributes to this approximate expression of the Zeldovich factor, although its primary impact on nucleation arises through the free energy barrier. It is important to note that $Z$ is dimensionless. The use of $( \press_{s}^{\mu} - \press_{l} )$ in the thermodynamic formalism is recommended; this formulation was proposed by Gibbs and integrates naturally into the thermodynamics of curved interfaces \cite{gibbs1948collected,kashchiev_2020}. Although it is not difficult to evaluate $( \press_{s}^{\mu} - \press_{l} )$, its value is sometimes computed directly with an approximation: Assuming that the solid is incompressible (which implies that its density does not change with pressure), one can show that $( \press_{s}^{\mu} - \press_{l} ) \simeq \rho_{s} \Delta \mu$, where $\Delta \mu$ denotes the difference in chemical potential between a bulk liquid and a bulk solid, at the pressure of the liquid phase $\press_{l}$. By making this substitution, two equations presented earlier in this review are obtained, with the derivation's origins now clearer:

\begin{equation}
        \Delta G_{crit} = \frac{16  \pi (\gamma)^3} {3  (\rho_{s} \Delta \mu)^2      }
    \label{eqt:final_10yes}
\end{equation}

\begin{equation}
        \Delta G = \gamma \area - \vol_{s} \rho_{s} \Delta \mu = \gamma \area - N_{s}  \Delta \mu
        \label{eqt:final_1yes}
\end{equation}
where $N_s$ represents the number of solid particles in the largest solid cluster. The key takeaway from this section is that $J$ can be estimated accurately if $\Delta G_{crit}$ is determined correctly. $\Delta G_{crit}$ is equivalent to one third of the product of $\gamma_C$ and $\area_C$ of the critical cluster, and this relationship is exact.

In umbrella sampling, as well as in metadynamics, $\Delta G_{crit}$ is computed directly without relying on specific thermodynamic definitions. In the seeding technique, the point at which a cluster becomes critical is determined, which remains independent of the chosen order parameter. To estimate the radius of the spherical cluster at the surface of tension (i.e., $\rad_C$), at which the formalism holds true, it is essential to use a robust order parameter. This should yield a value for $N_{crit}$ (the number of solid particles in the critical cluster), leading to accurate estimates of $\rad_C$ through the relationship:
\begin{equation}
\rad_C =  ( 3 N_{crit} / (4 \pi  \rho_s ))^{1/3}.
\label{eqt:final_R}
\end{equation}

Although $f^{+}$ should be determined in computer simulations, a quite accurate estimate can be obtained in the case of the freezing of a pure substance as $f^{+}= 24 D N_{crit}^{2/3}/\lambda^2$ where $D$ is the diffusion coefficient of the molecules in the liquid phase and $\lambda$ is of the order of one molecular diameter. 

\begin{figure}
    \centering
    \includegraphics[width=0.44\linewidth]{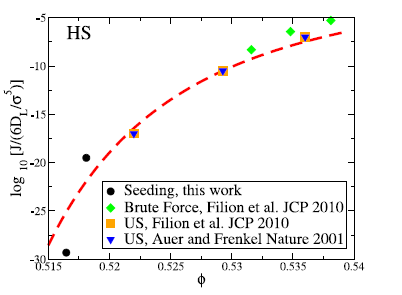} a)
        \includegraphics[width=0.44\linewidth]{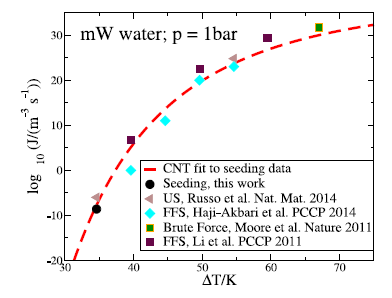} b)
    \caption{ a) Nucleation rates of a) HS and b) mW obtained from seeding compared to results obtained from brute force simulations, Umbrella sampling, and forward flux sampling. Reproduced with permission from Ref.\citep{montero2022thermodynamics} a) and Ref.\cite{seeding_2016} b). Copyright 2022 and 2016 by the American Institute of Physics. } 
    \label{J_seeding}
\end{figure}

Let us conclude by presenting results for the nucleation rate. In \cref{J_seeding} a), we show the estimates of the nucleation rate for hard spheres, while \cref{J_seeding} b) displays the results for water using the mW model. These estimates are obtained from \cref{eqt:J}, leveraging computer simulations to determine $\rho_l$, $f^{+}$, and a suitable order parameter for estimating the size of the critical solid cluster, $N_{crit}$, and consequently $\rad_C$. The results are compared with those obtained from umbrella sampling, brute-force simulations, and forward flux sampling. As illustrated, \cref{eqt:J} accurately estimates $J$ for these two systems. It is important to note that the value of $\gamma$ varies with the radius of curvature, as previously shown in \cref{variacion_gamma_radio}.

In summary, when the capillarity approximation is disregarded (which is often ill-defined for solids, as even for planar interfaces at coexistence, the value of $\gamma$ depends on the specific plane), and when utilizing input from simulations, good estimates of $J$ can be obtained. For many systems, having the correct value of $\gamma$ is crucial for producing reliable estimates of $J$. Ultimately, the key lies in employing an order parameter that provides an informed estimate of the radius at the surface of tension of the critical cluster, $\rad_C$.

\section{Concluding Remarks}
This review mainly focused on the determination of the solid-liquid interfacial properties in Molecular Dynamics simulations and its application to different systems, such as curved interfaces and nucleation theory. Despite being a long standing problem (it is about 150 years since Gibbs took it up) interfaces involving solids continue to pose several challenges, both on the theoretical and computational side. These challenges are probably the main reason why the determination of surface free energies in solid-liquid systems can still not be considered a routine calculation, as is the case for liquid-liquid systems. The challenges here are given by the fact that a more sophisticated treatment is needed for solid-liquid interfaces. These sophistications are exemplified by the fact that those attempting to calculate the surface free energy for solids so often turn to Thermodynamic Integration methods. In turn, the fact that such methods are employed means that there is not something like \textit{the} method to calculate the solid-liquid interfacial free energy (as in the liquid-liquid counterpart) but there are several different methods, each with their own merits and difficulties, which need to be chosen based on the particular problem considered. 

It therefore comes as no surprise that there is little penetration of such analysis and methodologies for the determination of surface free energies in solids, preferring more direct (but also less precise) methodologies. The need to learn different methodologies (e.g., thermodynamic integration) and adapt a piece of software for the particular systems considered is a great source of inertia for the proliferation of such techniques. A newcomer in the field that wants to determine the interfacial free energy of a liquid-liquid system will find several resources for this (relatively simple) calculation, which is now routinely done in widely-used MD software packages. The same newcomer facing the problem of determining the interfacial free energy for a solid-liquid system faces a steep learning curve and a plethora of different software, each tweaked for the calculation in particular systems. As others have already pointed out \citep{Koehler2020} (and the authors of this review agree), knowledge dissemination is important it must include the creation of a well-documented and maintained piece of software available to the community. For this reason, part of the scientific endeavour in this field should be devoted to making the use of these methodologies easier for the community. In this spirit, the authors of this review (working on different models related to solid-liquid interfaces) published documented software completed with examples on how to use the methodologies presented here (see the Mold technique \citep{Tejedor2024} with repository available at \citep{Mold}, Cleaving \citep{DiPasquale2024} with repository available at \citep{cleaving} and the Einstein crystal model with repository available at \citep{EinsCry}). 

A topic we did not touch on in this review is the description of solid-solid interfaces. The reason does not lie in the fact that they are not interesting, as they have several important applications, from solid-state batteries \citep{Xu2018batteries,Lou2021} to geological applications \citep{Knight2010}. However, as the passage from the study of liquid-liquid interfaces to solid-liquid ones is dark and full of terrors, moving to solid-solid systems further increases the complications in modelling these systems, particularly for heterointerfaces which would warrant a review on their own. And, this review is already long enough.

The task to determine interfacial properties for solid-solid systems using MD simulations is still in its infancy (the interested reader can find some examples in \citep{Frolov2015,Frolov2016}), but we are sure that the methods and ideas presented here will be the foundations on which new models for the treatment of these more complicated systems will be built.  

\section*{Acknowledgments}

We acknowledge support from the CECAM and CCP5 through the CECAM/CCP5 sandpit grant (EP/V028537/1) awarded to Nicodemo Di Pasquale and Lorenzo Rovigatti

\appendix

\section{Thermodynamic Integration}\label{sec:TI}

Because the direct methods presented in \cref{sec:dirmeth} are mainly based on the Thermodynamic Integration technique, we include here a brief section to recall the main features of this methodology leaving all the details to the excellent references available \citep{Chipot2007,Frenkel2001}.

As we have introduced in \cref{eq:helmgam}, $\gamma$ (under certain conditions) is the difference in the Helmholtz free energy between the state of the system with an interface and the state of the system without it. Unfortunately, we cannot directly calculate the Helmholtz free energy from the MD simulation, as the free energy is not the average of some function of the phase space, but it is related to the canonical partition function $Q(N,V,T)$ according to:
\begin{equation}
    \helm = -k_B T \ln Q(N,V,T)
\end{equation}
with
\begin{equation}\label{eq:defQ}
    Q(N,V,T)=\frac{1}{\Lambda^{3N} N!}\int \exp{\left( -\frac{U(\mathbf{r})}{k_B T}\right)}  \mbox{d}\mathbf{r}
\end{equation}
where $U$ is the interaction potential among the $N$ atoms in the systems, $\mathbf{r}=\{\mathbf{r}_1,\ldots,\mathbf{r}_N\}$ is their position in the physical space, and $\Lambda$ is de Broglie thermal wavelength.
The free energy therefore represents the volume (in the phase-space) accessible to the system, a quantity, as we said, that cannot be directly calculated. 

Even if $\helm$ cannot be directly derived from calculations, we can instead estimate its derivatives. For example, pressure is the derivative of $\helm$ with respect to the volume of the system at constant $N$ and $T$, and pressure is a perfectly legitimate quantity that can be determined in MD simulations.
This latter observation is the insight that allows us to determine the free energy of a system. 

Let us assume that the Helmholtz free energy depends on a generic parameter $\lambda$, and that the value of $\helm$ in which we are interested corresponds to a certain value $\lambda_{fin}$. In this case we can write:
\begin{equation}\label{eq:TIdef}
   \helm(\lambda_{fin}) - \helm(\lambda_{init}) = \int_{\lambda_{init}}^{\lambda_{fin}} \left(\frac{\partial \helm}{\partial \lambda}\right) \mbox{d} \lambda
\end{equation}
where $\helm(\lambda_{init})$ is the value of the Helmholtz free energy at another point identified by $\lambda_{init}$.

In order to use \cref{eq:TIdef} to compute the value of the Helmholtz free energy at $\lambda_{fin}$, two conditions should be met:
\begin{enumerate}
    \item[i] We know at least one value of Helmholtz free energy, indicated here as $\helm(\lambda_{init})$
    \item[ii] We can build a \textit{thermodynamic path} connecting the two states identified by  $\lambda_{init}$ and the state identified by $\lambda_{fin}$
\end{enumerate}
For each point along this path, we can calculate the value of $\left(\frac{\partial \helm}{\partial \lambda}\right)$ (which is a quantity accessible by MD simulations) and, therefore, we numerically evaluate the integral in \cref{eq:TIdef}. 

This is the basic philosophy of thermodynamic integration. However, it may still seem too abstract, so we will specialize the discussion to our problem. \Cref{eq:helmgam} only requires the difference of the Helmholtz free energy to determine $\gamma$. That means that, with reference to \cref{eq:TIdef} we can write:
\begin{equation}\label{eq:TIgamma}
    \gamma = \int_{\lambda_{init}}^{\lambda_{fin}} \left(\frac{\partial \helm}{\partial \lambda}\right) \mbox{d} \lambda
\end{equation}
where now the initial and final points are (generally) the system without and with an interface respectively. We now have to determine the term within the integral in \cref{eq:TIgamma}.  The actual explicit expression of this latter term depend on the thermodynamic path chosen to move from $\lambda_{init}$ to $\lambda_{fin}$, the choice of which is discussed in \cref{sec:dirmeth}, where each methodology presented represents a different thermodynamic path to create a new interface. In general, the parameter $\lambda$ does not need to be a physical quantity (as, e.g., the pressure), but it can also be a parameter on which the interactions among atoms depend: $U(\mathbf{r};\lambda)$. This allows us to write:
\begin{align}\label{eq:ensAv}
    \left(\frac{\partial \helm}{\partial \lambda}\right) & =  \left(\frac{\partial }{\partial \lambda}\right) \left( -k_B T \ln Q(N,V,T; \lambda) \right) = -k_B T  \frac{\partial }{\partial \lambda}\log\left[ \int \exp{\left( -\frac{1}{k_B T}U(\lambda)\right)}\mbox{d}\mathbf{r} \right] \nonumber \\
    & = \frac{\int \left(\frac{\partial U(\lambda)}{\partial \lambda}\right)\exp{\left( -\frac{1}{k_B T}U(\lambda)\right)}\mbox{d}\mathbf{r}}{\int \exp{\left( -\frac{1}{k_B T}U(\lambda)\right)}\mbox{d}\mathbf{r}} = \left\langle  \frac{\partial U(\lambda)}{\partial \lambda} \right\rangle_\lambda
\end{align}
where we dropped the dependence from $\mathbf{r}$ in $U(\mathbf{r};\lambda)$. Therefore, the integrand in \cref{eq:TIgamma} is the ensemble average of the derivative of the interaction potential $U$ with respect to the parameter $\lambda$. Note that the ensemble average itself in \cref{eq:ensAv} depends on $\lambda$ and therefore does not commute with the integral in \cref{eq:TIgamma}. The explicit expression of the ensemble average in \cref{eq:ensAv} is not known. In order to solve the problem in \cref{eq:TIgamma} an MD simulation is performed for a subset of values of $\lambda$ between the end points, and the integral is then calculated numerically, often using a quadrature approximation.

\bibliographystyle{unsrtnat}
\bibliography{bibliography_May23,todo_biblio1,todo_biblio2,carlos_pablo_final,clfrefs,jhhrefs,felipe}
\end{document}